\documentclass[twocolumn, tighten,times, longauthor]{aastex7}
\usepackage{natbib}
\usepackage{rotating}
\usepackage{graphicx}
\usepackage{float}
\usepackage{listings}
\usepackage{xcolor}
\usepackage{gensymb}

\usepackage{booktabs}
\usepackage{multirow}
\usepackage{amsmath}
\usepackage{amssymb}
\usepackage{dsfont}
\usepackage[percent]{overpic}

\shorttitle{LSTM-AE Applied to Stellar Photometry}
\shortauthors{B. D. Hutchinson et al.}

\received{March 8, 2025}
\revised{July 8, 2025}
\accepted{July 23, 2025}
\submitjournal{AJ}

\begin{document}
\title{A {Wavelength-Aware} Unsupervised Learning Approach for Large, Multicolor, Photometric Surveys }

\author[0009-0007-9953-1774]{Bradley D. Hutchinson}
\affiliation{Department of Astronomy, Indiana University Bloomington, 107 S Indiana Ave, Bloomington, IN 47405, USA}
\email{bradleydhutchinson@gmail.com}

\author[0000-0002-3007-206X]{Catherine A. Pilachowski}
\affiliation{Department of Astronomy, Indiana University Bloomington, 107 S Indiana Ave, Bloomington, IN 47405, USA}
\email{cpilacho@indiana.edu}

\author[0000-0002-8878-3315]{Christian I. Johnson}
\affiliation{Space Telescope Science Institute, 3700 San Martin Drive, Baltimore, MD 21218, USA}
\email{chjohnson1@stsci.edu}

\begin{abstract}
Observational astronomy has undergone a significant transformation driven by large-scale surveys, such as the Panoramic Survey Telescope and Rapid Response System (Pan-STARRS) Survey, the Sloan Digital Sky Survey (SDSS), and the Gaia Mission. These programs yield large, complex datasets that pose significant challenges for conventional analysis methods, and as a result, many different machine learning techniques are being tested and deployed. We introduce a new approach to analyzing multiband photometry by using a \textit{long-short term memory autoencoder} (LSTM-AE). This model {provides input-dependent reweighting across passbands on a star-by-star basis, enabling it} to encode patterns present in the stars' spectral energy distributions (SEDs) into a two-dimensional latent space. We showcase this by using Pan-STARRS \textit{grizy} mean magnitudes, and we use globular clusters, labels from SIMBAD, Gaia DR3 parallaxes, and PanSTARRS images to aid our analysis and understanding of the latent space. For  3,112,259 stars in an annulus around the North Galactic Cap, 99.51\% have their full SED shape reconstructed\textemdash that is the absolute difference between the observed and the model predicted magnitude in every band\textemdash within five hundredths of a magnitude. We show that the model likely denoises photometric data, potentially improving the quality of measurements. Lastly, we show that the detection of rare stellar types can be performed by analyzing poorly reconstructed photometry.
\end{abstract}


\section{Introduction} \label{sec:intro}
Observational astronomy is experiencing a transformative shift propelled by large-scale surveys that generate datasets of extraordinary volume and complexity. This change challenges traditional photometric analysis methods, necessitating innovative analytical approaches to navigate the deluge of data. The Sloan Digital Sky Survey \citep[SDSS;][]{Almeida_2023}, the Gaia mission \citep{refId0}, the Two Micron All Sky Survey (2MASS), and the Panoramic Survey Telescope and Rapid Response System \citep[Pan-STARRS;][]{chambers2019panstarrs1surveys} Survey are already pushing the boundaries of existing techniques. Upcoming missions such as the Rubin and Roman Observatories are set to escalate this challenge, with projected daily data collections in the 10-20 terabyte regime, underscoring a critical need for novel data analysis methodologies.

In response, many diverse machine learning methods are being explored to manage and interpret these voluminous datasets effectively. Recent advances, particularly in \textit{convolutional neural networks} (CNNs), have demonstrated significant potential for classifying astronomical objects based on photometric images \citep{10.1093/mnras/stad255}. This paper showcased a CNN-based classification network, SCNet, which distinguished between seven stellar classes using photometric images alone, highlighting the possibility of image-based stellar classification without spectral data. Furthermore, the development of \textit{long-short term memory} (LSTM) networks has shown promise in various predictive tasks, including the prediction of solar flares using parameters dervied from vector magnetograms \citep{Liu_2019}. The use of \textit{autoencoders} have become specifically useful in astronomy for their ability to run on unlabeled data, as demonstrated by \cite{Gheller_2021}, who employed convolutional deep denoising autoencoders to denoise synthetic images of radio telescopes to detect diffused radio sources.

In this context, we describe a new tool for analyzing multiband photometry of stars by integrating LSTM with autoencoders (LSTM-AE) based on \textit{grizy} photometry from Pan-STARRS1 (PS1). This approach diverges from traditional classification methods by leveraging the LSTM-AE's ability to represent the photometry into a two-dimensional latent space, efficiently managing large photometric datasets while preserving the intricate relationships dictated by a star's spectral energy distribution.

In this paper, we show that this latent space offers easy and efficient data interpretation, and we demonstrate the model's capabilities of dimensionality reduction, reconstruction, anomaly detection, and potentially denoising. Since this model can be deployed after training on a subset of data, these capabilities may be valuable for integration into the pipelines of future surveys such as Rubin. As a point of comparison, we compare the LSTM-AE's performance to a standard autoencoder, which has previously been applied to multiband photometry to aid in anomaly detection \citep{Quispe-Huaynasi_2025}.

\section{Data Selection}
\subsection{PanSTARRS}
We have selected PS1 to showcase our methodology because it offers a catalog for source types based on probabilities predicted by neural networks, called PS1-STRM \citep{2021MNRAS.500.1633B}. This allows us to easily remove non-stellar objects while still having an estimate of the possible contamination in our dataset. 

For a first analysis, we chose the field $60^{\circ} \leq b \leq 80^{\circ}$ and $0^{\circ} \leq l \leq 360^{\circ}$, surrounding the North Galactic Cap. Reddening is minimal, and the field is host to a handful of globular clusters--including M3, M53, NGC5466, NGC5053, and NGC4147--which, for stars in those clusters, will remove the distance component that causes the degeneracy between distance and intrinsic luminosity.

The algorithm described by \cite{2021AJ....161....6L} and  \cite{White_2022} that improves Pan-STARRS1 astrometry using Gaia DR3 has been applied to the PS1 database. Astrometry issues near the pole (defined in the PS1 DR1 caveats) are mostly corrected using the algorithm, but new issues arise from the lack of Gaia measurements in certain regions, causing data loss. 

For our dataset, we require a series of cuts:

\begin{itemize}
  \item $Prob_{Star} \geq 99\%$\footnote{Objects with $Prob_{Star} \geq 70\%$ that are within 0.15\degree of a globular cluster are ran through the model post-training to allow us to see the main-sequence turn off in the latent space.}. Our focus is on analyzing photometry of stellar objects; this cut minimizes the presence of non-stellar objects in our dataset.
  \item Stars without photometry for all five photometric bands are removed. This is necessary for the input of the model.
  \item Stars without at least five measurements made in each band are removed to ensure an accurate mean measurement.
  \item Out of all measurements made that contribute to the mean PSF magnitude in each band, 99\% of them must contain zero flags to ensure an accurate mean measurement.
  \item Stars with extreme or non-physical colors are removed to minimize large photometric errors and non-stellar objects. Stars with $g-r \leq 2$, $-1 < g-i < 4$, $-0.5 < z-y < 1.0$, $i-y > -0.75$ are kept.
  \item Errors in the mean photometry in each band must be $\sigma_{grizy} \leq 0.05$. This allows us to have precise measurements to focus on anomaly detection of rare stellar types, not detection of noisy stars. While out of the scope of this paper,  we note that perfect data can be used as a base and then one can manually introduce noise in training to see if and how effectively the model denoises. This is known as a \textit{denoising autoencoder} \citep{Vincent}. In our case, the model could learn noise implicitly, which is reviewed later in Section 5.2.
\end{itemize}

The cuts result in 3,114,181 stars with precise \textit{grizy} photometry. 99.74\% of the remaining stars have precise astrometry (errors within 100 milliarcseconds), which will allow us to match with Gaia and SIMBAD.  In addition, all stars are matched to Gaia DR3 using a search radius of 0.25".  If more than one object is found in that search radius, the star is not allowed in the Gaia subset.  Using Gaia parallaxes, we rejected stars in our Gaia subset beyond 3 kpc or  $\sigma_d > 0.1$ kpc\footnote{These stars are included in the full dataset.}. This is due to larger Gaia DR3 errors at far distances and allows us to have stronger constraints on the stellar types observed at different distances within our magnitude-limited dataset.

\subsection{SIMBAD Matching}
Stars with $\sigma_{\alpha, \delta} < 100$ milliarcseconds are matched with SIMBAD \citep{Wenger_2000} by using a search radius of one arc-second, yielding 70,196 objects. 1,922 non-stellar objects are discarded\footnote{SIMBAD matched stars were based on a previous set of color cuts, leaving one object, a galaxy, not matched and thus not removed.}, leaving 3,112,259 total PS1 stars ready for analysis.

\subsection{Dereddening}

All bands are corrected for reddening using the extinction map from \cite{Schlegel_1998}.  $R_v = 3.1$ is chosen as it corresponds to the standard interstellar extinction law for diffuse Milky Way dust, along with extinction coefficients $k_{g} = 3.172,k_{r} = 2.271,k_{i} = 1.682,k_{z} =1.322,k_{y} = 1.087 $ from \cite{2011ApJ...737..103S}, where they recalibrated the Schlegel, Finkbeiner, and Davis (SFD) extinction map \citep{1998ApJ...500..525S} by measuring reddening with SDSS spectra.  The extinction coefficient is chosen due to the high Galactic latitudes. E(B-V) values range from 1.98$\times$10$^{-3}$ to 9.99$\times$10$^{-2}$, with a mean of 1.89$\times$10$^{-2}$, leading to small corrections to the photometric magnitudes.

\section{Long-Short Term Memory Autoencoder}

\subsection{Long-Short Term Memory (LSTM)}
\label{sec:LSTM}

\begin{figure}[t]
    \centering
    \includegraphics[width=0.45\textwidth]{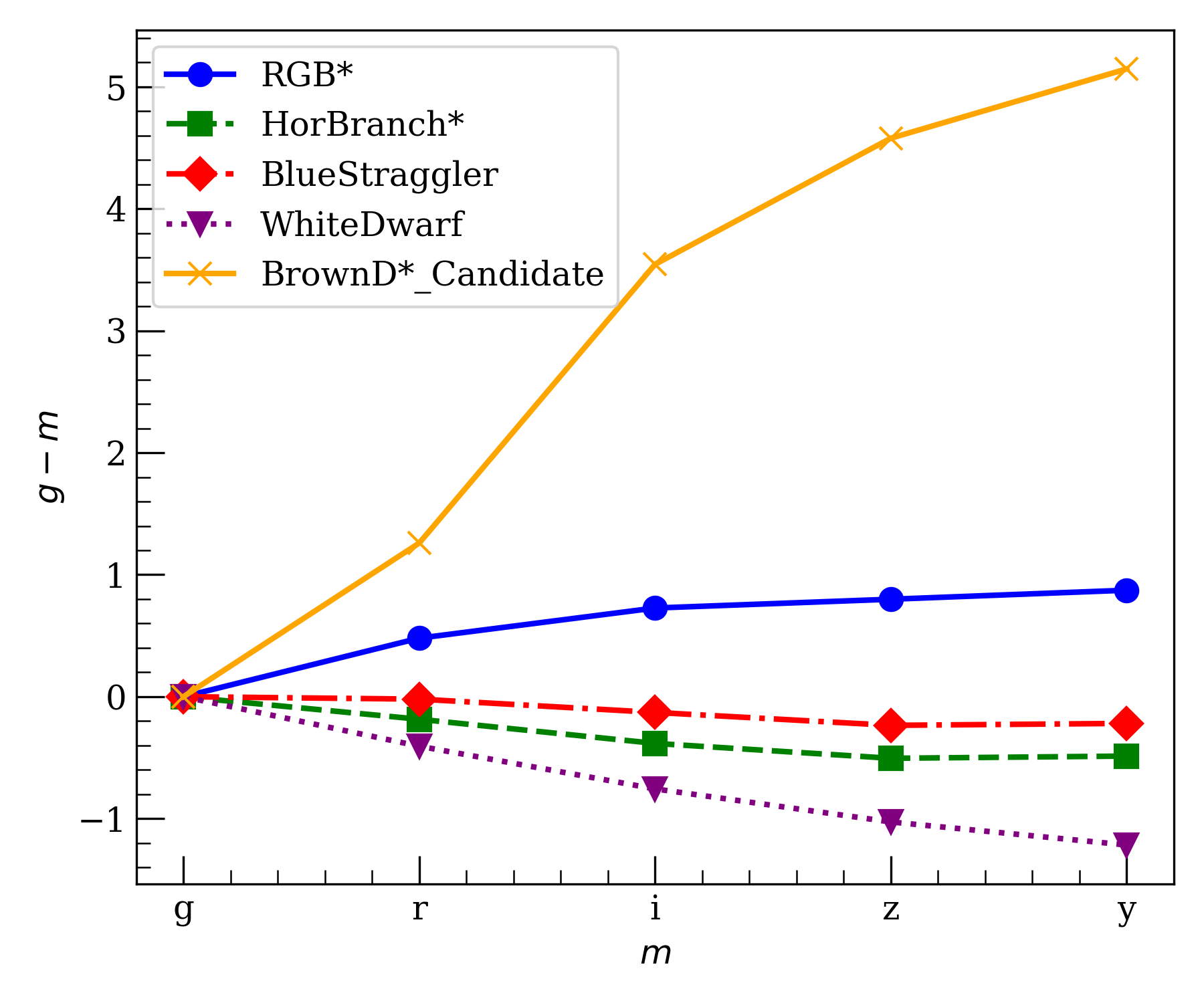}
    \caption{SED shapes of randomly chosen stars from five different stellar types matched from SIMBAD across different photometric bands, highlighting the {wavelength-dependent structure} of the data. The SED shape is normalized to $g$ for all bands. For example, the figure shows that the white dwarf is brightest in $g$ (SED peaked in or before $g$), while the brown dwarf is brightest in $y$ (SED peaked in or beyond $y$).}
    \label{fig:sed_shapes}
\end{figure}

The shape of a star's SED, along with the wavelengths and strengths of various absorption and emissions lines, encodes information about its atmospheric properties such as temperature, surface gravity, and metallicity.  These physical properties can be extracted from an SED using the relative flux measurements of a star when viewed through filters having different effective wavelength ranges. A star's change in brightness over different bands and the steepness of these changes from shorter to longer wavelengths are deeply related to the underlying physical processes and properties that govern its behavior, as shown in Figure \ref{fig:sed_shapes}. {Therefore, it motivates the use of wavelength-aware models that can adaptively emphasize specific passbands for each star.}

\textit{Long-Short Term Memory} \citep[LSTM;][]{6795963} networks, a variant of \textit{recurrent neural networks} \citep[RNNs;][]{Rumelhart1986LearningIR}, address the challenge of learning long-term dependencies.  Traditional RNNs often struggle with retaining earlier information in a sequence, a problem known as the vanishing and exploding gradient problem \citep{279181}.
LSTMs incorporate memory cells that are capable of storing information for long periods of time, that is for many timesteps.  The cell state $\mathcal{C}$ is the core component of the LSTM, acting as its long-term memory. It carries information throughout the sequence of inputs, allowing the network to make informed decisions based on both recent and previous data. The flow of information within the LSTM unit is regulated by gating mechanisms that determine which information to add, remove, or retain in order to minimize the loss function.  We denote time-step {as} $t$, where the gates are defined as following:

\begin{enumerate}
    \item Input Gate $i_t$: Determines which new information is significant enough to be incorporated into the cell state. Significant information contributes to the minimization of the loss function.
    \item Forget Gate $f_{t}$: Ensures the network's memory stays focused and relevant by filtering out information deemed irrelevant. Irrelevant information does not contribute to the minimization of the loss function.
    \item Output Gate $o_{t}$: Dictates what part of the cell state should contribute to the output at each timestep. This selective mechanism ensures that only important information deduced from recent and past data stored in the cell state influences the network's predictions.
\end{enumerate}

We now formulate this mathematically:

\begin{equation*}
i_t = \sigma (W_i[h_{t-1}, m_t] + b_i)
\end{equation*}
\begin{equation*}
f_t = \sigma (W_f[h_{t-1}, m_t] + b_f)
\end{equation*}
\begin{equation*}
o_t = \sigma (W_o[h_{t-1}, m_t] + b_o)
\end{equation*}
\begin{equation*}
\tilde{\mathcal{C}_{t}} = tanh(W_\mathcal{C}[h_{t-1}, m_t] + b_\mathcal{C})
\end{equation*}
\begin{equation*}
\mathcal{C}_t = f_t \odot \mathcal{C}_{t-1} + i_t \odot \tilde{\mathcal{C}_{t}}
\end{equation*}
\begin{equation*}
h_t = o_t \odot tanh(\mathcal{C}_t
)\end{equation*}

where $\sigma$ is the \textit{sigmoid activation function}, $W_\alpha \;(\alpha =i, f, o,\mathcal{C})$ is the weight {matrix} with the respective {gate or candidate state}, $h_{t-1}$ is the output of the previous LSTM block from the previous timestep, $m_t$ is the input at the current timestep, $b_\alpha$ is the bias for the respective gate, $\tilde{\mathcal{C}_t}$  is the candidate for cell state at timestep $t$, $\tanh$ is the \textit{hyperbolic tangent activation function}, $\mathcal{C}_t$ is the cell state at current timestep $t$, $\mathcal{C}_{t-1}$ is the cell state from the previous timestep, and $h_t$ is the output of the current LSTM block.

All gates use \textit{sigmoid activation} as it outputs a value between 0 and 1, corresponding to either blocking or allowing information to pass through the gate. At any passband, $\mathcal{C}_t$ takes into the account the information to forget from the previous timestep plus what information to allow from the current timestep. {In our case, we do not unroll the network across passbands. The model processes the five magnitudes jointly in a single step: a timestep of one with a total of five features. This makes $h_{-1} = \mathcal{C}_{-1} = 0$, the term $f_0 \odot \mathcal{C}_{-1}$ vanishes, and both $i$ and $o$ become functions of the entire photometric vector in the first hidden layer of the encoder.}


This formulation of LSTM for multiband photometry provides two primary advantages:
\begin{enumerate}

    \item {The gates compute input-dependent, per-sample reweighting, allowing the model to emphasize the most informative wavelength regions for each star.  A white dwarf will tend to up-weight blue-sensitive features driven by $g$ and $r$ and down-weight the redder bands, while a cool main-sequence (MS) star will shift its weights towards $i$ and $y$. This capability is important as photometric datasets are imbalanced, dominated by a MS majority. For a rare blue straggler, high weights will be assigned to both the blue and red ends to capture its SED shape. The gates are able to mitigate majority-class bias, capturing distinct SED shapes across diverse stellar types in a heavily imbalanced dataset.}
    \item Photometric data are susceptible to noise. The gating mechanisms allow the LSTM to learn what information is important to retain and what information is not, making it robust to noise. This is important for training a model on real, observational data that is riddled with measurement uncertainties and distortions.
\end{enumerate}

In the case of unlabeled photometric data, training a standard LSTM is not feasible because of the absence of
labels necessary for supervised learning. However, LSTM  units can be used in an unsupervised manner when paired with an autoencoder.

\subsection{Autoencoder (AE)}

An \textit{autoencoder} \citep{Lecun1987,Bourlard1988,Hinton1993,bank2021autoencoders} is a neural network architecture used in unsupervised learning to efficiently encode data as a method for dimensionality reduction and then to decode it (reconstruct the input from the encoded representation) in order to detect anomalies. It consists of two main components: an encoder and a decoder. The encoder's task is to compress the input data into a lower-dimensional space, called the latent space, aiming to capture its principle features. The decoder then attempts to reconstruct the input data from this latent space representation. The objective is to discover a compressed, yet informative representation of the data through the minimization of a loss function that measures the discrepancy between the original and reconstructed outputs.  

 In the context of large observational surveys, where the dataset is multiband photometry, AE's can help distill the complex, non-linear relationships into a more manageable form. This is what separates AE's from other dimensionality reduction techniques, like Principal Component Analysis \citep[PCA;][]{MACKIEWICZ1993303}, which assumes linear relationships between features. Furthermore, the presence of rare stars with uncommon photometric properties will have larger reconstruction errors than their common counterparts, allowing for anomaly detection.

\subsection{Long-Short Term Memory Autoencoder (LSTM-AE)}
We now introduce our model of choice: the \textit{long-short term memory autoencoder} (LSTM-AE). This model combines {a wavelength-aware LSTM} with the unsupervised learning, dimensionality reduction, and anomaly detection prowess of AEs, creating a powerful tool for analyzing multiband photometric data.

The LSTM-AE architecture is designed to capture the intrinsic patterns within the photometry and encode them into a lower-dimensional space. It then reconstructs the original photometry from this space, and monitors itself based on how well it reconstructs the input. This offers a handful of advantages:
\begin{enumerate}
    \item Unsupervised learning: The ability to operate without labels allows the LSTM-AE to freely explore the data, potentially uncovering new insights into stellar phenomena.
    \item Pattern recognition: It excels at recognizing and learning from {wavelength-dependent structure in multiband photometry (see Section \ref{sec:LSTM}), permitting detailed separation of stellar populations based on the underlying SED shapes.}
    \item Dimensionality reduction: By encoding photometry into a lower-dimensional  space, the model simplifies the data, facilitating easier visualization, interpretation, and analysis.
    \item Anomaly detection: The model's reconstruction error can serve as a metric for identifying anomalies in the data, flagging stars that deviate significantly from known patterns.
    \item Denoising: Since the training set is real observational data, the small noise present in the dataset can act as a form of implicit regularization, introducing natural perturbations in the training set, which in tandem with the contraction of the higher input space to a lower input space (another form of regularization, as this prevents the learning of the identity function) pushes the model to learn a robust representation of the dataset that is relevant to astrophysical properties of the stars. This makes it a \textit{regularized autoencoder} \citep{bengio2014representationlearningreviewnew}, and the output of the model, here the reconstructed photometry, could be denoised as the regularization forces the encoder to focus on the underlying shapes of the SEDs, filtering out noise present in the dataset without the need of a \textit{denoising autoencoder} \citep{Vincent}. The addition of LSTM units also aid in denoising, as described in Section 3.1.

\end{enumerate}

\section{Model}
\subsection{Implementation}
The implementation of the model we chose was done through \textit{Keras}  \citep{chollet2015keras}, a high-level API for the Google-developed open-source library \textit{TensorFlow} \citep{tensorflow2015-whitepaper} that is used to develop and deploy machine learning models using Python. It provides simplicity, flexibility, scalability, and optimization, all of which are crucial for astronomers implementing and deploying models on large datasets.  The entire code on the implementation of Keras + Tensorflow is located in the appendix.

\subsection{Data Preprocessing}

Since neural networks typically perform better when the data are scaled near zero, as the step size in gradient descent depends on the scale of the features, we choose to scale our data by dividing all magnitudes by 10, as the range across photometric bands is similar. This is why more complex scaling techniques are not necessary here. Furthermore,  by using either \textit{StandardScaler} or \textit{MinMaxScaler} \citep{pedregosa2018scikitlearnmachinelearningpython} performance may worsen due to the presence of outliers in the dataset, as one would be scaling the magnitudes for each photometric band separately.

\subsection{Architecture}

The model takes on the standard architecture of an AE, with the encoder and decoder both consisting of multiple hidden layers. The encoder takes the input photometry {as an n-dimensional vector at a single timestep}, where we denote the number of bands as $n$, and passes it through three LSTM layers, consisting of four, three, and two nodes respectively.  The two nodes are referred to as the bottleneck which produces the low-dimensional representation of the data, commonly known as the latent space. The latent space is the result of the model recognizing stellar features via patterns present in multiband photometry and representing it into a compact space. The decoder is then used to reconstruct the magnitudes using a mirrored structure of the encoder, consisting of three LSTM layers, three, four, and $n$ nodes, where $n$ must match the input of the model. The LSTM units in the encoder share memory with their mirrored counterparts in the decoder, allowing the model to quickly learn the mapping from the latent space to the reconstructed magnitudes.

Each LSTM layer uses an \textit{Exponential Linear Unit (elu)}  activation function,\footnote{This activation function specifically replaces the $\tanh$ activation function in our mathematical formulation in Section 3.1. This was decided through manual testing of both activation functions on the dataset.}  introducing non-linearity into the model, along with speeding up learning and alleviating the vanishing gradient problem \citep{clevert2016fastaccuratedeepnetwork}. In addition, \textit{Early Stopping} \citep{Prechelt2012} is used as a form of regularization to prevent overfitting, stopping the model once the validation loss has not improved for five epochs. The model can run for up to 200 epochs with a learning rate (lr) of 0.0001 and batch size of 512\footnote{All combinations of lr = 0.01, 0.001, 0.0001, batch-size = 128, 256, 512, loss function of either mean-square error or mean-absolute error were tested as a parameter search, in which we selected lr = 0.0001 and batch-size = 512 due to its convergence and stability.} . \textit{Mean-squared error} (MSE) is used as our loss function:
\begin{equation*}
MSE  = \frac{1}{N_\lambda} \sum_{\lambda=1}^{N_\lambda}\frac{1}{n}\sum_{i=1}^n(m_{i,\lambda} - m^\prime_{i,\lambda})^2,
\end{equation*}
where $N_\lambda$ is the number of passbands, here $N_\lambda=5$, and $(m_{i,\lambda} - m^\prime_{i,\lambda})$
is the difference for a given star $i$ in a certain passband $\lambda$ between the PS observed value and the LSTM-AE's predicted value.
Lastly, we use the \textit{Adam} optimizer, which can adapt parameter learning rates and has been shown to enable faster convergence compared to other optimization methods \citep{kingma2017adammethodstochasticoptimization}.

We run the model five times with random train/validation/test splits and weight initializations to ensure consistent convergence and results, and show the run in this paper that both generalizes well and has an easy to interpret latent space. Since the only meaning behind the two dimensions that make up the latent space is the distances between points, the actual axes have no physical meaning. Each unique run is expected to have completely different numbers that represent the embedding of the SED shape and potentially differing global structure, but which stars end up close together should remain broadly similar across runs.  When presenting the latent space in figures for this paper, we rotate the axes so their meaning is more easily understood by the reader. 
\section{Model Verification \& Generalization}

\subsection{Loss}

\begin{figure}[!t]
    \centering
    \includegraphics[width=0.45\textwidth]{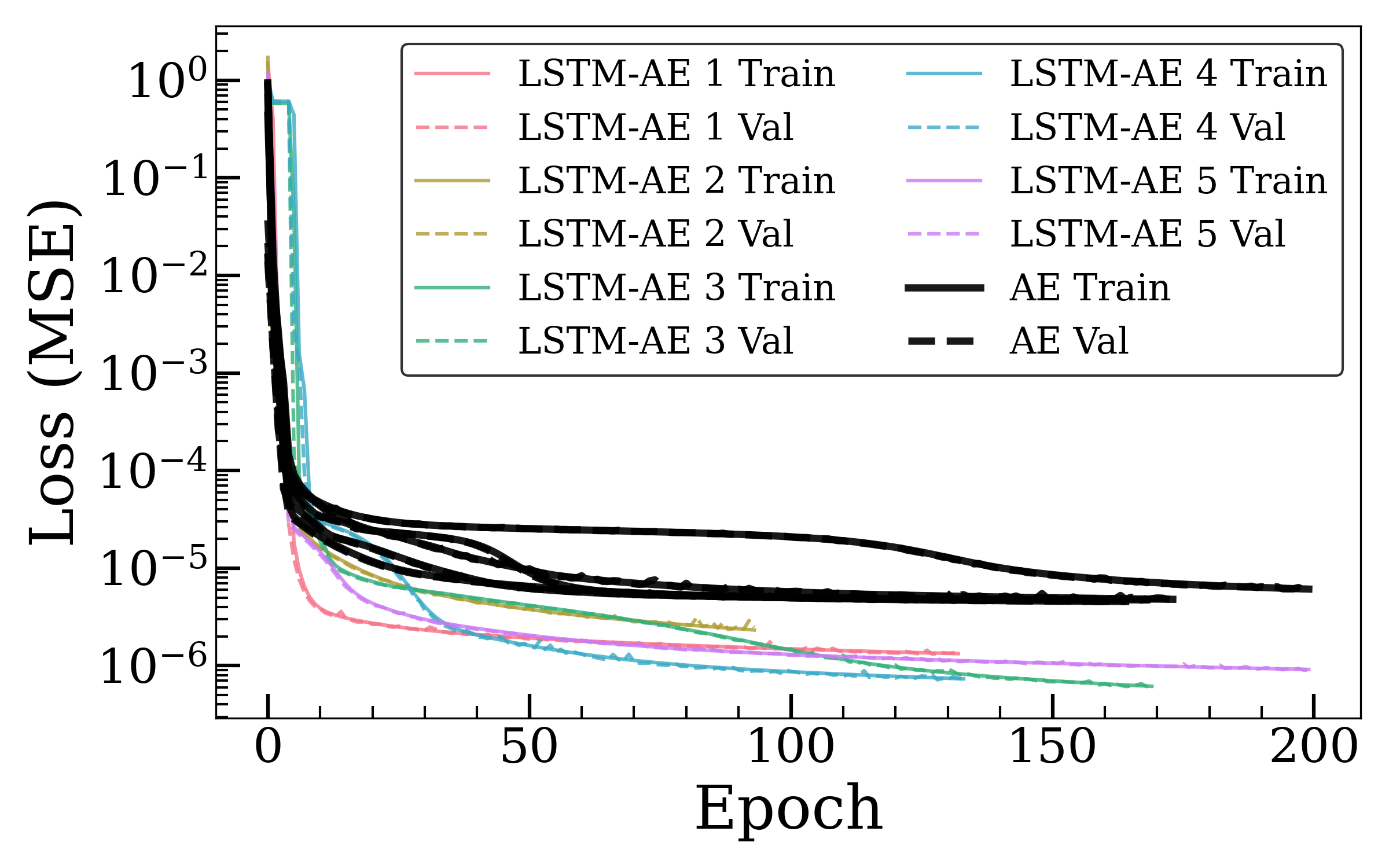}
    \caption{Loss vs. epoch curves are shown for the five LSTM-AE runs in color. Five runs of a standard autoencoder with the same architecture (excluding LSTM units) and hyper-parameters are shown in black, highlighting the superior performance of an LSTM-AE over an AE.}
    \label{fig:loss_curves}
\end{figure}

\begin{deluxetable}{lccc}
    \tablecaption{Experimental results showing the training loss associated with the epoch with the lowest validation loss, the lowest validation loss achieved in each run, and the test loss for different runs.\label{tab:loss_results}}
    \tablehead{
    \colhead{\textbf{Run}} & \colhead{\textbf{Train Loss}} & \colhead{\textbf{Val Loss}} & \colhead{\textbf{Test Loss}} 
    }
    \startdata
    1 & 1.35$\times$10$^{-6}$ & 1.32$\times$10$^{-6}$ & 1.31$\times$10$^{-6}$ \\
    2 & 2.43$\times$10$^{-6}$ & 2.35$\times$10$^{-6}$ & 2.34$\times$10$^{-6}$ \\
    3 & 6.34$\times$10$^{-7}$ & 6.11$\times$10$^{-7}$ & 6.10$\times$10$^{-7}$ \\
    4 & 7.50$\times$10$^{-7}$ & 7.17$\times$10$^{-7}$ & 7.16$\times$10$^{-7}$ \\
    5 & 9.15$\times$10$^{-7}$ & 9.00$\times$10$^{-7}$ & 9.08$\times$10$^{-7}$ \\
    \enddata
    \end{deluxetable}

\begin{figure}[!t]
    \centering
    \includegraphics[width=0.49\textwidth]{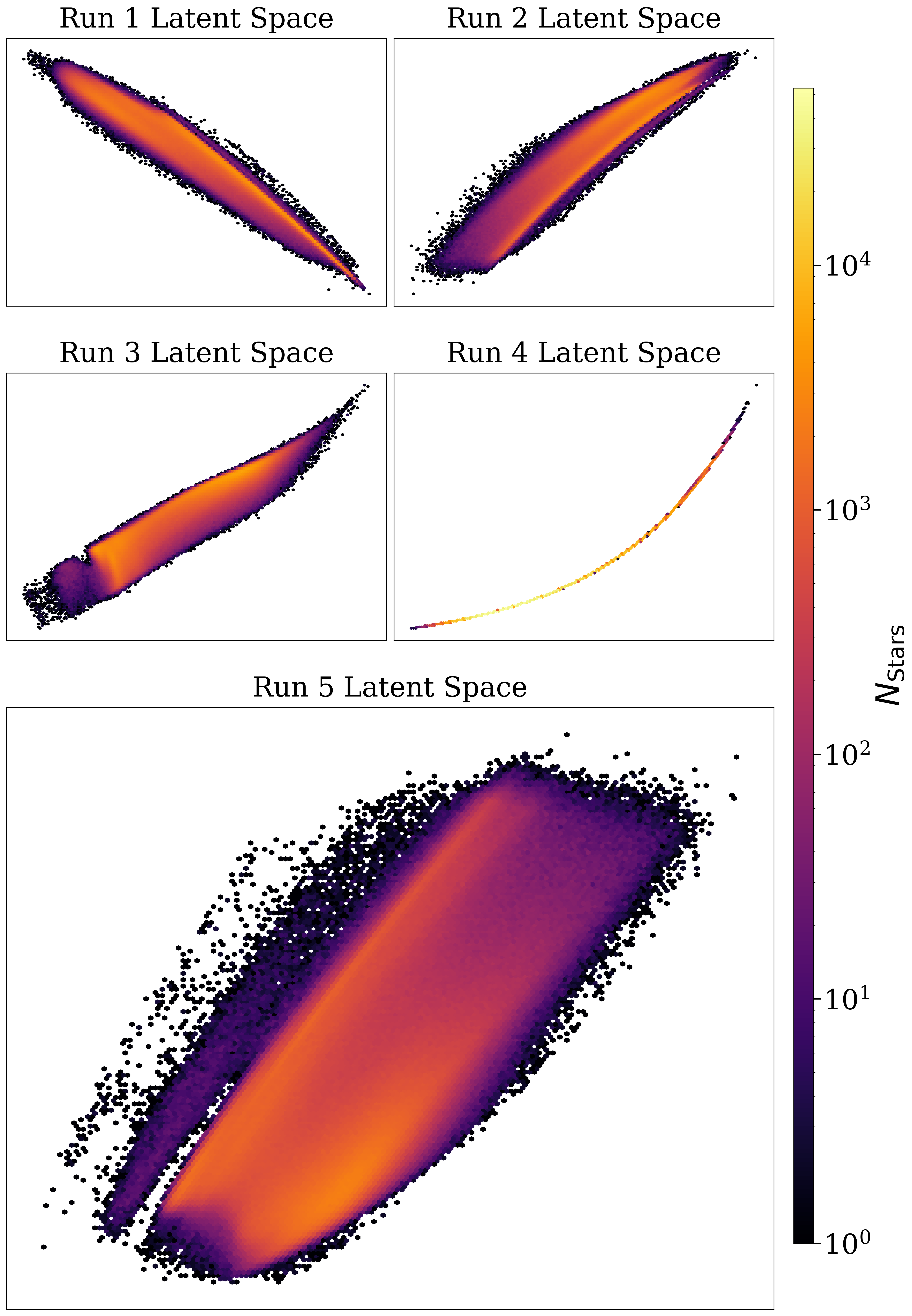}
    \caption{Latent spaces are shown for the five LSTM-AE runs, highlighting the different latent spaces produced by different runs, and how run 5 is the easiest to interpret due to its clear global structure.}
    \label{fig:latent_spaces}
\end{figure}

Figure \ref{fig:loss_curves} shows the loss and validation loss vs. epoch for the training process of both the LSTM-AE and a standard AE, displaying convergence, a low loss, and that the LSTM-AE consistently outperforms the AE by approximately an order of magnitude. Table \ref{tab:loss_results} provides numerical values for the test and final train and validation losses. The variations between losses across runs fluctuate, but the variations are small, which is typical of minor stochasticity rather than instability of the model. The test and validation losses are very similar to the training losses, mostly being slightly below, meaning that the model has no significant overfitting and has robust generalization performance. However, all five runs produce different latent spaces with varying levels of ease of interpretability, as seen in Figure \ref{fig:latent_spaces}. Since we are concerned with accurate reconstruction of photometry, strong generalization, and using the latent space for easy analysis, we choose to show the results of \textit{Run 5} for the remainder of this paper\footnote{If accurate reconstruction and strong generalization are the primary concerns, then the superior model is the one that displays convergence and possesses the lowest validation loss.}.

\subsection{Reconstruction \& Denoising}

\begin{deluxetable}{lccccl}
\tablecaption{The percentage of stars for which the absolute difference between the predicted and observed magnitude exceeds 0.01, 0.025, 0.05 and 0.1 magnitudes in each filter band.\label{tab:stars_band_differences}}

\tablehead{
\colhead{\textbf{Absolute Difference}} & \multicolumn{4}{c}{\textbf{\% of Stars Within Threshold}}\\
\colhead{$|m - m^{\prime}|$} & \colhead{$\leq 0.01$} & \colhead{$\leq 0.025$} & \colhead{$\leq 0.05$} & \colhead{$\leq 0.1$}
} 
\startdata
$|g - g^{\prime}|$   & 82.283 & 99.086 & 99.947 & 99.996 \\ 
$|r - r^{\prime}|$   & 85.441 & 99.358 & 99.978  & 99.998 \\ 
$|i - i^{\prime}|$   & 64.887 & 95.385 & 99.619 & 99.968\\ 
$|z - z^{\prime}|$   & 79.151 & 98.384 & 99.879 & 99.990 \\ 
$|y - y^{\prime}|$   & 74.819 & 97.944  & 99.867 & 99.991 \\ 
\enddata
\end{deluxetable}

\begin{figure}
    \centering
    \includegraphics[width=0.45\textwidth]{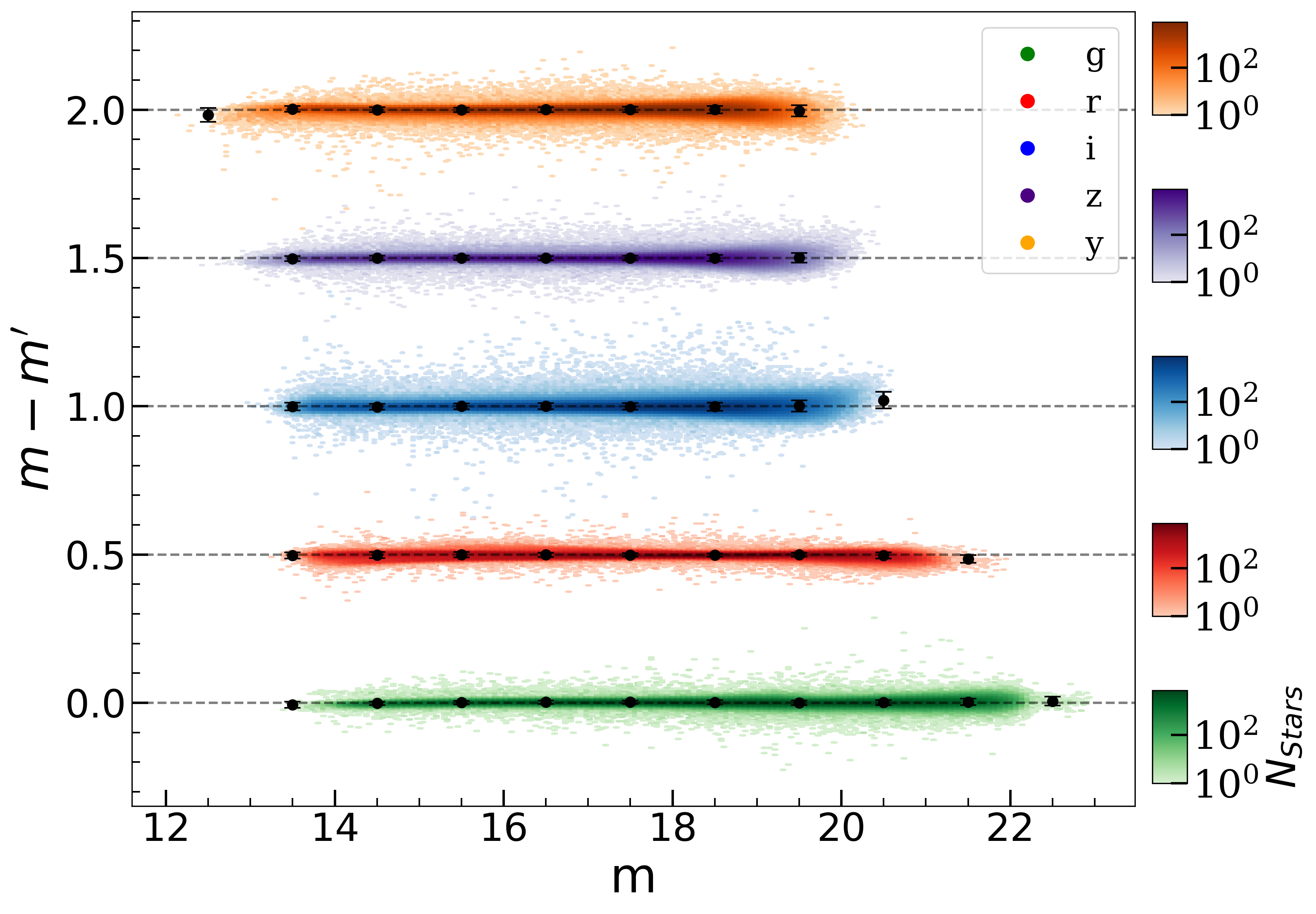}
    \caption{Differences between the reconstructed and observed magnitudes are shown as a function of observed magnitude for all five filter passbands considered here. Each sequential band is shifted vertically by a value of one-half for display purposes. Standard deviation of the differences are shown in bins of 1 magnitude, with showing the value on the plot at the bin center. For example, the standard deviation in the difference for all stars with magnitude between 16 and 17 is shown at 16.5.}
    \label{fig:diff_vs_mag}
\end{figure}

\begin{figure}
    \centering
    \includegraphics[width=0.45\textwidth]{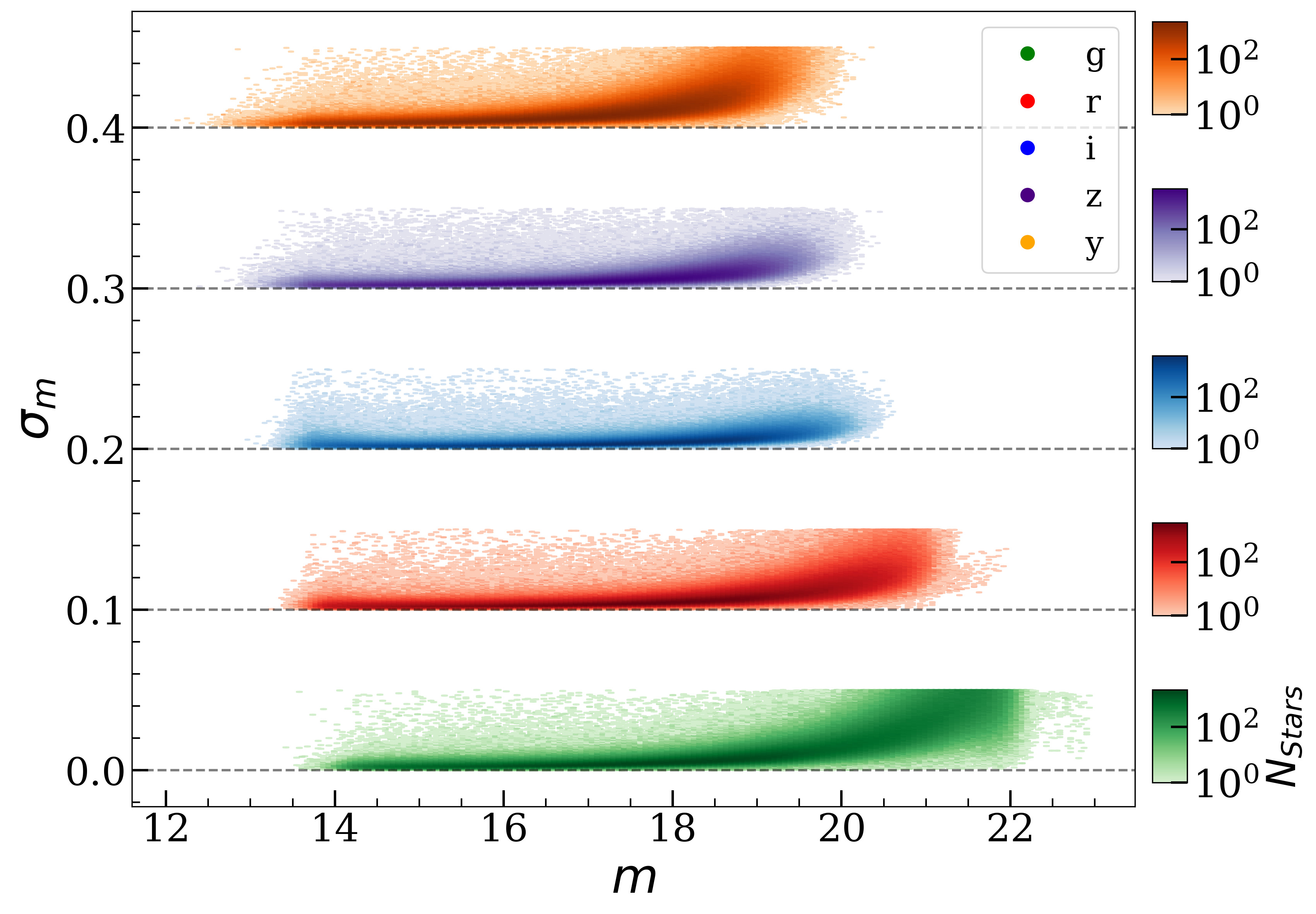}
    \caption{Uncertainties on PS measurement of magnitudes are shown as a function of observed magnitude for all five filter passbands considered here. Each sequential band is shifted vertically by a value of one-tenth for display purposes.}
    \label{fig:uncer_vs_mag}
\end{figure}

\begin{figure}
    \centering
    \includegraphics[width=0.45\textwidth]{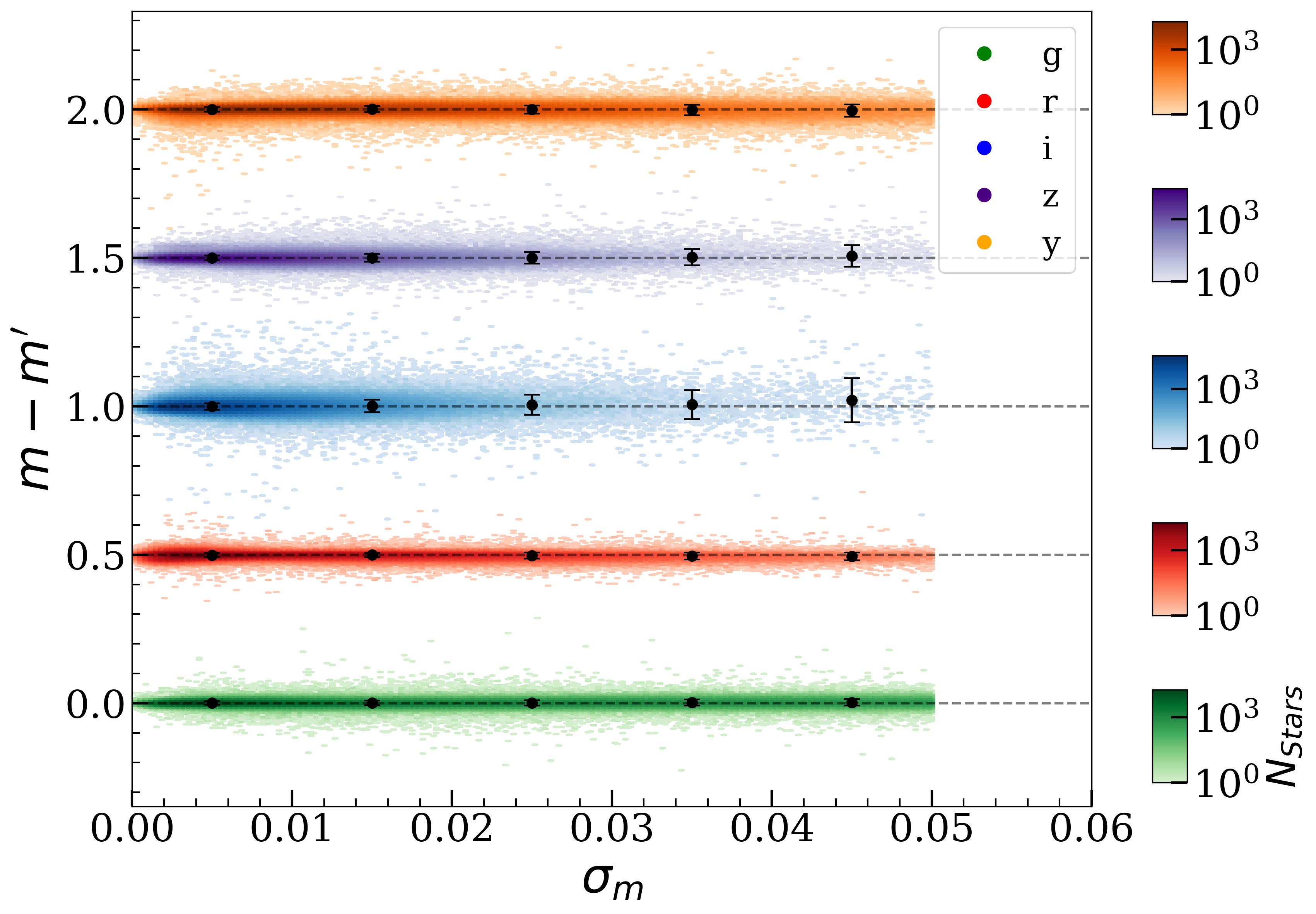}
    \caption{Differences between the reconstructed and observed magnitudes are shown as a function of uncertainty on the PS measurement for all five filter passbands considered here. Each sequential band is shifted vertically by a value of one-half for display purposes. Standard deviation of the differences are shown in bins of 0.01 in uncertainty, with showing the value on the plot at the bin center. For example, the standard deviation in the difference for all stars with uncertainty between 0.01 and 0.02 is shown at 0.015.}
    \label{fig:diff_vs_uncer}
\end{figure}

\begin{figure}
    \centering
    \includegraphics[width=0.45\textwidth]{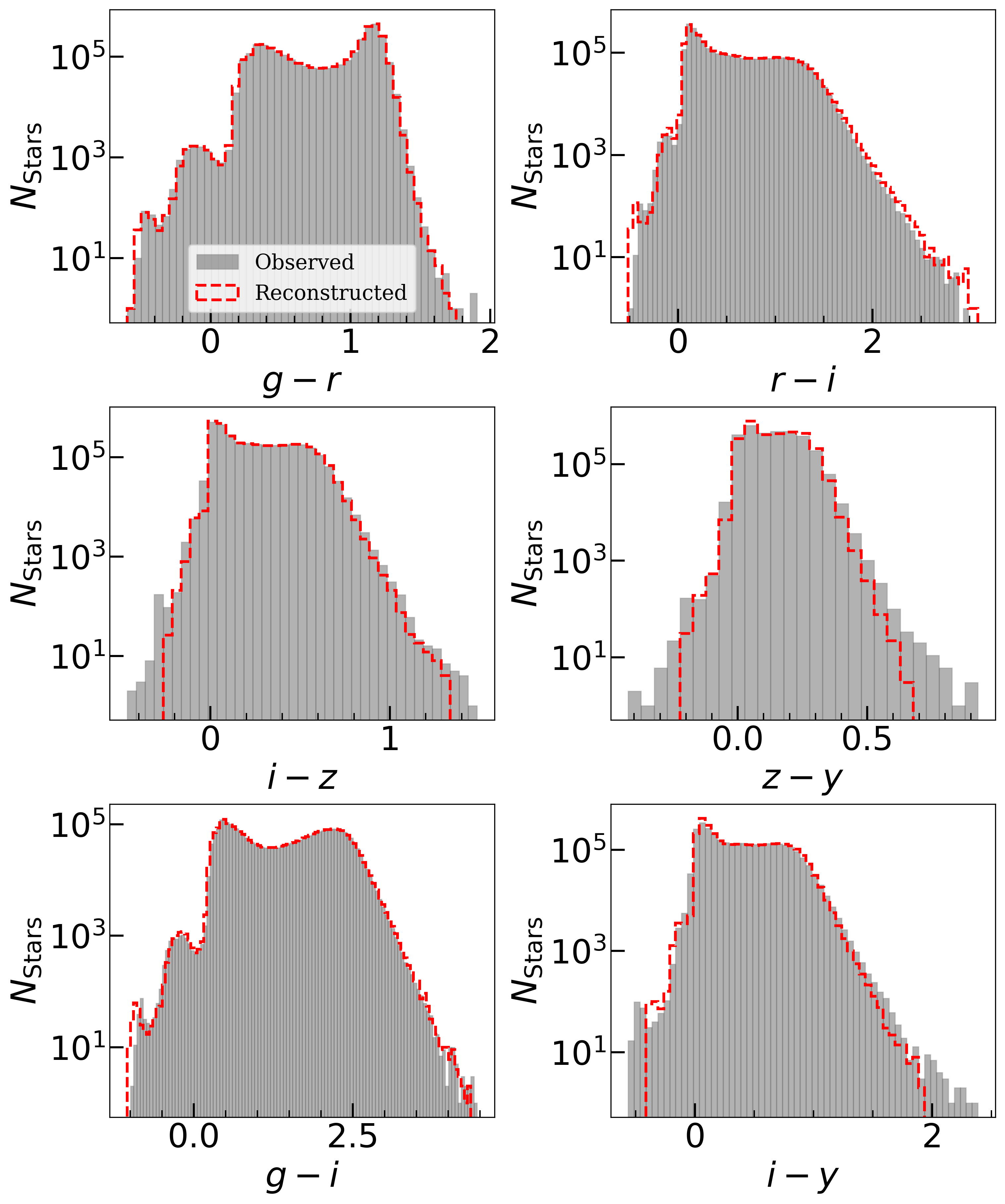}
    \caption{Histograms of the PS1 observed color in solid gray and the LSTM-AE reconstructed color in dashed red.}
    \label{fig:rec_colors}
\end{figure}

\begin{figure*}
    \centering
    \includegraphics[width=0.95\textwidth]{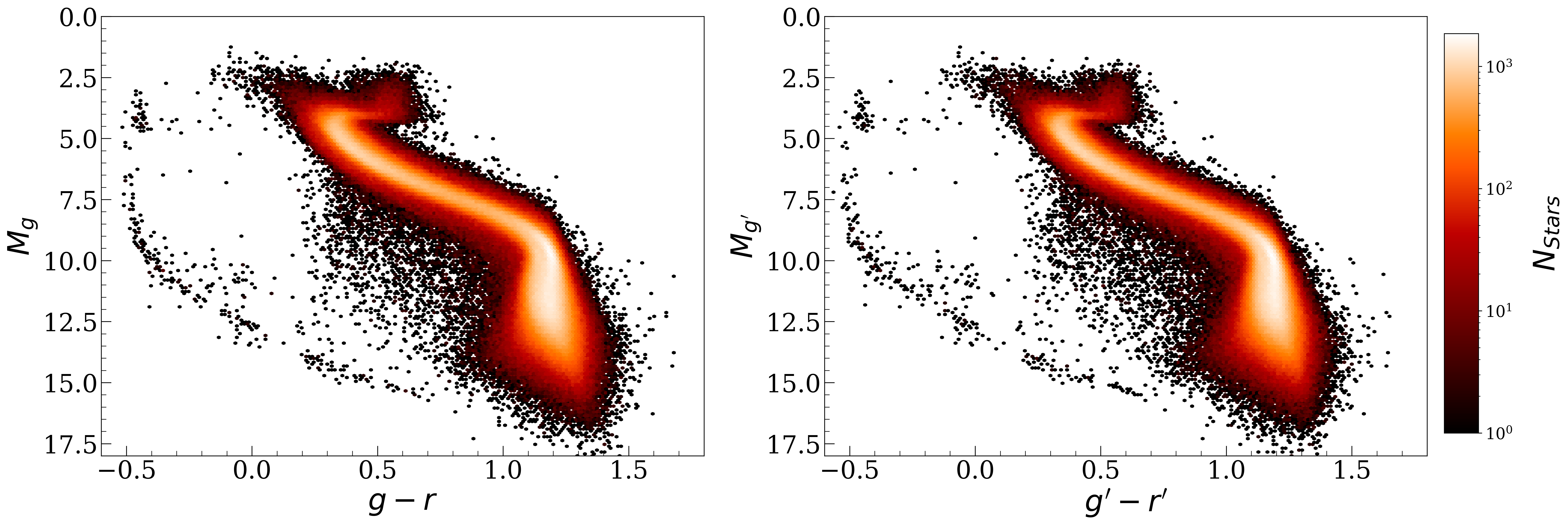}
    \caption{Hertzsprung-Russel (H-R) diagram of both PS1 observed magnitudes (left) and reconstructed magnitudes from LSTM-AE (right).}
    \label{fig:rec_hrdiagram}
\end{figure*}

We define reconstruction as a metric that measures the discrepancy between the input magnitudes and the magnitudes predicted by the model. These reconstructed magnitudes arise from a mapping from the two-dimensional latent space back to the five-dimensional photometry. Table \ref{tab:stars_band_differences} shows how many stars are beyond certain magnitude thresholds when taking the absolute difference between the reconstructed magnitudes $m^\prime$ and the true magnitudes $m$.  In general, the model has high reconstruction accuracy, with an overwhelmingly majority of the dataset (99.51\%) being reconstructed within a small magnitude threshold (0.05). Notably, the $r$ band is reconstructed best, with 99.978\% of all stars having this band reconstructed within 0.05 magnitudes. 

The percentage of stars with 0.01 uncertainty or less in their observed magnitude in $grizy$ is 53.810\%, 74.446\%, 92.690\%, 84.171\%, and 57.626\% respectively. Comparing this with Table \ref{tab:stars_band_differences}, the LSTM-AE predicts a much larger portion of the dataset within 0.01 magnitudes of the recorded measurement for a majority of the bands. By comparing Figures \ref{fig:diff_vs_mag}, \ref{fig:uncer_vs_mag}, and \ref{fig:diff_vs_uncer}, one can see there is little to no correlation between the neural networks prediction and the uncertainty in the observed magnitude. This is why a fixed reconstruction metric threshold is chosen for detecting anomalies in Section 7, since a dynamic anomaly threshold based on increasing uncertainty in magnitudes does not provide a sufficiently significant advantage. As shown in Figure \ref{fig:diff_vs_uncer}, the standard deviation of $m-m^\prime$  in the bands $gry$ stays roughly the same as measurement uncertainty increases, which is a sign of denoising, and potentially providing more accurate measurements in these bands.
Note that the LSTM-AE does not always regress to the mean at faint and bright ends, which are considered the outliers based on the setup of the model. Figure \ref{fig:diff_vs_mag} shows that the model tends to predict a star to be brighter than what PS predicts at both the brightest and faintest end of $z$. Furthermore, the LSTM-AE trends to overestimate the magnitude (predict it to be brighter than the observed) at the faint end of $i$ and $z$, while stars at the faint end of $r$ are underestimated. One possible explanation is that the corrections made to stars near the bright and faint limits of PS are slightly too strong or too weak, leading to these trends in the models prediction. Regardless, such patterns at the faint and bright limits are expected as no cuts were made to address the limits of PS1.

Figure \ref{fig:rec_colors} displays the LSTM-AE's ability to reconstruct various colors. The LSTM-AE can reconstruct complex distributions present in the dataset, and importantly, reconstructs $g-i$ the best given its large range compared to other color indices, where its large range shows that it is the most significant proxy for effective temperature and metallicity in this dataset.
Having shown that both magnitudes and colors are reconstructed with great precision, we use the Gaia subset to show a Hertzsprung-Russel (H-R) diagram created from the LSTM-AE reconstructed magnitudes and compare it to the original H-R diagram. Figure \ref{fig:rec_hrdiagram} confirms our previous analysis, and shows that the majority of stars are reconstructed well across a wide range of stellar types, from brown dwarfs to main sequence stars, blue stragglers, white dwarfs , and even UV bright stars. Note that the limit of up to $M_g \approx 2$ is due to the combination of only going out to three kpc with Gaia and the bright limits of PS1.

\section{Latent Space}
\subsection{General Structure}
\begin{figure}
    \centering
    \includegraphics[width=0.45\textwidth]{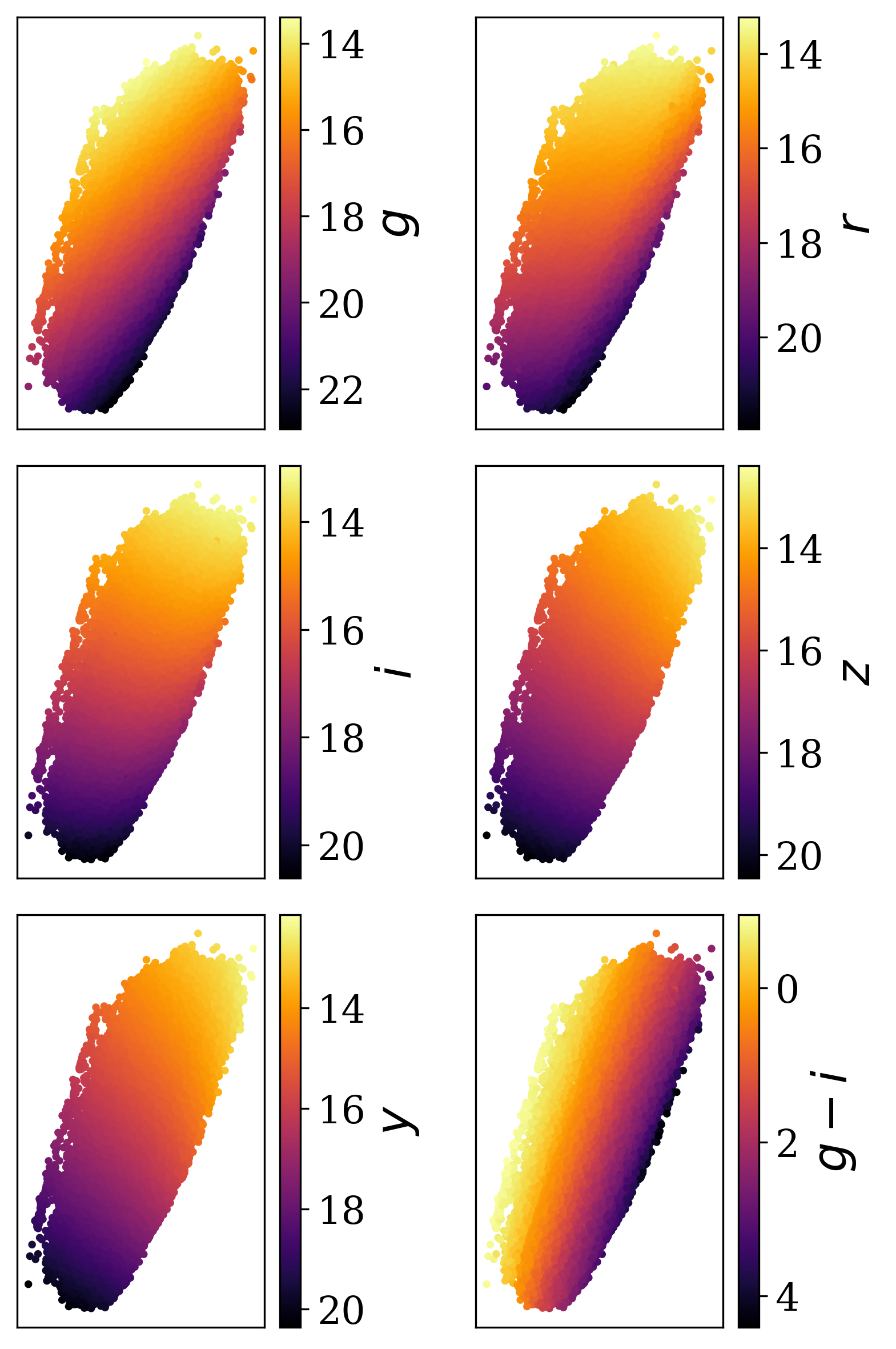}
    \caption{These panels show the latent space colored by all five photometric bands and one color index, displaying clear gradients of apparent brightness and color.}
    \label{fig:latent_space_colored}
\end{figure}  

\begin{figure}
    \centering
    \includegraphics[width=0.45\textwidth]{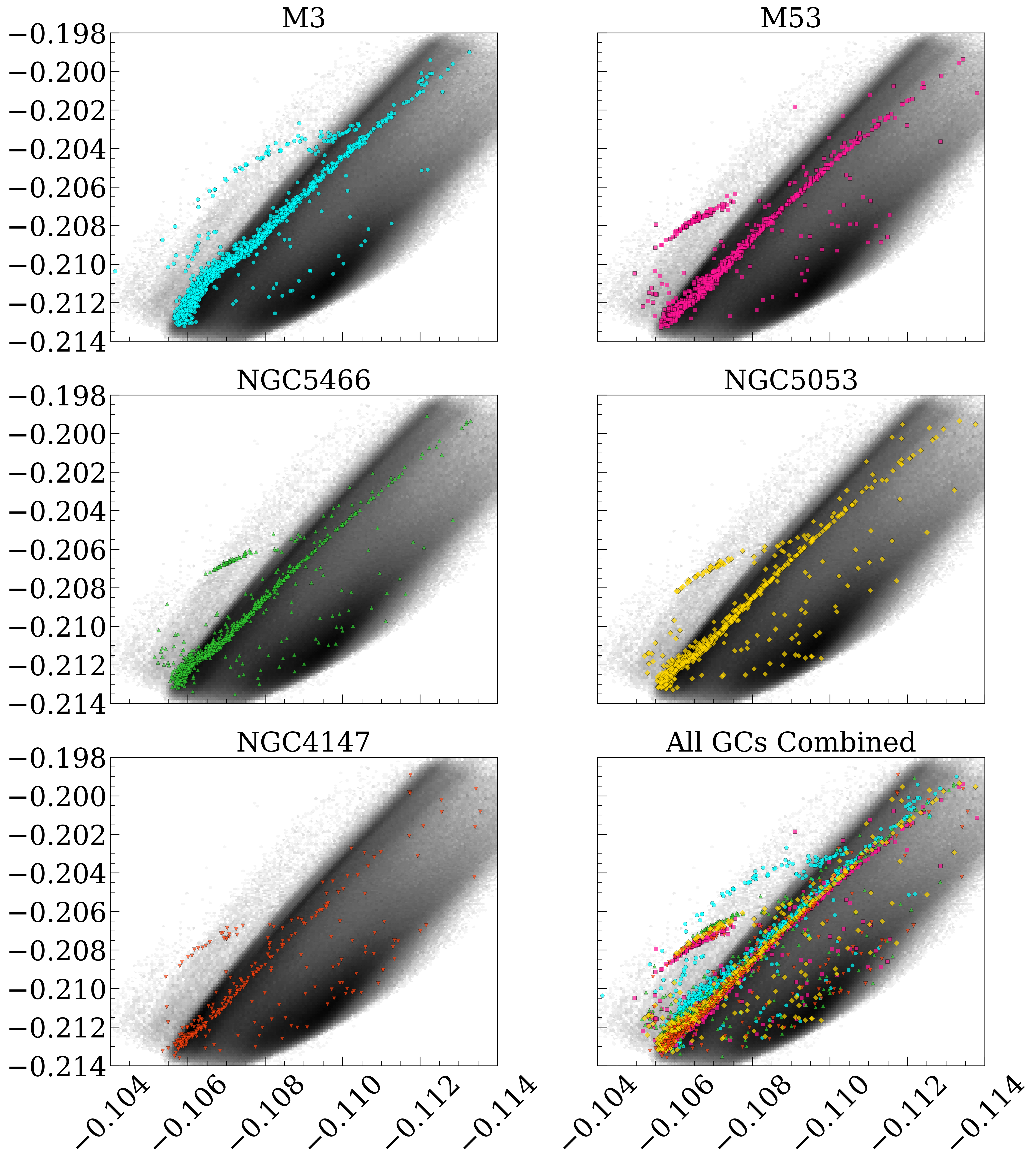}
 \caption{The latent space of all stars are shown in a hexbin plot with cmap \textit{Greys}, and overlapped with stars within 0.15 of a degree of the centers of five galactic globular clusters, with no membership cuts being made.}
    \label{fig:gcs}
\end{figure}

We start our understanding of the latent space by interpreting how it represents the input photometry into two dimensions. Figure \ref{fig:latent_space_colored} shows the latent space colored by \textit{grizy} and $g-i$, displaying clear gradients in both magnitudes and color. The gradients between magnitudes and color are near orthogonal to each other, giving rise to the (slightly slanted) left-to-right gradient temperature, with hot stars on the left and cool stars on the right, and (slightly slanted) up-to-down gradient representing apparent brightness.

Figure \ref{fig:gcs} shows the latent space of stars within 0.15 of a degree from the center of five globular clusters (GCs). The overall position of each GC in the latent space depends on distance, with the closest GC lying highest in the latent space (M3) and the furthest GCs being the lowest (M53 \& NGC 4147).  This is expected as we are dealing with apparent magnitudes, and since M3 is closest it will appear brightest, and thus highest in the latent space. While analyzing the latent space without labeled stellar types and using GCs to remove the distance-luminosity degeneracy provides insight, much more can be obtained by examining where the labeled subsets described in Section 2 fall within the latent space.

\subsection{Post Label Analysis}
\begin{figure}
    \centering
    \includegraphics[width=0.45\textwidth]{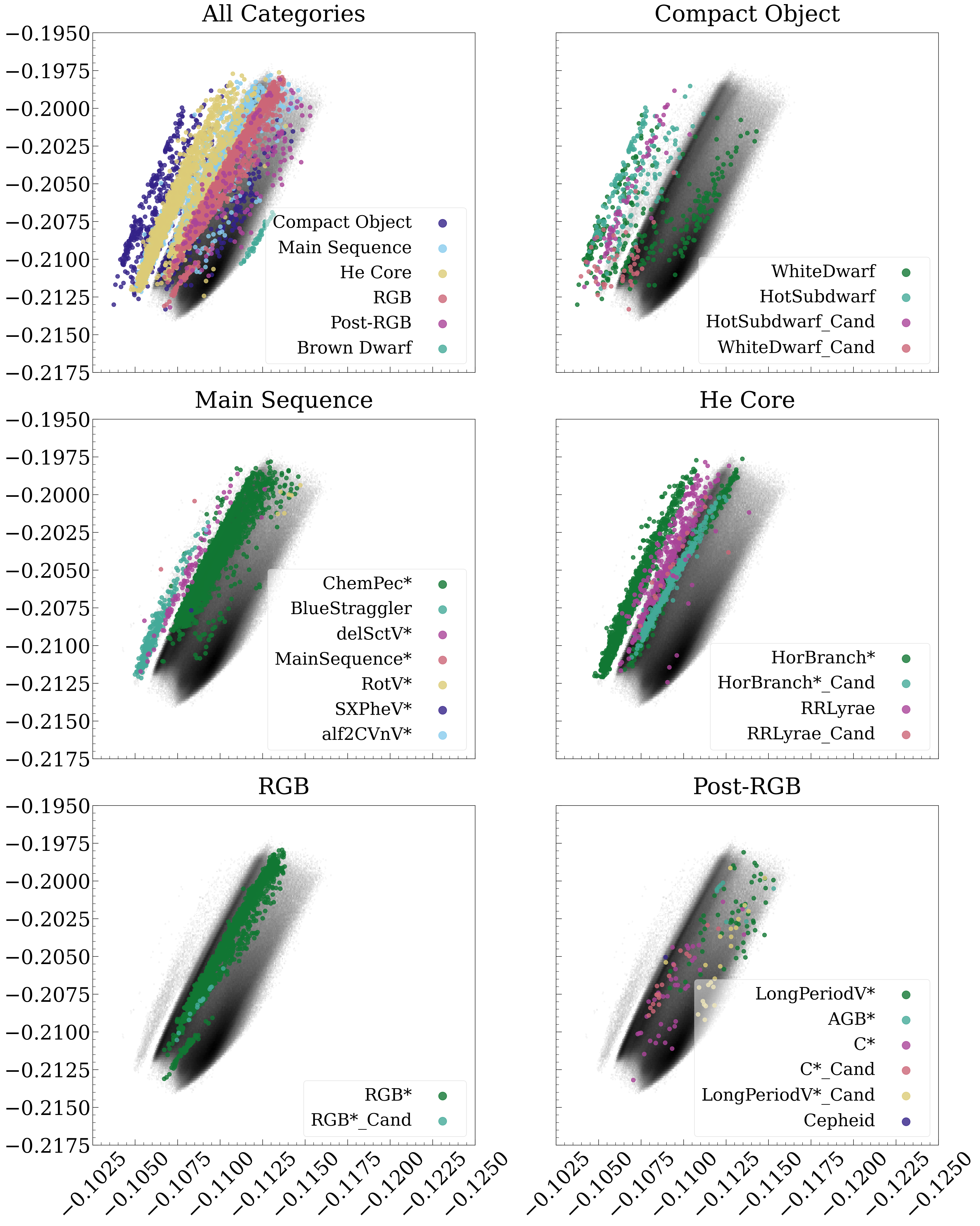}
    \caption{The latent space of all stars are shown in a hexbin plot with cmap \textit{Greys}. The various colored symbols indicate the locations in latent space for stars with various labels determined by matching with SIMBAD. SIMBAD labels are categorized into Compact Object, Main Sequence, He Core, RGB, and Post RGB, with plots of the subtypes also shown.}
    \label{fig:categories}
\end{figure}  

\begin{figure}
    \centering
    \includegraphics[width=0.45\textwidth]{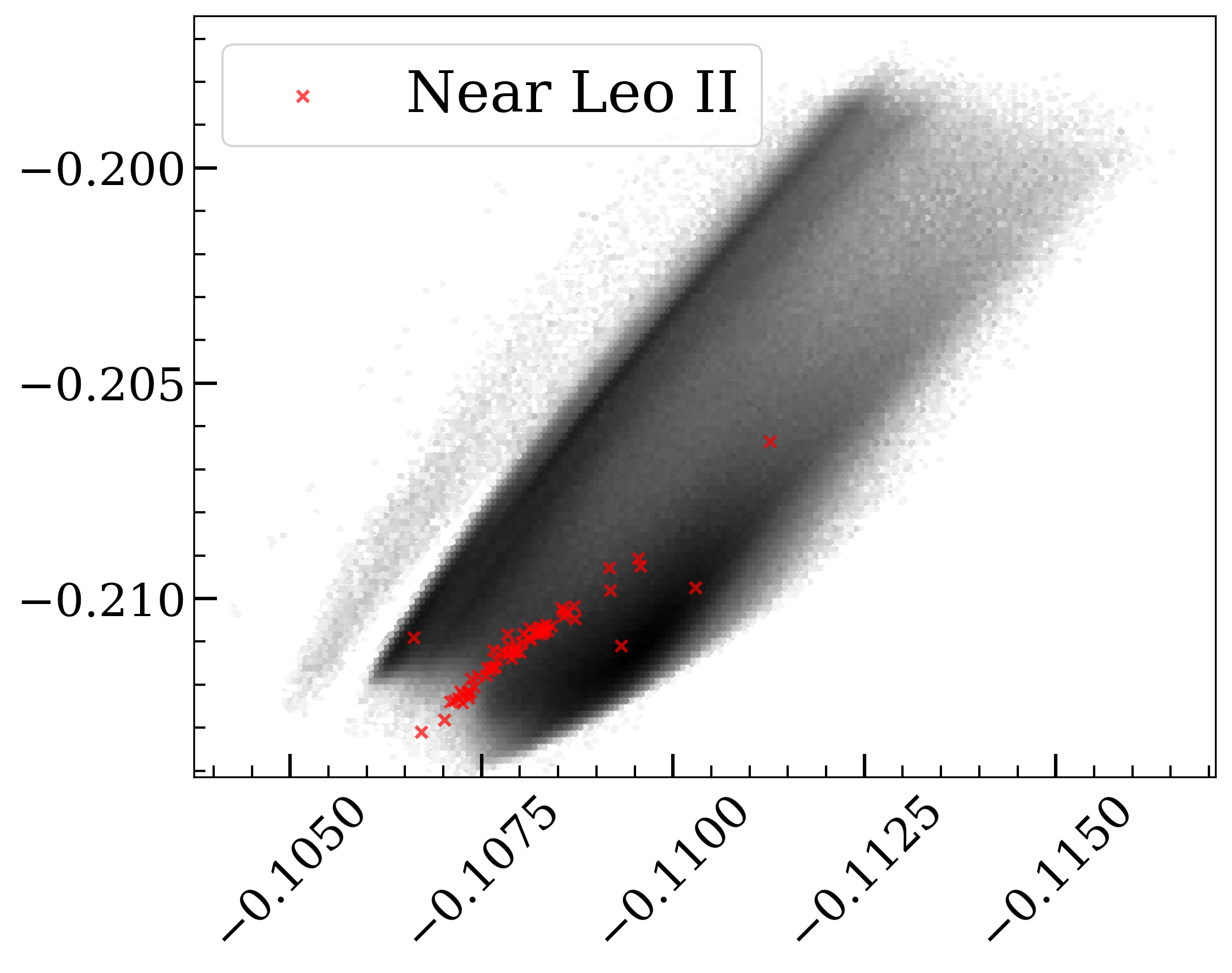}
    \caption{The latent space of all stars are shown in a hexbin plot with cmap \textit{Greys}. Stars within 0.05$\degree$ of Leo II are shown in red, highlighting the RGB of Leo II.}
    \label{fig:leo2}
\end{figure}

\begin{figure}
    \centering
    \includegraphics[width=0.45\textwidth]{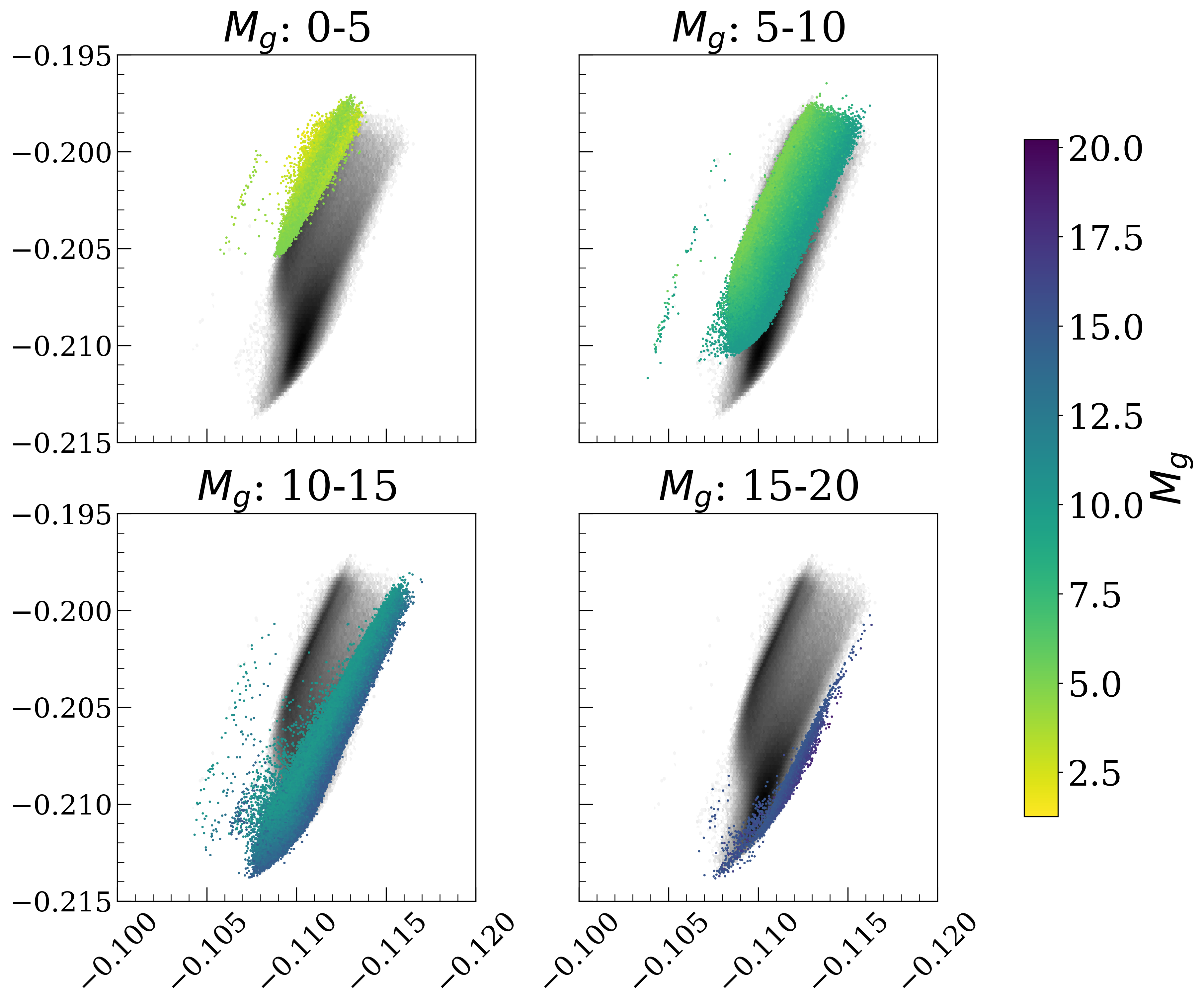}    \caption{The latent space of all stars in the Gaia subset are shown in a hexbin plot with cmap \textit{Greys}.  In addition, stars matched with Gaia parallaxes are overlapped and colored by absolute magnitude, $M_g$.}
    \label{fig:absg}
\end{figure}

\begin{figure}
    \centering
    \includegraphics[width=0.45\textwidth]{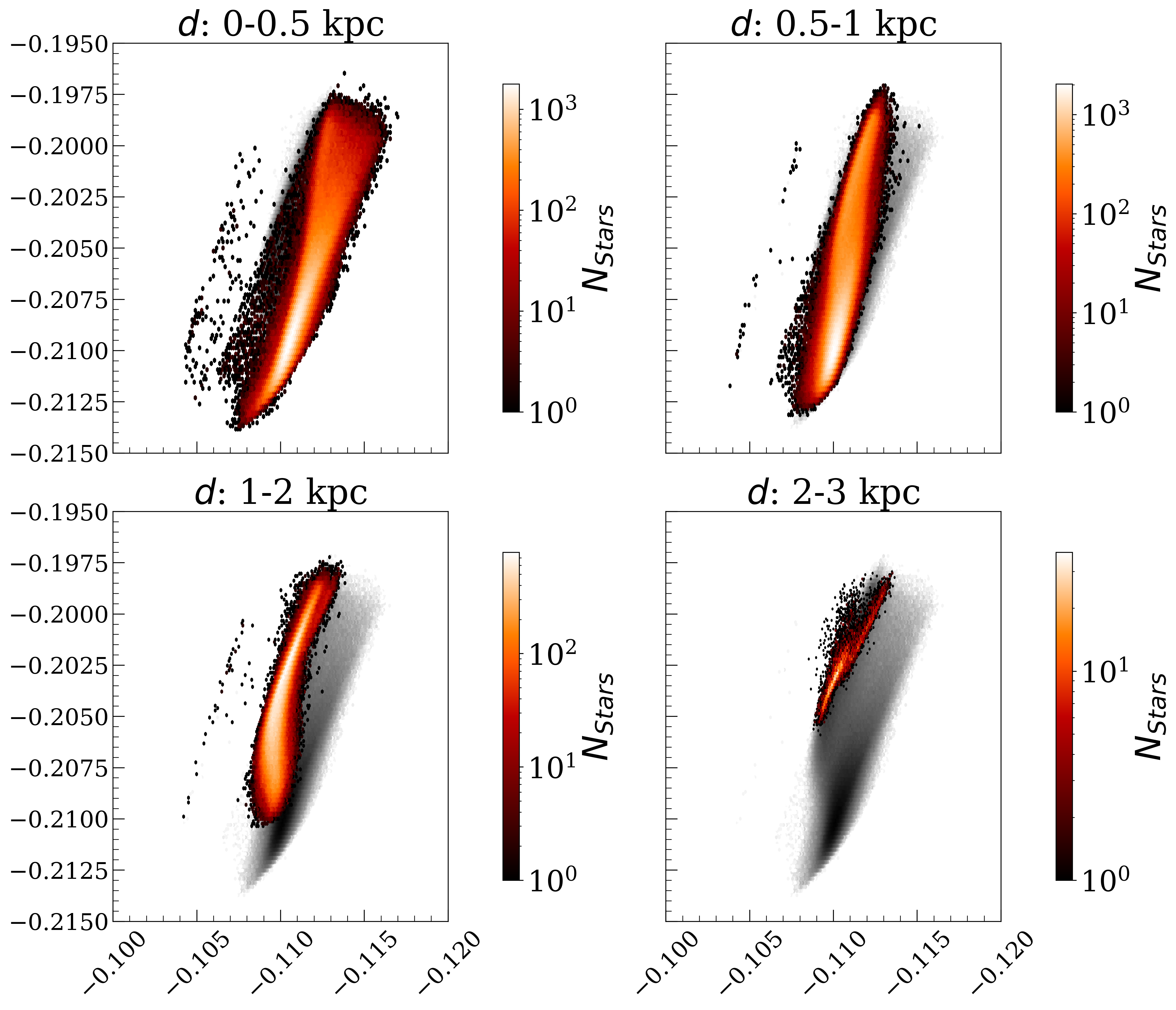}    \caption{The latent space of all stars in the Gaia subset are shown in a hexbin plot with cmap \textit{Greys}.  In addition, stars matched with Gaia parallaxes are overlapped and colored by distance $d$, in kiloparsec ranges, highlighting population changes as distance increases.}
    \label{fig:lsdistance}
\end{figure}

Figures \ref{fig:categories} again shows the PS1 stars in the latent space, but now with labels added for stars identified in SIMBAD.

The gradient in temperature matches our deduction from Figure \ref{fig:latent_space_colored}, as stars generally become cooler moving left-to-right across the figure, and reveals that the distance-luminosity degeneracy is still present, as expected. The small cluster of giants in the lower portion of the latent space is confirmed to be the dwarf galaxy Leo II, as these stars cluster in galactic coordinates at the location of Leo II as shown in Figure \ref{fig:leo2}. This shows that our method may be combined with spatial clustering algorithms to reduce the false positive rate when searching for co-eval populations, stellar streams, and other similar populations.

Further analyzing Figure \ref{fig:categories}, one can see blue stragglers are distinguished from the main sequence, overlapping with the blue horizontal branch (HB) stars, because of the distance-luminosity degeneracy present. The gradient in temperature is further evidenced by hot subdwarfs on the far left and brown dwarfs on the far right. The He core subplot reveals the different stellar types in the latent space being distinguished by temperature, as from left to right are blue HB, RRLyrae gap, and red HB stars. 
Post-RGB stars vary over most the latent space, except on the extreme hot, blue end, which matches the large range of temperatures they can span.

We use the $g-i$ gradient in Figure \ref{fig:latent_space_colored} with Figure \ref{fig:absg} to further see where specific stellar populations fall. In agreement with previous discussion, brown dwarfs fall on the furthest right side of the latent space (the cool end), while white dwarfs fall on the furthest left (the hot end). Subgiants are revealed at the upper center of the latent space, and a population of potential UV bright stars at the upper left of the latent space (hot and bright). We extend our use of the Gaia subset to show population changes of stars in Milky Way shown by the latent space. Figure \ref{fig:lsdistance} shows how the population of stars changes with distance. Low-mass disk stars dominate within half a kiloparsec, with substantial numbers of white dwarfs. At larger distances intermediate-mass main sequence stars, subgiants, and lower giant branch stars dominate, as expected in a magnitude limited simple.

\section{Anomaly Detection}

\begin{figure}
    \centering
    \includegraphics[width=0.45\textwidth]{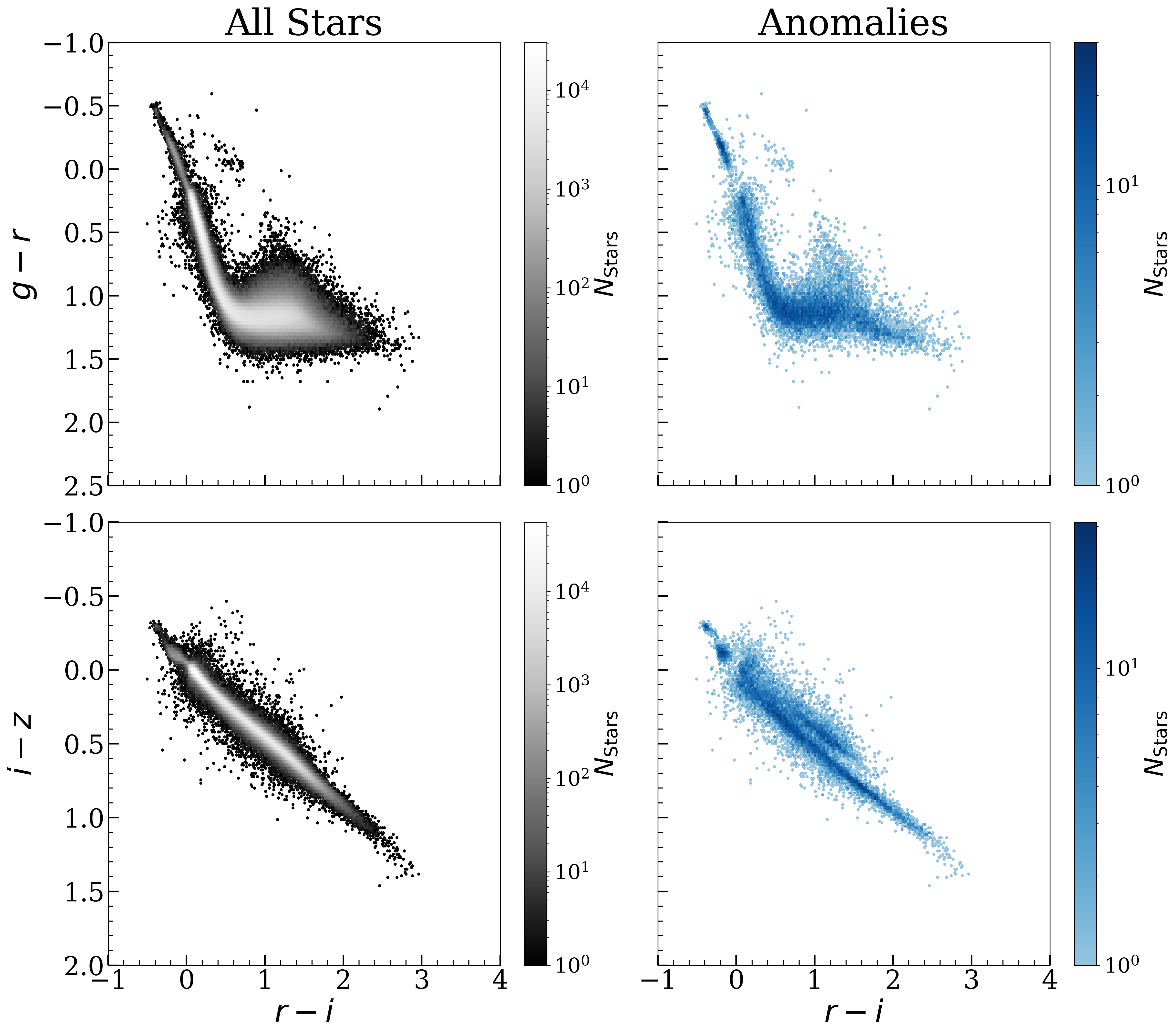}    \caption{Color-color plot comparison showing all stars in a hexbin plot with cmap \textit{Greys} along  with anomalous stars in a hexbin plot with cmap \textit{Blues} , highlighting the detection of several distinct stellar populations. Anomalous here defined as any star in which $|m-m^\prime| > 0.05$.
    \label{fig:colorcomparison}}
\end{figure}

\begin{figure*}[t]
    \begin{minipage}[c]{0.23\textwidth}
        \centering
        \begin{overpic}[width=0.45\textwidth]{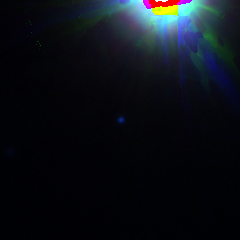}
            \put(5,5){\color{white}\bfseries\small giy}
        \end{overpic}%
        \includegraphics[width=0.45\textwidth]{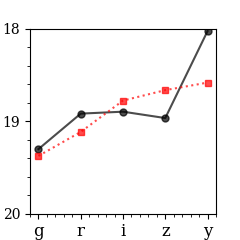}
    \end{minipage}%
    \begin{minipage}[c]{0.23\textwidth}
        \centering
        \begin{overpic}[width=0.45\textwidth]{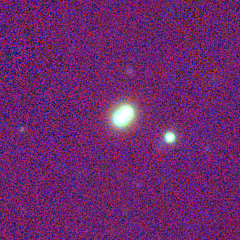}
            \put(5,5){\color{white}\bfseries\small giy}
        \end{overpic}%
        \includegraphics[width=0.45\textwidth]{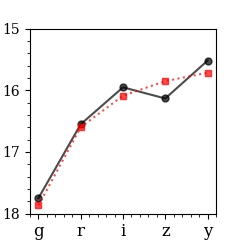}
    \end{minipage}%
    \begin{minipage}[c]{0.23\textwidth}
        \centering
        \begin{overpic}[width=0.45\textwidth]{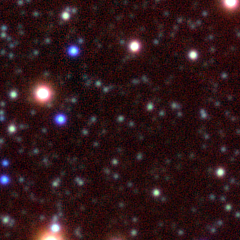}
            \put(5,5){\color{white}\bfseries\small giy}
        \end{overpic}%
        \includegraphics[width=0.45\textwidth]{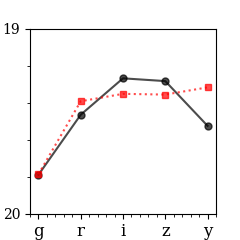}
    \end{minipage}%
    \begin{minipage}[c]{0.23\textwidth}
        \centering
        \begin{overpic}[width=0.45\textwidth]{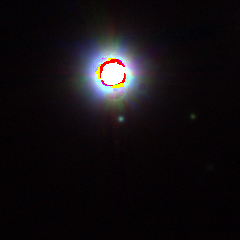}
            \put(5,5){\color{white}\bfseries\small giy}
        \end{overpic}%
        \includegraphics[width=0.45\textwidth]{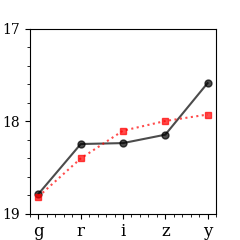}
    \end{minipage}

    \begin{minipage}[c]{0.23\textwidth}
        \centering
        \begin{overpic}[width=0.45\textwidth]{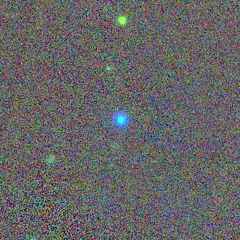}
            \put(5,5){\color{white}\bfseries\small giy}
        \end{overpic}%
        \includegraphics[width=0.45\textwidth]{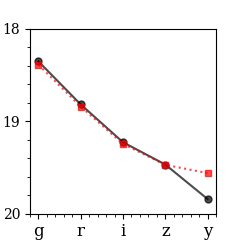}
    \end{minipage}%
    \begin{minipage}[c]{0.23\textwidth}
        \centering
        \begin{overpic}[width=0.45\textwidth]{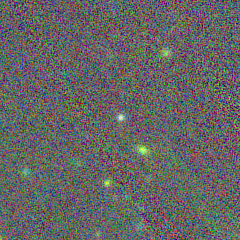}
            \put(5,5){\color{white}\bfseries\small giy}
        \end{overpic}%
        \includegraphics[width=0.45\textwidth]{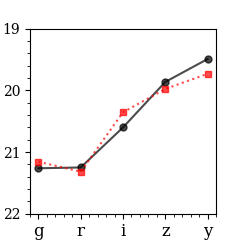}
    \end{minipage}%
    \begin{minipage}[c]{0.23\textwidth}
        \centering
        \begin{overpic}[width=0.45\textwidth]{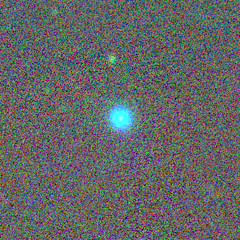}
            \put(5,5){\color{white}\bfseries\small giy}
        \end{overpic}%
        \includegraphics[width=0.45\textwidth]{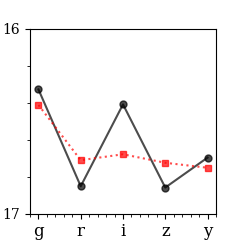}
    \end{minipage}%
    \begin{minipage}[c]{0.23\textwidth}
        \centering
        \begin{overpic}[width=0.45\textwidth]{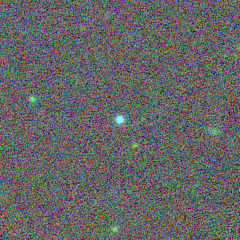}
            \put(5,5){\color{white}\bfseries\small giy}
        \end{overpic}%
        \includegraphics[width=0.45\textwidth]{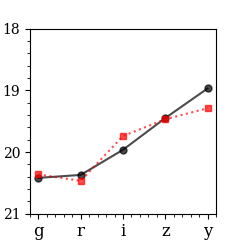}
    \end{minipage}

    \begin{minipage}[c]{0.23\textwidth}
        \centering
        \begin{overpic}[width=0.45\textwidth]{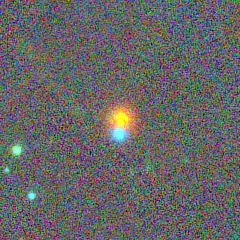}
            \put(5,5){\color{white}\bfseries\small giy}
        \end{overpic}%
        \includegraphics[width=0.45\textwidth]{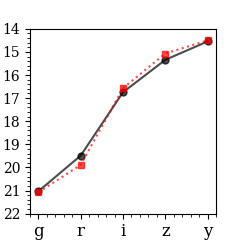}
    \end{minipage}%
    \begin{minipage}[c]{0.23\textwidth}
        \centering
        \begin{overpic}[width=0.45\textwidth]{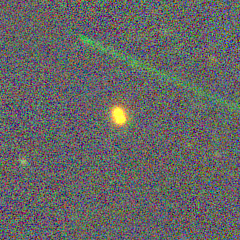}
            \put(5,5){\color{white}\bfseries\small giz}
        \end{overpic}%
        \includegraphics[width=0.45\textwidth]{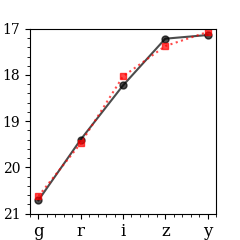}
    \end{minipage}%
    \begin{minipage}[c]{0.23\textwidth}
        \centering
        \begin{overpic}[width=0.45\textwidth]{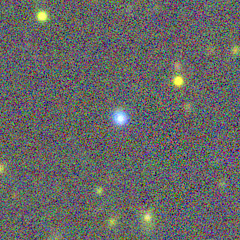}
            \put(5,5){\color{white}\bfseries\small giz}
        \end{overpic}%
        \includegraphics[width=0.45\textwidth]{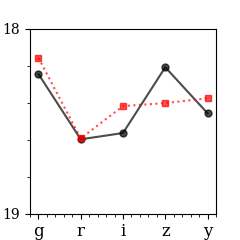}
    \end{minipage}%
    \begin{minipage}[c]{0.23\textwidth}
        \centering
        \begin{overpic}[width=0.45\textwidth]{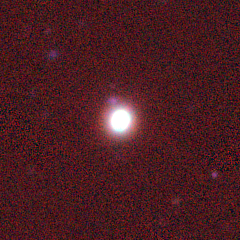}
            \put(5,5){\color{white}\bfseries\small giz}
        \end{overpic}%
        \includegraphics[width=0.45\textwidth]{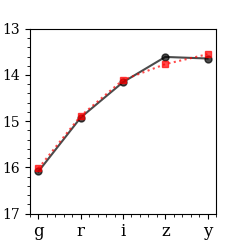}
    \end{minipage}

    \begin{minipage}[c]{0.23\textwidth}
        \centering
        \begin{overpic}[width=0.45\textwidth]{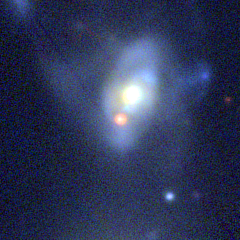}
            \put(5,5){\color{white}\bfseries\small gri}
        \end{overpic}%
        \includegraphics[width=0.45\textwidth]{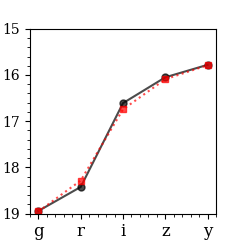}
    \end{minipage}%
    \begin{minipage}[c]{0.23\textwidth}
        \centering
        \begin{overpic}[width=0.45\textwidth]{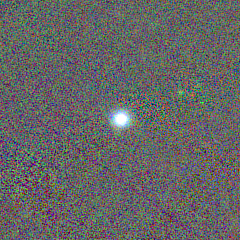}
            \put(5,5){\color{white}\bfseries\small gri}
        \end{overpic}%
        \includegraphics[width=0.45\textwidth]{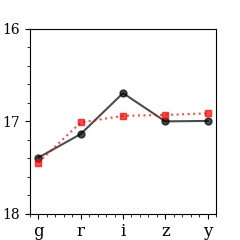}
    \end{minipage}%
    \begin{minipage}[c]{0.23\textwidth}
        \centering
        \begin{overpic}[width=0.45\textwidth]{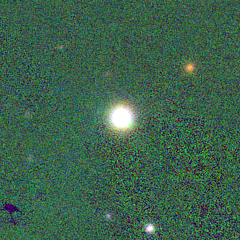}
            \put(5,5){\color{white}\bfseries\small gri}
        \end{overpic}%
        \includegraphics[width=0.45\textwidth]{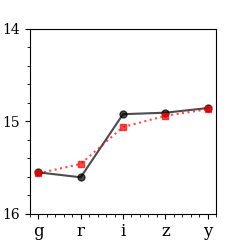}
    \end{minipage}%
    \begin{minipage}[c]{0.23\textwidth}
        \centering
        \begin{overpic}[width=0.45\textwidth]{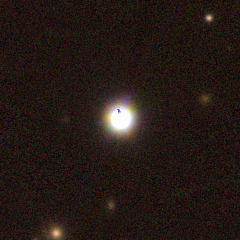}
            \put(5,5){\color{white}\bfseries\small riz}
        \end{overpic}%
        \includegraphics[width=0.45\textwidth]{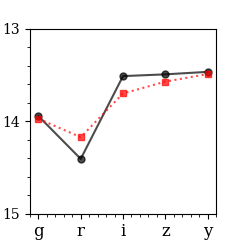}
    \end{minipage}

    \caption{PS1 images (What bands an image is colored by is on the lower left of each image in white) alongside observed photometry highlighting SED shape of detected anomalies. Observed SED shape in black, reconstructed SED shape in red. Anomalies here are defined as stars with $|m-m^\prime| \geq 0.15$. Manually selected subset to highlight the diverse SED shapes detected via poor LSTM-AE reconstruction. Image cutouts are 0.25" per pixel with a size of 240 pixels. All anomalies are shown in the appendix.
    }
    \label{fig:images_special}
\end{figure*}

\begin{figure}
    \centering
    \includegraphics[width=0.45\textwidth]{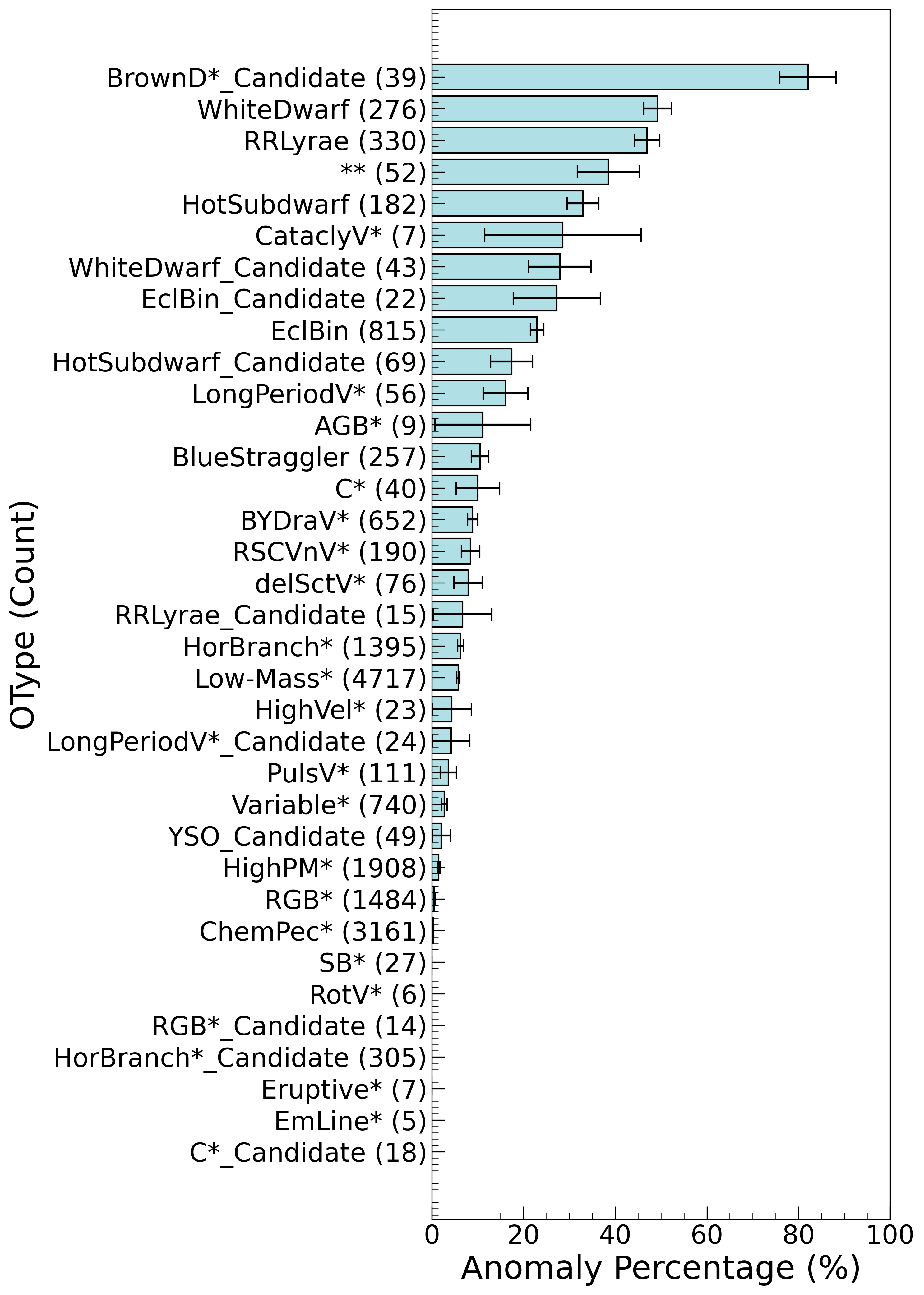}
    \caption{For a given object type (OType) from SIMBAD that has at least five stars, the percentage of stars in that OType that are detected as anomalies ($|m-m^\prime | > 0.05$) are shown. SIMBAD designations "Star" and "GlobCluster" are not included, and the symbol "*" refers to star, with "**" meaning double or multiple star. }
    \label{fig:anomaly_types}
\end{figure}

\begin{figure}
    \centering
    \includegraphics[width=0.45\textwidth]{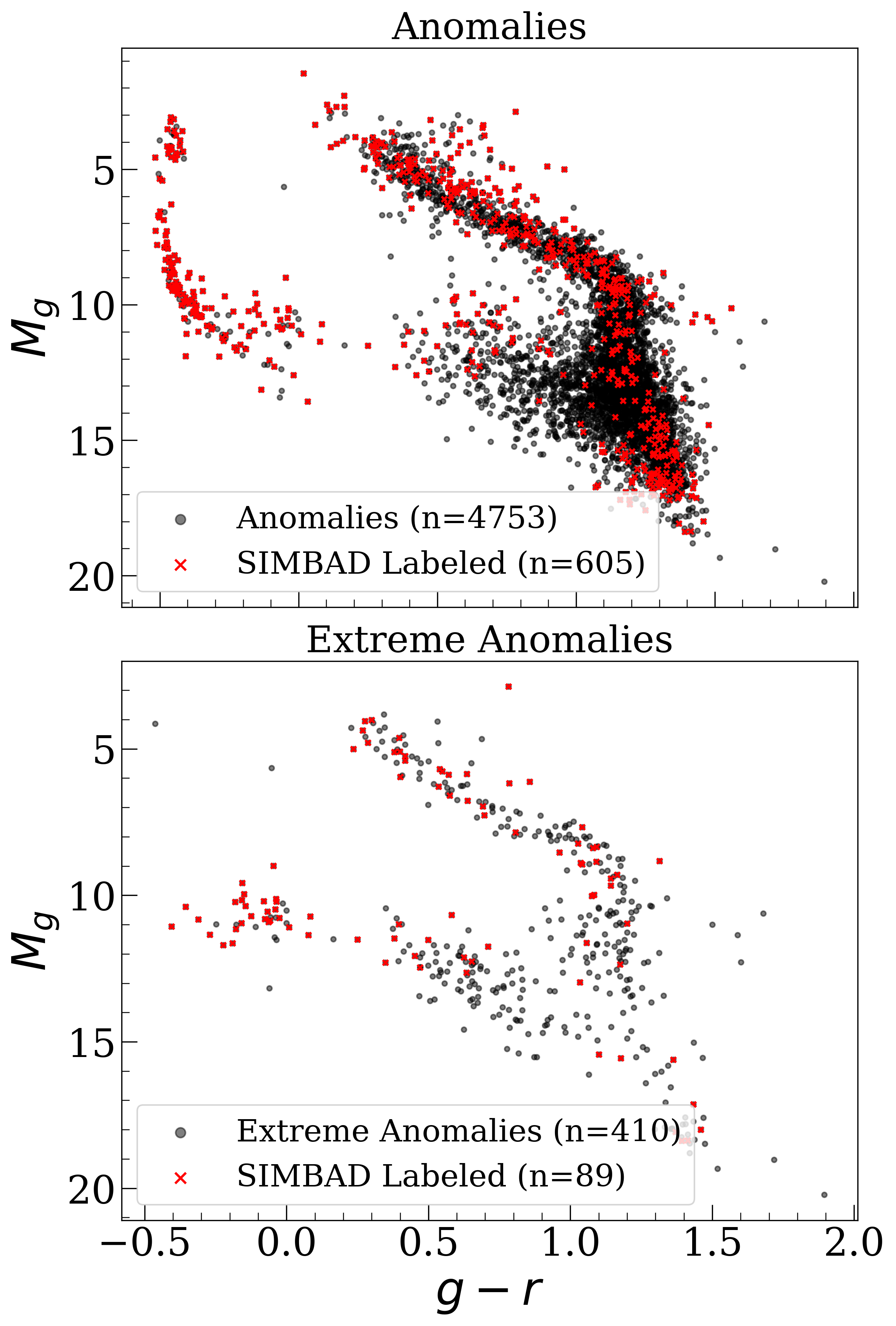}
    \caption{H-R diagram of detected anomalies, $|m-m^\prime | > 0.05$, and extreme anomalies, $|m-m^\prime | \geq 0.1$, highlighting that a majority of the detected anomalies in the Gaia subset are not in the SIMBAD database and unlikely to have a classification.}
    \label{fig:hr_anomalies}
\end{figure}

\begin{figure}
    \centering
    \includegraphics[width=0.45\textwidth]{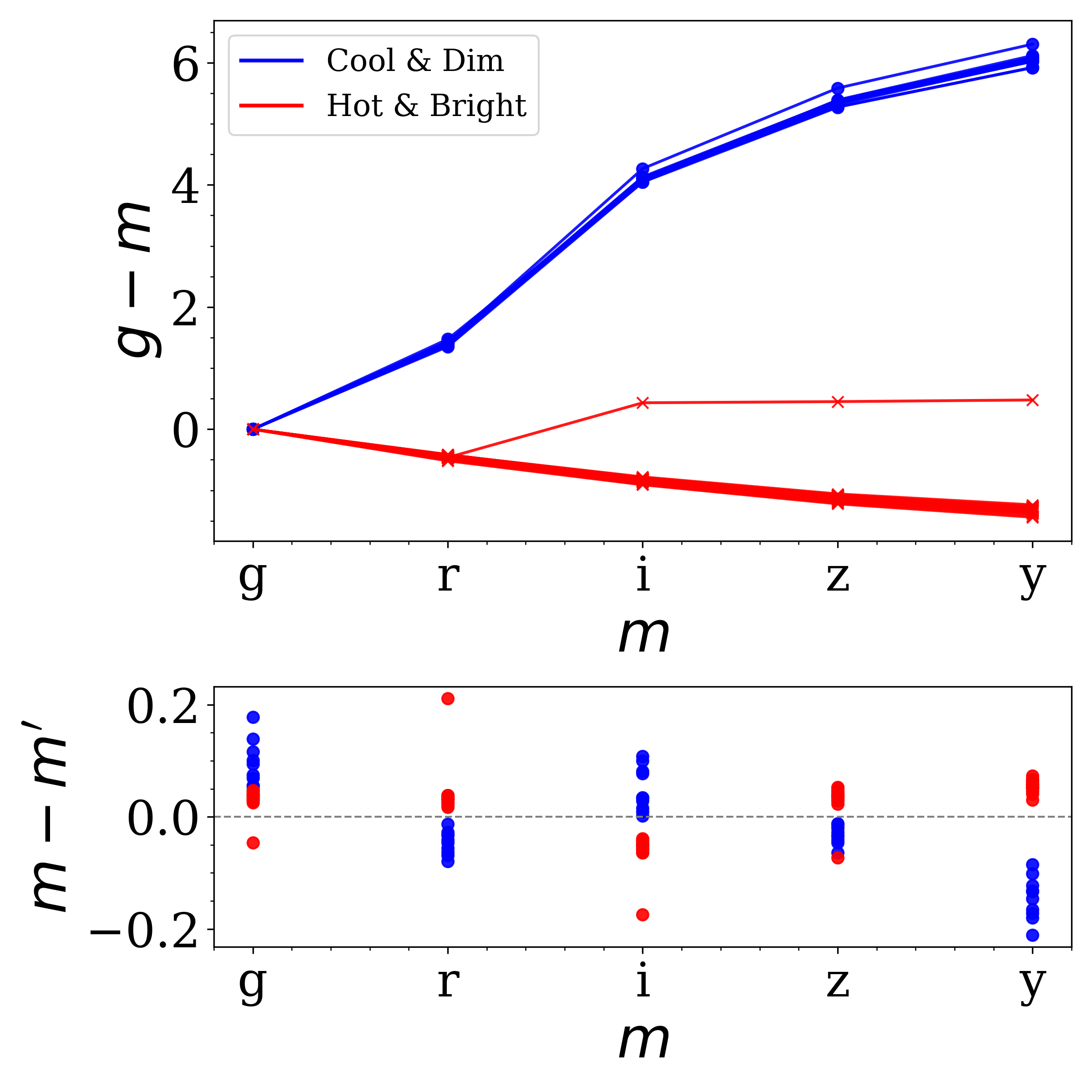}
    \caption{SED shapes and LSTM-AEs prediction accuracy metric, $m-m^\prime$, of both cool and dim ($1.3 \leq g-r \leq 1.5, 18 \leq M_g \leq 19$) and hot and bright ($-0.6 \leq g-r \leq -0.4, 3 \leq M_g \leq 5$) stars from the Gaia anomaly subset, showcasing that both types can be separated solely from using $m-m^\prime$.}
    \label{fig:cool_and_hot}
\end{figure}

\begin{figure}
    \centering
    \includegraphics[width=0.45\textwidth]{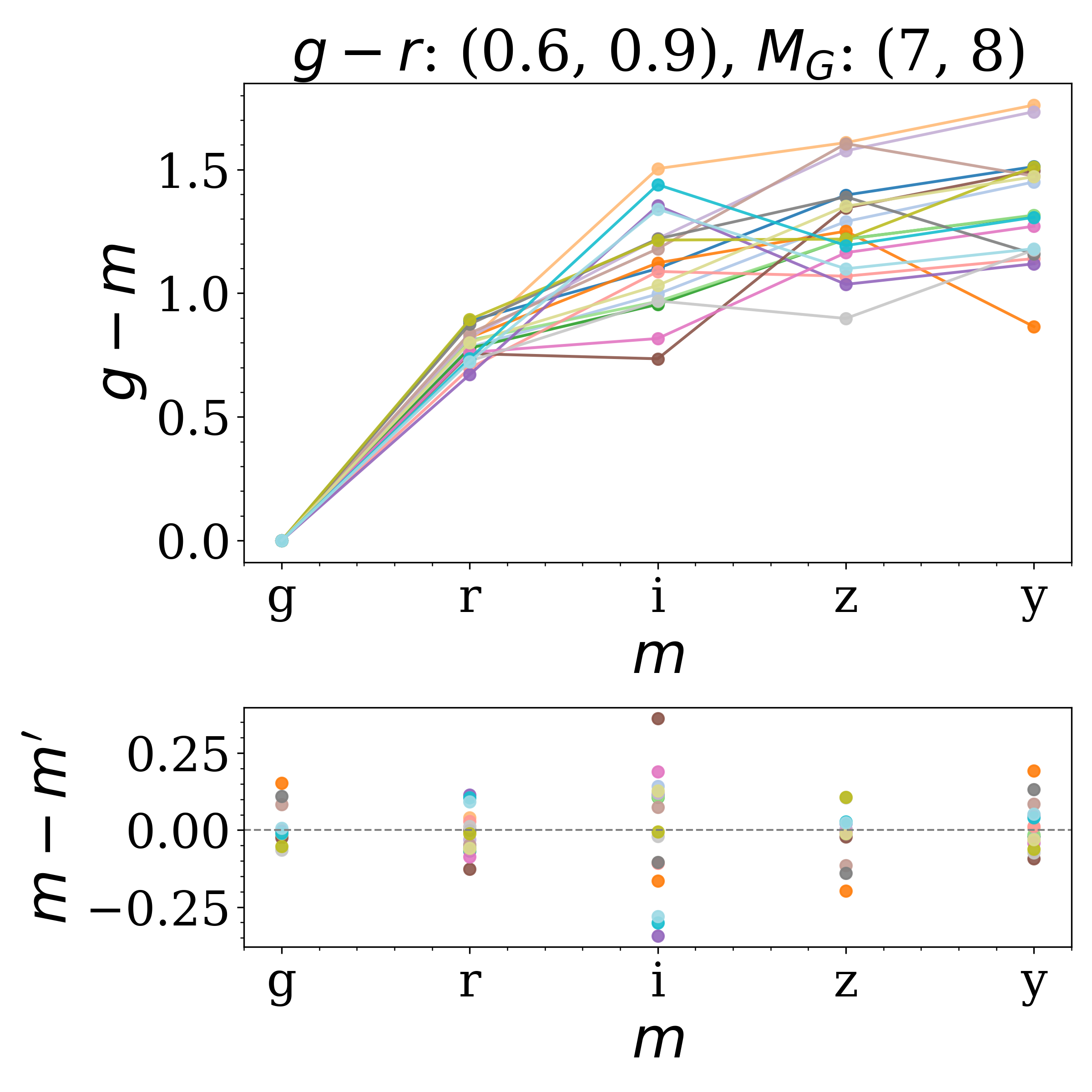}
    \caption{SED shapes and LSTM-AEs prediction accuracy metric, $m-m^\prime$, of a selected group of MS stars ($0.6 \leq g-r \leq 0.9, 7 \leq M_g \leq 8$), showing how in a region of similar colors, one can use $m-m^\prime$ to seperate the populations.}
    \label{fig:MS}
\end{figure}

Figure \ref{fig:colorcomparison} shows two color-color plots in which stars flagged as anomalous are marked in blue. In both plots it is clear that this methodology is able to grab interesting stars in traditional color-color plots, with specific structures, including both clusters of stars in various regions and a strip at $r-i \approx 1.2$, $i-z \approx 0.40$. This is especially useful as it shows that no analysis using color-indices is needed; simply by running a dataset through such a model one is able to reveal distinct, rare stellar populations with minimal effort. 

Figure \ref{fig:images_special} shows the PS1 images alongside the SED shapes of a few selected anomalies. This displays the diversity of anomalies; the model does not detect only a few types of deviations in a stars' SED, and not all anomalies are binaries. Some anomalies would be difficult to detect traditionally, as many SED shapes do not appear unusual until they are compared with their reconstructed counterpart, as seen with the double star with a large gradient in color between the two stars (column one, row three), or the star in the foreground of a galaxy\footnote{The SDSS image of this object was checked, confirming that this is not a supernova.} (column one, row four). In cases where images are not revealing, the SED shapes are not well-behaved and show major deviations from what the model predicts, like a significant increase in the $i$ or $y$ band magnitude (column two, row four \& column one, row one) or brightness peaks in multiple bands (column 3, row 2).

0.49\% of the dataset is determined as anomalous, meaning they are not reconstructed within 0.05 magnitudes in every band. Only 7.74\% of anomalies have labels in SIMBAD, indicating that the majority of anomalies are unlikely to have a classification. Figure \ref{fig:anomaly_types} shows the percentage of stars in an object type (OType) that gets flagged as anomalous within our SIMBAD subset.  Notably, a tiny fraction of RGB and ChemPec MS stars are anomalies, and for most OTypes, there is a small likelihood that a star with that OType will be anomalous.  This suggests that the SED shapes of most object types are reconstructed well by the LSTM-AE. Instead, other factors such as binary phenomena, intrinsic characteristics, minor positional errors, or background and foreground contamination are more likely to cause changes in the SED shape, leading certain stars in an OType to be classified as anomalous.

Figure \ref{fig:hr_anomalies} returns to the Gaia subset, showing the stellar types that are flagged as anomalous on an H-R diagram. A majority of the anomalies do not have labels in SIMBAD. Potential UV bright stars and the hottest white dwarfs are reconstructed within 0.1 magnitudes; it is primarily MS stars here that are anomalies. Figures \ref{fig:rec_hrdiagram} and \ref{fig:hr_anomalies} show that by using the model's predictions to produce a reconstructed H-R diagram, one can easily determine how stars are reconstructed based on their evolutionary phase, along with determining the stellar types of rare stars flagged by the model as anomalous.

Returning to the solely Gaia extreme anomalies, we now consider their rough SED shapes and analyze in detail their reconstruction. Both Figures \ref{fig:cool_and_hot} and \ref{fig:MS} show, with illustrating difficulties, how specific stellar types have distinct reconstruction errors across \textit{grizy}, which can be used to separate populations. Nearly all cool \& dim ($1.3 \leq g-r \leq 1.5, 18 \leq M_g \leq 19$) and hot \& bright ($-0.6 \leq g-r \leq -0.4, 3 \leq M_g \leq 5$) stars have similar SED shapes\footnote{The outlier hot \& bright star is approx. a magnitude beyond the conservative bright limits reported by PS in \textit{grizy}, so this is likely erroneous data.}, and the LSTM-AE over or underestimates their magnitudes in all the bands in such a way one can distinguish these two populations. This extends to the more complicated example with the MS stars, as stars for which $g^\prime$ is predicted to be dimmer than $g$ are only the stars that peak in $z$, and $i$ is the largest probe for distinguishing these populations that fall within this absolute magnitude and color range ($0.6 \leq g-r \leq 0.9, 7 \leq M_g \leq 8$).

\section{Conclusion}

We have successfully developed a methodology to easily analyze multiband photometry through dimensionality reduction and detect anomalies by using a \textit{long-short term memory autoencoder} (LSTM-AE). While future work is required to test this models scalability, it is expected to have the ability to handle large amounts of data. The LSTM-AE compresses multiband photometry into a two-dimensional latent space without the need of predefined labels, and we used GCs, labels from SIMBAD, parallaxes from Gaia, and PS images to understand the latent space and demonstrated how analyzing the latent space facilitates efficient data interpretation. Since the model must reconstruct {all} magnitudes, it implicitly reconstructs all colors as well. This means that the latent space representation captures more information than traditional analysis methods that use a collection of CMDs and color-color plots, as those do not probe the entire SED shape. The model demonstrates high accuracy in reconstructing the input photometry, and we show that anomalous stellar types can be easily detected by analyzing poorly reconstructed photometry. Future work is required to clearly distinguish physical anomalies from observational artifacts.

The model has generalization capabilities, making it a useful model to deploy in the pipeline of large-scale photometric surveys for dimensionality reduction, anomaly detection, and potentially denoising purposes. While outside the scope of this paper, AEs possess the ability to perform data imputation \citep{Imputation_Pereira}. If future studies demonstrate that an LSTM-AE can reliably fill in missing photometric bands for stars with well-behaved SEDs, it would be advantageous for surveys to train and deploy such a model on their data.  Since the proposal of the model arose from the need to {exploit the wavelength-dependent nature of the data,} it is clear that the LSTM-AE can be applied to many other surveys and other types of astronomical objects in which multiband photometry reveals underlying physics of the objects that are being analyzed, like galaxies. 

\begin{acknowledgments}

We thank the anonymous referee for their valuable input that helped strengthen this manuscript.
B.H gratefully acknowledges funding for this project through the Kirkwood Endowment and support from the Cox Scholars Program.

This research was supported in part by Lilly Endowment, Inc., through its support for the Indiana University Pervasive Technology Institute. 

The Pan-STARRS1 Surveys (PS1) and the PS1 public science archive have been made possible through contributions by the Institute for Astronomy, the University of Hawaii, the Pan-STARRS Project Office, the Max-Planck Society and its participating institutes, the Max Planck Institute for Astronomy, Heidelberg and the Max Planck Institute for Extraterrestrial Physics, Garching, The Johns Hopkins University, Durham University, the University of Edinburgh, the Queen's University Belfast, the Harvard-Smithsonian Center for Astrophysics, the Las Cumbres Observatory Global Telescope Network Incorporated, the National Central University of Taiwan, the Space Telescope Science Institute, the National Aeronautics and Space Administration under Grant No. NNX08AR22G issued through the Planetary Science Division of the NASA Science Mission Directorate, the National Science Foundation Grant No. AST–1238877, the University of Maryland, Eotvos Lorand University (ELTE), the Los Alamos National Laboratory, and the Gordon and Betty Moore Foundation.

This work has made use of data from the European Space Agency (ESA) mission
{\it Gaia} (\url{https://www.cosmos.esa.int/Gaia}), processed by the {\it Gaia}
Data Processing and Analysis Consortium (DPAC,
\url{https://www.cosmos.esa.int/web/Gaia/dpac/consortium}). Funding for the DPAC
has been provided by national institutions, in particular the institutions
participating in the {\it Gaia} Multilateral Agreement.

Some/all of the data presented in this paper were obtained from the Multimission Archive at the Space Telescope Science Institute (MAST). STScI is operated by the Association of Universities for Research in Astronomy, Inc., under NASA contract NAS5-26555. Support for MAST for non-HST data is provided by the NASA Office of Space Science via grant NAG5-7584 and by other grants and contracts.

This research has made use of the SIMBAD database,
operated at CDS, Strasbourg, France.

This work made use of the following software packages: \texttt{astropy} \citep{astropy:2013, astropy:2018, astropy:2022}, \texttt{matplotlib} \citep{Hunter:2007}, \texttt{numpy} \citep{numpy}, \texttt{pandas} \citep{mckinney-proc-scipy-2010, pandas_13819579}, \texttt{python} \citep{python}, \texttt{scipy} \citep{2020SciPy-NMeth}, \texttt{Cython} \citep{cython:2011}, \texttt{h5py} \citep{collette_python_hdf5_2014}, \texttt{JAX} \citep{jax2018github}, \texttt{seaborn} \citep{Waskom2021}, and \texttt{tensorflow} \citep{tensorflow_13989084}.

Software citation information aggregated using \texttt{\href{https://www.tomwagg.com/software-citation-station/}{The Software Citation Station}} \citep{software-citation-station-paper, software-citation-station-zenodo}.
\end{acknowledgments}

\begin{contribution}
    \textbf{Conceptualization:} \\ B. D. Hutchinson, C. A. Pilachowski \\
    \textbf{Data curation:} \\ B. D. Hutchinson \\
    \textbf{Formal analysis:} \\ B. D. Hutchinson \\
    \textbf{Investigation:} \\ B. D. Hutchinson \\
    \textbf{Methodology:} \\ B. D. Hutchinson, C. A Pilachowski, C. I. Johnson \\
    \textbf{Software:} \\ B. D. Hutchinson \\
    \textbf{Supervision:} \\ C. A. Pilachowski, C. I. Johnson\\
    \textbf{Validation:} \\ B. D. Hutchinson, C. A. Pilachowski, C. I. Johnson \\
    \textbf{Visualization:} \\ B. D. Hutchinson \\
    \textbf{Writing --- original draft:} \\ B. D. Hutchinson \\
    \textbf{Writing --- review \& editing:} \\ B. D. Hutchinson, C. A. Pilachowski, C. I. Johnson \\
\end{contribution}

\bibliographystyle{aasjournal}
\bibliography{main}

\appendix

\section{Tensorflow + Keras Implementation}

\begin{lstlisting}
# python 3.11.4
import os
# pandas version 2.2.3
import pandas as pd
# scikit-learn version 1.6.0
from sklearn.model_selection import train_test_split
# keras version 3.7.0
from keras.models import Model
from keras.layers import LSTM, Dense, RepeatVector, Input
from keras.callbacks import EarlyStopping, Callback
from tensorflow.keras.initializers import HeNormal
from tensorflow.keras.optimizers import Adam
# numpy version 2.0.2
import numpy as np
# tensorflow version 2.18.0
import tensorflow as tf

class LSTMAutoencoder:
    def __init__(self, input_shape):

        # Define the input layer
        inputs = Input(shape=(input_shape, 5))
        
        x, state_h1, state_c1 = LSTM(4, activation='elu', kernel_initializer=HeNormal(),
                                     recurrent_initializer="orthogonal", return_state=True,
                                     return_sequences=True)(inputs)

        x, state_h2, state_c2 = LSTM(3, activation='elu', kernel_initializer=HeNormal(),
                                     recurrent_initializer="orthogonal", return_state=True, 
                                     return_sequences=True)(x)
        
        
        # Latent Space
        x, state_h3, state_c3 = LSTM(2, activation='elu', kernel_initializer=HeNormal(),
                                     recurrent_initializer="orthogonal", return_state=True,
                                     return_sequences=False)(x)
        
        # Store the encoder model for later use
        self.encoder = Model(inputs, [x, state_h1, state_c1, state_h2, state_c2, state_h3, state_c3])
        
        # Repeat vector
        x = RepeatVector(input_shape)(x)
        
        # Decoder
        x = LSTM(3, activation='elu', kernel_initializer=HeNormal(),recurrent_initializer="orthogonal",
                 return_sequences=True)(x, initial_state=[state_h2, state_c2])

        x = LSTM(4, activation='elu', kernel_initializer=HeNormal(),recurrent_initializer="orthogonal", 
                 return_sequences=True)(x, initial_state=[state_h1, state_c1])
    
        outputs = LSTM(5, activation='elu', kernel_initializer=HeNormal(),recurrent_initializer="orthogonal",
                       return_sequences=True)(x)
        
        # Create the full model
        self.model = Model(inputs, outputs)
        self.model.compile(optimizer=Adam(learning_rate=0.0001), loss='mse')

    def train(self, train_data, test_data, epochs, batch_size, callbacks=None):
        return self.model.fit(train_data, train_data, epochs=epochs, batch_size=batch_size,
                              validation_data=(test_data, test_data), verbose=1, callbacks=callbacks)

    def predict(self, data):
        return self.model.predict(data)

    def extract_latent_space(self, data):
        latent_space = self.encoder.predict(data)
        return latent_space[0]  # Return the latent space representation only
    
    def reconstruct(self, data):
        return self.model.predict(data)
    

def prepare_data(autoencoder, data, scaler, dataset_type, y_data):
    # Extract latent space representations
    latent_representations = autoencoder.extract_latent_space(data)
    
    # Reconstruct magnitudes and reverse decimal scaling (multiply by 10)
    reconstructed_magnitudes = autoencoder.reconstruct(data).reshape(-1, 5)
    reconstructed_magnitudes_unscaled = np.round(reconstructed_magnitudes * 10, 3)
    
    # Create a DataFrame
    df = pd.DataFrame(reconstructed_magnitudes_unscaled, columns=['g_prime', 'r_prime', 'i_prime', 'z_prime', 'y_prime'])
    df[['prime_x', 'prime_y']] = latent_representations
    
    # Add metadata to the DataFrame
    df['dataset_type'] = dataset_type
    df['ra'] = y_data['ra'].values
    df['dec'] = y_data['dec'].values
    df['l'] = y_data['l'].values
    df['b'] = y_data['b'].values
    df['objID'] = y_data['objID'].values
    
    return df

if __name__ == '__main__':
    DATA_DIR = "path/to/data"
    PLOT_DIR = "path/to/plots"
    HEXBIN_PLOT_DIR = os.path.join(PLOT_DIR, 'hexbin_plots')

    try:
        file_path = os.path.join(DATA_DIR, 'grizy_photometry.csv')
        full_data = pd.read_csv(file_path)[['g', 'r', 'i', 'z', 'y','objID','ra','dec','l','b']]
        print(f"Loaded {len(full_data)} stars from {file_path}")
    except FileNotFoundError:
        print(f"Error: The file {file_path} was not found.")
        exit(1)
    except pd.errors.EmptyDataError:
        print(f"Error: The file {file_path} is empty.")
        exit(1)
    except Exception as e:
        print(f"An error occurred while reading the file: {e}")
        exit(1)

    features = ['g', 'r', 'i', 'z','y']

    print(f"Using features: {features}")
    X = full_data[features]
    y = full_data.drop(columns=features)

    run_results = []  # List to store results for all runs
  
    # Loop for 5 runs
    for run in range(5):
        print(f"Starting Run {run + 1}")

        # Create random splits with different random_state
        X_train, X_temp, y_train, y_temp = train_test_split(X, y, test_size=0.20, random_state=41 + run)
        X_val, X_test, y_val, y_test = train_test_split(X_temp, y_temp, test_size=0.20, random_state=41 + run)

        # Apply decimal scaling by dividing by 10
        X_train_scaled = np.round(X_train / 10, 3).values 
        X_val_scaled = np.round(X_val / 10, 3).values 
        X_test_scaled = np.round(X_test / 10, 3).values 

        train_data = X_train_scaled.reshape((X_train_scaled.shape[0], 1, X_train_scaled.shape[1]))
        val_data = X_val_scaled.reshape((X_val_scaled.shape[0], 1, X_val_scaled.shape[1]))
        test_data = X_test_scaled.reshape((X_test_scaled.shape[0], 1, X_test_scaled.shape[1]))

        print("Mean:", np.mean(X_train_scaled))
        print("Standard deviation:", np.std(X_train_scaled))

        # Initialize the model for each run
        autoencoder = LSTMAutoencoder(input_shape=1)
        
        early_stopping = EarlyStopping(monitor='val_loss', patience=5, mode='min', restore_best_weights=True)

        # Train the model
        history = autoencoder.train(train_data, val_data, epochs=200, batch_size=512, callbacks=[early_stopping])
        # Save the model after each run
        autoencoder.model.save(os.path.join(DATA_DIR, f'grizy_LSTMAE_run_{run + 1}.h5'))

        # Evaluate the model on the test data
        test_loss = autoencoder.model.evaluate(test_data, test_data)
        print(f"Run {run + 1} - Test Loss: {test_loss}")

        # Store results for this run
        run_results.append({
            'run': run + 1,
            'random_state': 41 + run,
            'test_loss': test_loss,
            'final_train_loss': history.history['loss'][-1],
            'final_val_loss': history.history['val_loss'][-1],
            'epochs': len(history.history['loss'])
        })

        # Prepare data for storing in CSV
        train_df = prepare_data(autoencoder, train_data, None, 'train', y_train)
        val_df = prepare_data(autoencoder, val_data, None, 'validation', y_val)
        test_df = prepare_data(autoencoder, test_data, None, 'test', y_test)
        complete_df = pd.concat([train_df, val_df, test_df], ignore_index=True)

        # Save the individual run data
        try:
            complete_df.to_csv(os.path.join(DATA_DIR, f'PS1_grizy_run_{run + 1}.csv'), index=False)
            pd.DataFrame(history.history).to_csv(os.path.join(DATA_DIR, f'PS1_th_grizy_run_{run + 1}.csv'), index=False)
        except Exception as e:
            print(f"Error saving CSV files for run {run + 1}: {e}")  

    # Save all run results to a single CSV file after all runs
    run_results_df = pd.DataFrame(run_results)
    run_results_df.to_csv(os.path.join(DATA_DIR, 'PS1_grizy_run_results.csv'), index=False)

    print("All runs completed. Results saved.")

\end{lstlisting}

\subsection{Supplemental Figures: Anomaly Images and SED Shapes}

\begin{figure*}[htbp]
    \centering
    \begin{minipage}[c]{0.23\textwidth}
        \centering
        \begin{overpic}[width=0.45\textwidth]{ 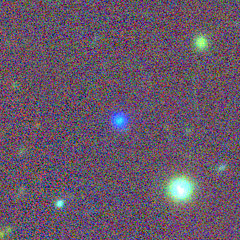 } 
            \put(5,5){\color{white}\bfseries\small giy }
        \end{overpic}%
        \includegraphics[width=0.45\textwidth]{ 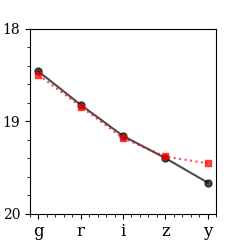 }
    \end{minipage}%
    \begin{minipage}[c]{0.23\textwidth}
        \centering
        \begin{overpic}[width=0.45\textwidth]{ 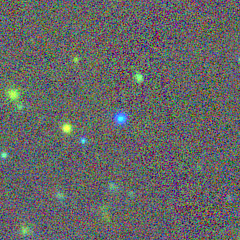 } 
            \put(5,5){\color{white}\bfseries\small giy }
        \end{overpic}%
        \includegraphics[width=0.45\textwidth]{ 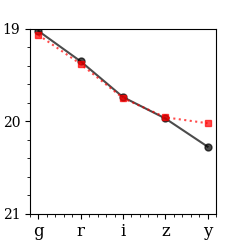 }
    \end{minipage}%
    \begin{minipage}[c]{0.23\textwidth}
        \centering
        \begin{overpic}[width=0.45\textwidth]{ 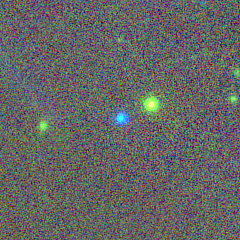 } 
            \put(5,5){\color{white}\bfseries\small giy }
        \end{overpic}%
        \includegraphics[width=0.45\textwidth]{ 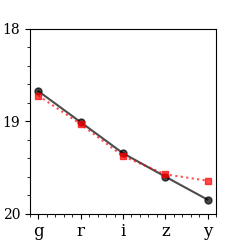 }
    \end{minipage}%
    \begin{minipage}[c]{0.23\textwidth}
        \centering
        \begin{overpic}[width=0.45\textwidth]{ anomaly_colored_by_giy_upper_159_y_184.1822_2.1433.png } 
            \put(5,5){\color{white}\bfseries\small giy }
        \end{overpic}%
        \includegraphics[width=0.45\textwidth]{ SED_upper_159.png }
    \end{minipage}%
    \\[1ex] 
    \begin{minipage}[c]{0.23\textwidth}
        \centering
        \begin{overpic}[width=0.45\textwidth]{ 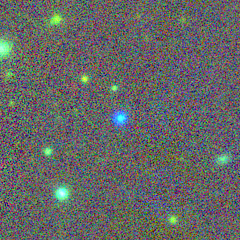 } 
            \put(5,5){\color{white}\bfseries\small giy }
        \end{overpic}%
        \includegraphics[width=0.45\textwidth]{ 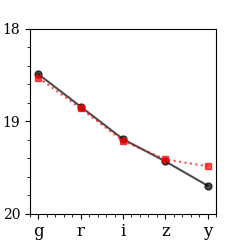 }
    \end{minipage}%
    \begin{minipage}[c]{0.23\textwidth}
        \centering
        \begin{overpic}[width=0.45\textwidth]{ 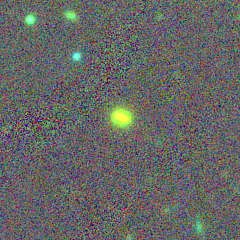 } 
            \put(5,5){\color{white}\bfseries\small giy }
        \end{overpic}%
        \includegraphics[width=0.45\textwidth]{ 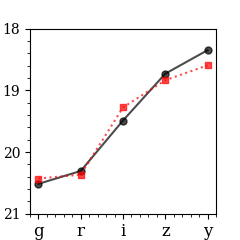 }
    \end{minipage}%
    \begin{minipage}[c]{0.23\textwidth}
        \centering
        \begin{overpic}[width=0.45\textwidth]{ 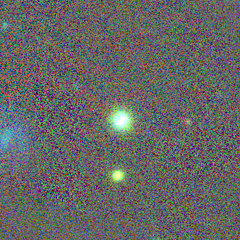 } 
            \put(5,5){\color{white}\bfseries\small giy }
        \end{overpic}%
        \includegraphics[width=0.45\textwidth]{ 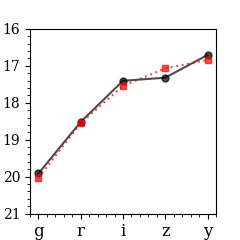 }
    \end{minipage}%
    \begin{minipage}[c]{0.23\textwidth}
        \centering
        \begin{overpic}[width=0.45\textwidth]{ 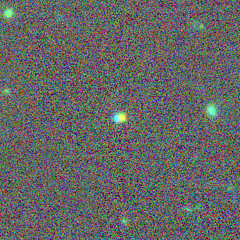 } 
            \put(5,5){\color{white}\bfseries\small giy }
        \end{overpic}%
        \includegraphics[width=0.45\textwidth]{ 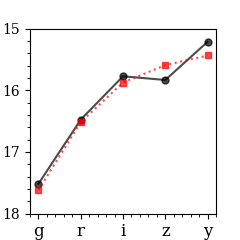 }
    \end{minipage}%
    \\[1ex] 
    \begin{minipage}[c]{0.23\textwidth}
        \centering
        \begin{overpic}[width=0.45\textwidth]{ 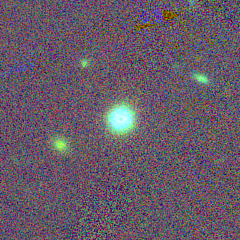 } 
            \put(5,5){\color{white}\bfseries\small giy }
        \end{overpic}%
        \includegraphics[width=0.45\textwidth]{ 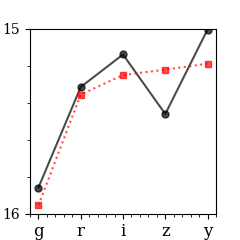 }
    \end{minipage}%
    \begin{minipage}[c]{0.23\textwidth}
        \centering
        \begin{overpic}[width=0.45\textwidth]{ 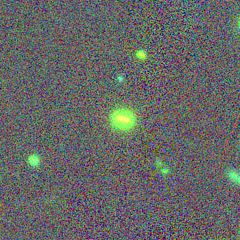 } 
            \put(5,5){\color{white}\bfseries\small giy }
        \end{overpic}%
        \includegraphics[width=0.45\textwidth]{ 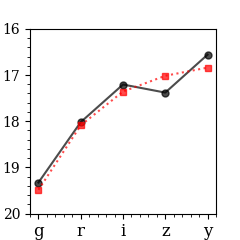 }
    \end{minipage}%
    \begin{minipage}[c]{0.23\textwidth}
        \centering
        \begin{overpic}[width=0.45\textwidth]{ 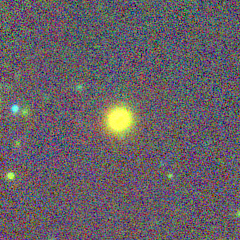 } 
            \put(5,5){\color{white}\bfseries\small giy }
        \end{overpic}%
        \includegraphics[width=0.45\textwidth]{ 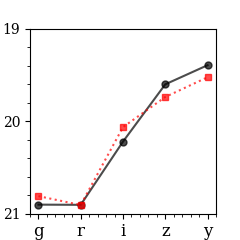 }
    \end{minipage}%
    \begin{minipage}[c]{0.23\textwidth}
        \centering
        \begin{overpic}[width=0.45\textwidth]{ 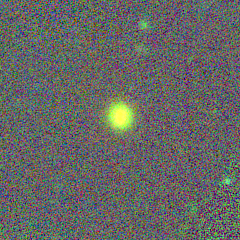 } 
            \put(5,5){\color{white}\bfseries\small giy }
        \end{overpic}%
        \includegraphics[width=0.45\textwidth]{ 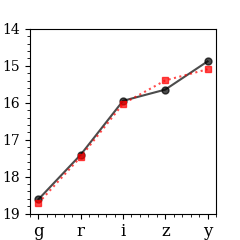 }
    \end{minipage}%
    \\[1ex] 
    \begin{minipage}[c]{0.23\textwidth}
        \centering
        \begin{overpic}[width=0.45\textwidth]{ 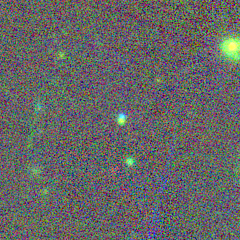 } 
            \put(5,5){\color{white}\bfseries\small giy }
        \end{overpic}%
        \includegraphics[width=0.45\textwidth]{ 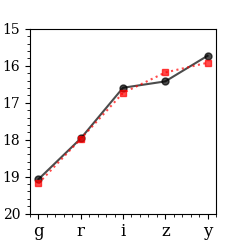 }
    \end{minipage}%
    \begin{minipage}[c]{0.23\textwidth}
        \centering
        \begin{overpic}[width=0.45\textwidth]{ anomaly_colored_by_giy_upper_132_z_165.0283_40.4012.png } 
            \put(5,5){\color{white}\bfseries\small giy }
        \end{overpic}%
        \includegraphics[width=0.45\textwidth]{ SED_upper_132.png }
    \end{minipage}%
    \begin{minipage}[c]{0.23\textwidth}
        \centering
        \begin{overpic}[width=0.45\textwidth]{ 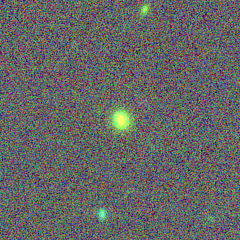 } 
            \put(5,5){\color{white}\bfseries\small giy }
        \end{overpic}%
        \includegraphics[width=0.45\textwidth]{ 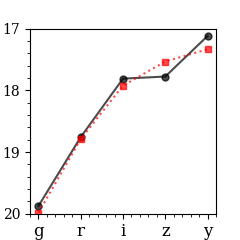 }
    \end{minipage}%
    \begin{minipage}[c]{0.23\textwidth}
        \centering
        \begin{overpic}[width=0.45\textwidth]{ 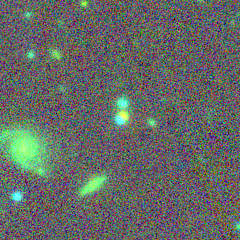 } 
            \put(5,5){\color{white}\bfseries\small giy }
        \end{overpic}%
        \includegraphics[width=0.45\textwidth]{ 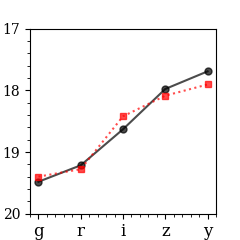 }
    \end{minipage}%
    \\[1ex] 
    \begin{minipage}[c]{0.23\textwidth}
        \centering
        \begin{overpic}[width=0.45\textwidth]{ 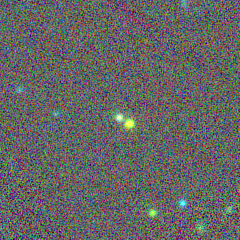 } 
            \put(5,5){\color{white}\bfseries\small giy }
        \end{overpic}%
        \includegraphics[width=0.45\textwidth]{ 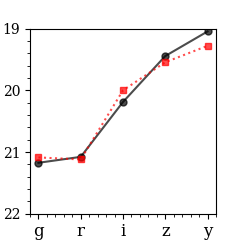 }
    \end{minipage}%
    \begin{minipage}[c]{0.23\textwidth}
        \centering
        \begin{overpic}[width=0.45\textwidth]{ 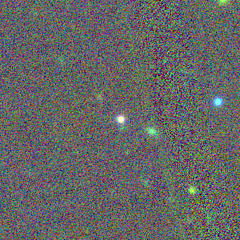 } 
            \put(5,5){\color{white}\bfseries\small giy }
        \end{overpic}%
        \includegraphics[width=0.45\textwidth]{ 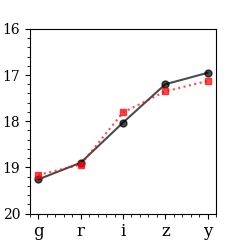 }
    \end{minipage}%
    \begin{minipage}[c]{0.23\textwidth}
        \centering
        \begin{overpic}[width=0.45\textwidth]{ 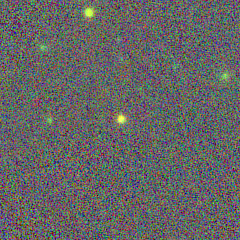 } 
            \put(5,5){\color{white}\bfseries\small giy }
        \end{overpic}%
        \includegraphics[width=0.45\textwidth]{ 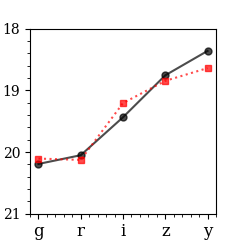 }
    \end{minipage}%
    \begin{minipage}[c]{0.23\textwidth}
        \centering
        \begin{overpic}[width=0.45\textwidth]{ anomaly_colored_by_giy_upper_26_g_165.5503_21.9261.png } 
            \put(5,5){\color{white}\bfseries\small giy }
        \end{overpic}%
        \includegraphics[width=0.45\textwidth]{ 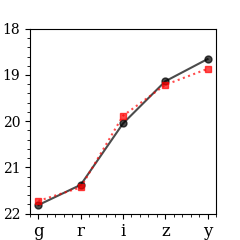 }
    \end{minipage}%
    \\[1ex] 
    \begin{minipage}[c]{0.23\textwidth}
        \centering
        \begin{overpic}[width=0.45\textwidth]{ 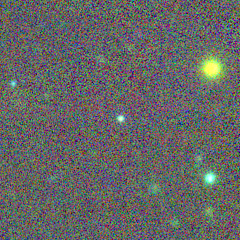 } 
            \put(5,5){\color{white}\bfseries\small giy }
        \end{overpic}%
        \includegraphics[width=0.45\textwidth]{ SED_upper_26.png }
    \end{minipage}%
    \begin{minipage}[c]{0.23\textwidth}
        \centering
        \begin{overpic}[width=0.45\textwidth]{ 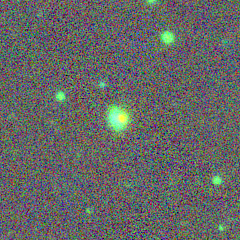 } 
            \put(5,5){\color{white}\bfseries\small giy }
        \end{overpic}%
        \includegraphics[width=0.45\textwidth]{ 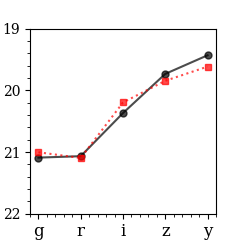 }
    \end{minipage}%
    \begin{minipage}[c]{0.23\textwidth}
        \centering
        \begin{overpic}[width=0.45\textwidth]{ 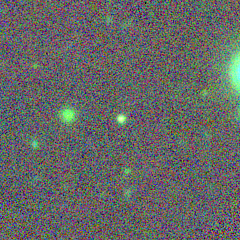 } 
            \put(5,5){\color{white}\bfseries\small giy }
        \end{overpic}%
        \includegraphics[width=0.45\textwidth]{ 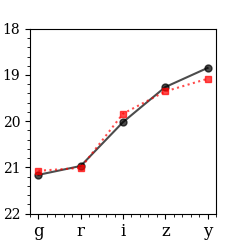 }
    \end{minipage}%
    \begin{minipage}[c]{0.23\textwidth}
        \centering
        \begin{overpic}[width=0.45\textwidth]{ 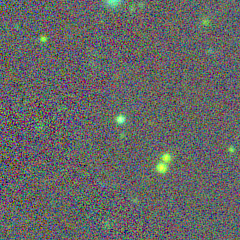 } 
            \put(5,5){\color{white}\bfseries\small giy }
        \end{overpic}%
        \includegraphics[width=0.45\textwidth]{ 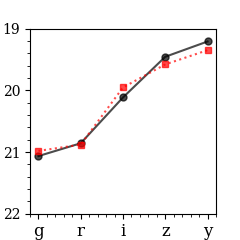 }
    \end{minipage}%
    \\[1ex] 
    \begin{minipage}[c]{0.23\textwidth}
        \centering
        \begin{overpic}[width=0.45\textwidth]{ 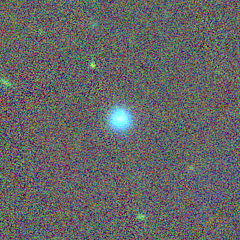 } 
            \put(5,5){\color{white}\bfseries\small giy }
        \end{overpic}%
        \includegraphics[width=0.45\textwidth]{ 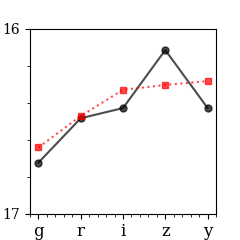 }
    \end{minipage}%
    \begin{minipage}[c]{0.23\textwidth}
        \centering
        \begin{overpic}[width=0.45\textwidth]{ 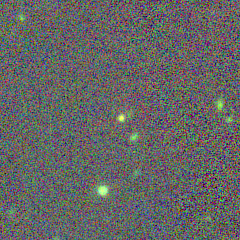 } 
            \put(5,5){\color{white}\bfseries\small giy }
        \end{overpic}%
        \includegraphics[width=0.45\textwidth]{ 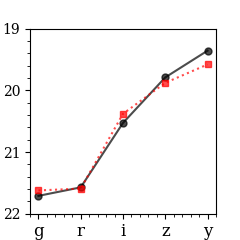 }
    \end{minipage}%
    \begin{minipage}[c]{0.23\textwidth}
        \centering
        \begin{overpic}[width=0.45\textwidth]{ 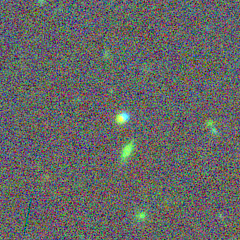 } 
            \put(5,5){\color{white}\bfseries\small giy }
        \end{overpic}%
        \includegraphics[width=0.45\textwidth]{ 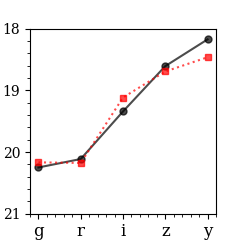 }
    \end{minipage}%
    \begin{minipage}[c]{0.23\textwidth}
        \centering
        \begin{overpic}[width=0.45\textwidth]{ 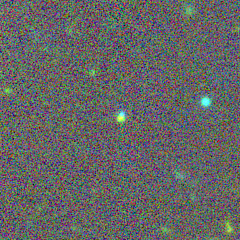 } 
            \put(5,5){\color{white}\bfseries\small giy }
        \end{overpic}%
        \includegraphics[width=0.45\textwidth]{ 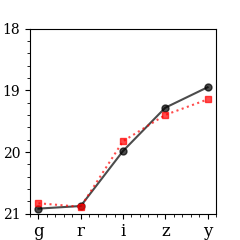 }
    \end{minipage}%
    \\[1ex] 
    \begin{minipage}[c]{0.23\textwidth}
        \centering
        \begin{overpic}[width=0.45\textwidth]{ 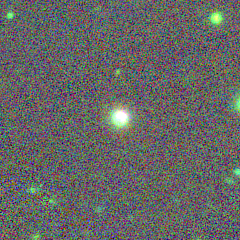 } 
            \put(5,5){\color{white}\bfseries\small giy }
        \end{overpic}%
        \includegraphics[width=0.45\textwidth]{ 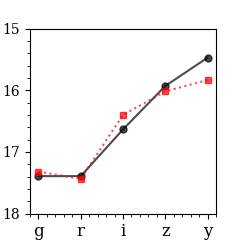 }
    \end{minipage}%
    \begin{minipage}[c]{0.23\textwidth}
        \centering
        \begin{overpic}[width=0.45\textwidth]{ 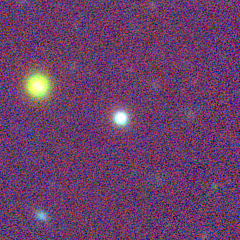 } 
            \put(5,5){\color{white}\bfseries\small giy }
        \end{overpic}%
        \includegraphics[width=0.45\textwidth]{ 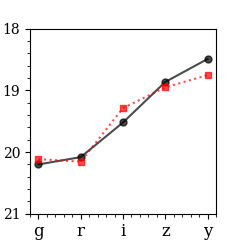 }
    \end{minipage}%
    \begin{minipage}[c]{0.23\textwidth}
        \centering
        \begin{overpic}[width=0.45\textwidth]{ 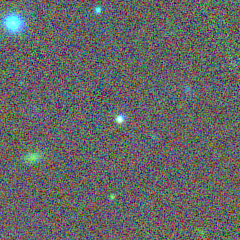 } 
            \put(5,5){\color{white}\bfseries\small giy }
        \end{overpic}%
        \includegraphics[width=0.45\textwidth]{ 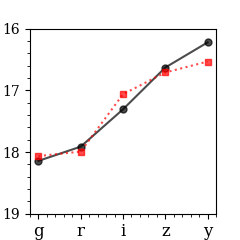 }
    \end{minipage}%
    \begin{minipage}[c]{0.23\textwidth}
        \centering
        \begin{overpic}[width=0.45\textwidth]{ 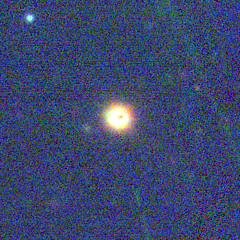 } 
            \put(5,5){\color{white}\bfseries\small gri }
        \end{overpic}%
        \includegraphics[width=0.45\textwidth]{ 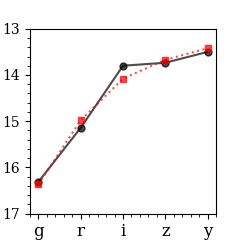 }
    \end{minipage}%
    \\[1ex] 
    \begin{minipage}[c]{0.23\textwidth}
        \centering
        \begin{overpic}[width=0.45\textwidth]{ anomaly_colored_by_gri_upper_69_r_167.942_16.4559.png } 
            \put(5,5){\color{white}\bfseries\small gri }
        \end{overpic}%
        \includegraphics[width=0.45\textwidth]{ SED_upper_69.png }
    \end{minipage}%
    \begin{minipage}[c]{0.23\textwidth}
        \centering
        \begin{overpic}[width=0.45\textwidth]{ 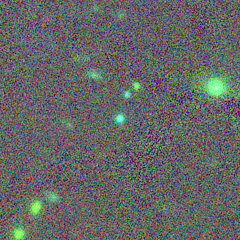 } 
            \put(5,5){\color{white}\bfseries\small giy }
        \end{overpic}%
        \includegraphics[width=0.45\textwidth]{ 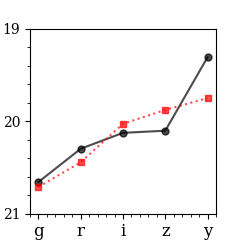 }
    \end{minipage}%
    \begin{minipage}[c]{0.23\textwidth}
        \centering
        \begin{overpic}[width=0.45\textwidth]{ anomaly_colored_by_giy_upper_110_z_185.0365_9.8834.png } 
            \put(5,5){\color{white}\bfseries\small giy }
        \end{overpic}%
        \includegraphics[width=0.45\textwidth]{ SED_upper_110.png }
    \end{minipage}%
    \begin{minipage}[c]{0.23\textwidth}
        \centering
        \begin{overpic}[width=0.45\textwidth]{ 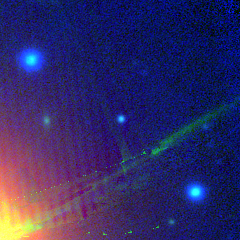 } 
            \put(5,5){\color{white}\bfseries\small giy }
        \end{overpic}%
        \includegraphics[width=0.45\textwidth]{ 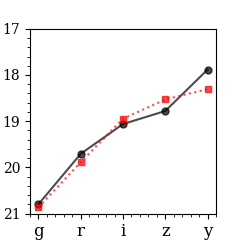 }
    \end{minipage}%
    \\[1ex] 
    \begin{minipage}[c]{0.23\textwidth}
        \centering
        \begin{overpic}[width=0.45\textwidth]{ 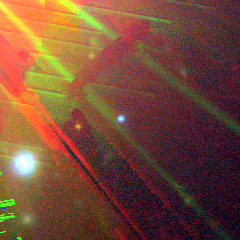 } 
            \put(5,5){\color{white}\bfseries\small giy }
        \end{overpic}%
        \includegraphics[width=0.45\textwidth]{ 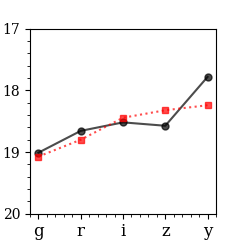 }
    \end{minipage}%
    \begin{minipage}[c]{0.23\textwidth}
        \centering
        \begin{overpic}[width=0.45\textwidth]{ 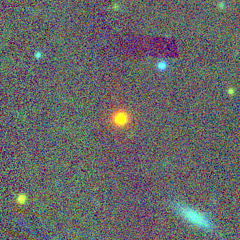 } 
            \put(5,5){\color{white}\bfseries\small giy }
        \end{overpic}%
        \includegraphics[width=0.45\textwidth]{ 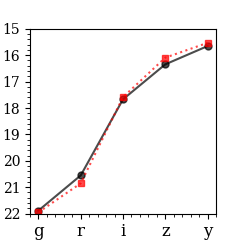 }
    \end{minipage}%
    \begin{minipage}[c]{0.23\textwidth}
        \centering
        \begin{overpic}[width=0.45\textwidth]{ 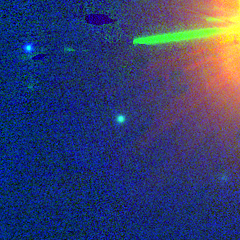 } 
            \put(5,5){\color{white}\bfseries\small giy }
        \end{overpic}%
        \includegraphics[width=0.45\textwidth]{ 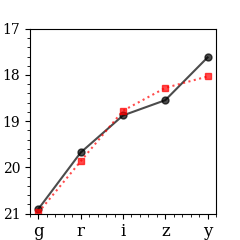 }
    \end{minipage}%
    \begin{minipage}[c]{0.23\textwidth}
        \centering
        \begin{overpic}[width=0.45\textwidth]{ 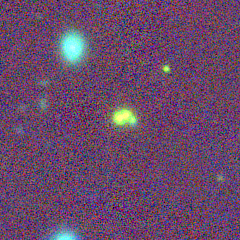 } 
            \put(5,5){\color{white}\bfseries\small giy }
        \end{overpic}%
        \includegraphics[width=0.45\textwidth]{ 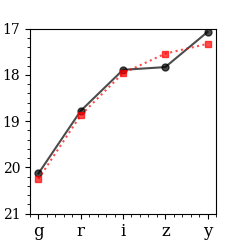 }
    \end{minipage}%
    \\[1ex] 
    \begin{minipage}[c]{0.23\textwidth}
        \centering
        \begin{overpic}[width=0.45\textwidth]{ 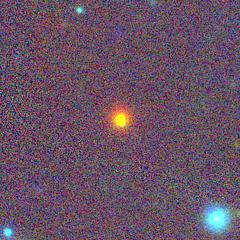 } 
            \put(5,5){\color{white}\bfseries\small giy }
        \end{overpic}%
        \includegraphics[width=0.45\textwidth]{ 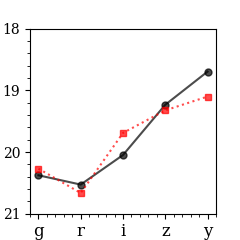 }
    \end{minipage}%
    \begin{minipage}[c]{0.23\textwidth}
        \centering
        \begin{overpic}[width=0.45\textwidth]{ 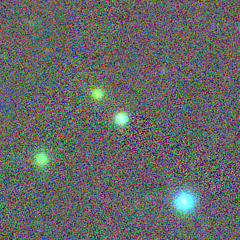 } 
            \put(5,5){\color{white}\bfseries\small giy }
        \end{overpic}%
        \includegraphics[width=0.45\textwidth]{ 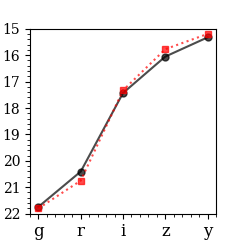 }
    \end{minipage}%
    \begin{minipage}[c]{0.23\textwidth}
        \centering
        \begin{overpic}[width=0.45\textwidth]{ 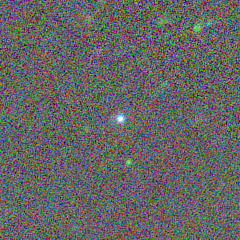 } 
            \put(5,5){\color{white}\bfseries\small giy }
        \end{overpic}%
        \includegraphics[width=0.45\textwidth]{ 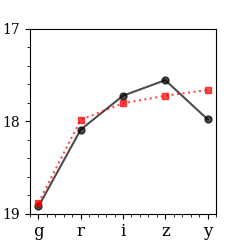 }
    \end{minipage}%
    \begin{minipage}[c]{0.23\textwidth}
        \centering
        \begin{overpic}[width=0.45\textwidth]{ 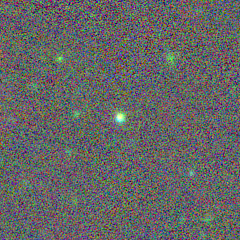 } 
            \put(5,5){\color{white}\bfseries\small giy }
        \end{overpic}%
        \includegraphics[width=0.45\textwidth]{ 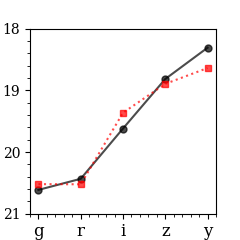 }
    \end{minipage}%
    \\[1ex] 
    \begin{minipage}[c]{0.23\textwidth}
        \centering
        \begin{overpic}[width=0.45\textwidth]{ 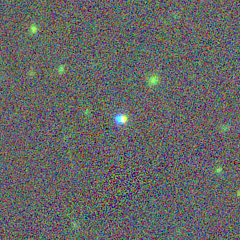 } 
            \put(5,5){\color{white}\bfseries\small giy }
        \end{overpic}%
        \includegraphics[width=0.45\textwidth]{ 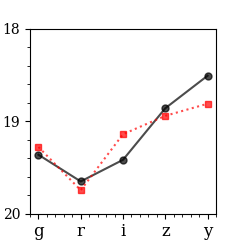 }
    \end{minipage}%
    \begin{minipage}[c]{0.23\textwidth}
        \centering
        \begin{overpic}[width=0.45\textwidth]{ 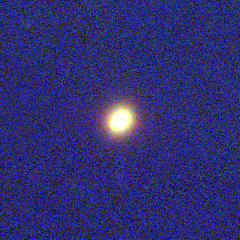 } 
            \put(5,5){\color{white}\bfseries\small giy }
        \end{overpic}%
        \includegraphics[width=0.45\textwidth]{ 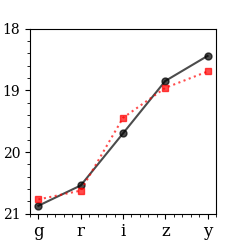 }
    \end{minipage}%
    \begin{minipage}[c]{0.23\textwidth}
        \centering
        \begin{overpic}[width=0.45\textwidth]{ 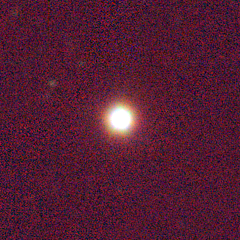 } 
            \put(5,5){\color{white}\bfseries\small giy }
        \end{overpic}%
        \includegraphics[width=0.45\textwidth]{ 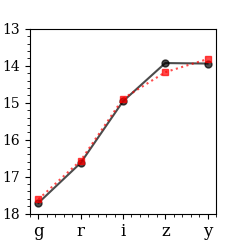 }
    \end{minipage}%
    \begin{minipage}[c]{0.23\textwidth}
        \centering
        \begin{overpic}[width=0.45\textwidth]{ 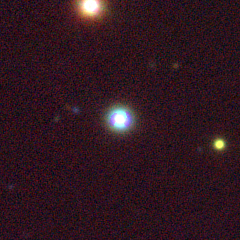 } 
            \put(5,5){\color{white}\bfseries\small giy }
        \end{overpic}%
        \includegraphics[width=0.45\textwidth]{ 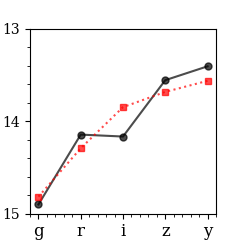 }
    \end{minipage}%
    \\[1ex] 
    \end{figure*}  

    \begin{figure*}[htbp]
        \centering
    \begin{minipage}[c]{0.23\textwidth}
        \centering
        \begin{overpic}[width=0.45\textwidth]{ 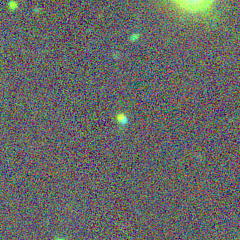 } 
            \put(5,5){\color{white}\bfseries\small giy }
        \end{overpic}%
        \includegraphics[width=0.45\textwidth]{ 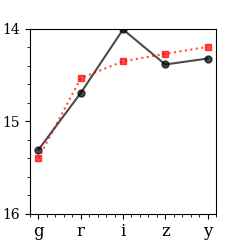 }
    \end{minipage}%
    \begin{minipage}[c]{0.23\textwidth}
        \centering
        \begin{overpic}[width=0.45\textwidth]{ 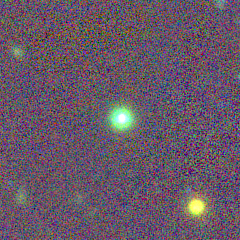 } 
            \put(5,5){\color{white}\bfseries\small giy }
        \end{overpic}%
        \includegraphics[width=0.45\textwidth]{ 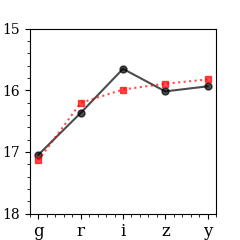 }
    \end{minipage}%
    \begin{minipage}[c]{0.23\textwidth}
        \centering
        \begin{overpic}[width=0.45\textwidth]{ 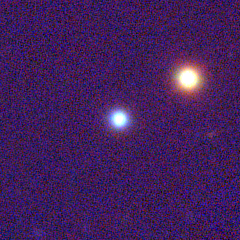 } 
            \put(5,5){\color{white}\bfseries\small giy }
        \end{overpic}%
        \includegraphics[width=0.45\textwidth]{ 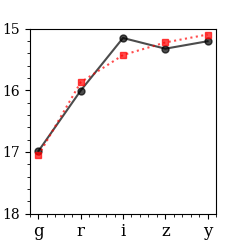 }
    \end{minipage}%
    \begin{minipage}[c]{0.23\textwidth}
        \centering
        \begin{overpic}[width=0.45\textwidth]{ 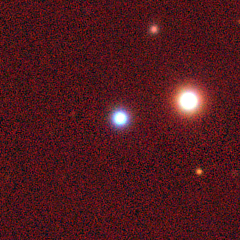 } 
            \put(5,5){\color{white}\bfseries\small giy }
        \end{overpic}%
        \includegraphics[width=0.45\textwidth]{ 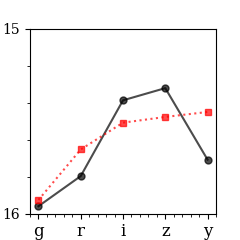 }
    \end{minipage}%
    \\[1ex] 
    \begin{minipage}[c]{0.23\textwidth}
        \centering
        \begin{overpic}[width=0.45\textwidth]{ 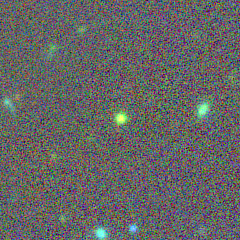 } 
            \put(5,5){\color{white}\bfseries\small giy }
        \end{overpic}%
        \includegraphics[width=0.45\textwidth]{ 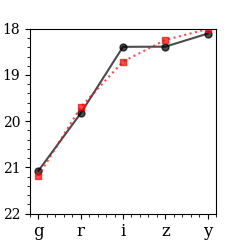 }
    \end{minipage}%
    \begin{minipage}[c]{0.23\textwidth}
        \centering
        \begin{overpic}[width=0.45\textwidth]{ 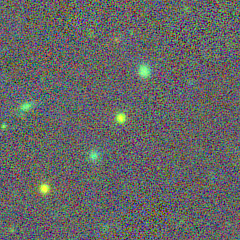 } 
            \put(5,5){\color{white}\bfseries\small giy }
        \end{overpic}%
        \includegraphics[width=0.45\textwidth]{ 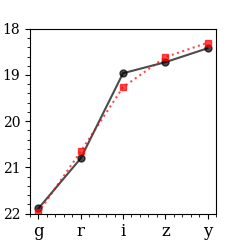 }
    \end{minipage}%
    \begin{minipage}[c]{0.23\textwidth}
        \centering
        \begin{overpic}[width=0.45\textwidth]{ 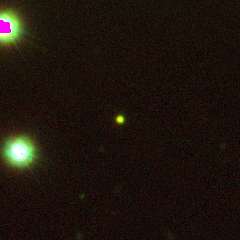 } 
            \put(5,5){\color{white}\bfseries\small giy }
        \end{overpic}%
        \includegraphics[width=0.45\textwidth]{ 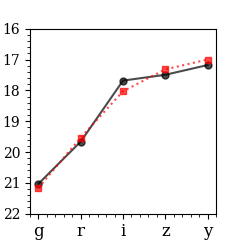 }
    \end{minipage}%
    \begin{minipage}[c]{0.23\textwidth}
        \centering
        \begin{overpic}[width=0.45\textwidth]{ 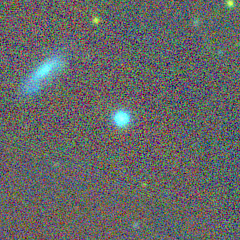 } 
            \put(5,5){\color{white}\bfseries\small giy }
        \end{overpic}%
        \includegraphics[width=0.45\textwidth]{ 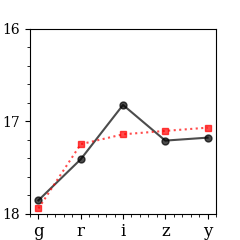 }
    \end{minipage}%
    \\[1ex] 
    \begin{minipage}[c]{0.23\textwidth}
        \centering
        \begin{overpic}[width=0.45\textwidth]{ 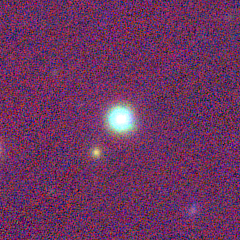 } 
            \put(5,5){\color{white}\bfseries\small giy }
        \end{overpic}%
        \includegraphics[width=0.45\textwidth]{ 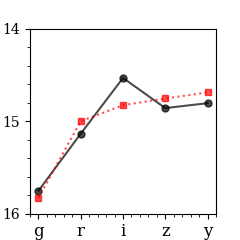 }
    \end{minipage}%
    \begin{minipage}[c]{0.23\textwidth}
        \centering
        \begin{overpic}[width=0.45\textwidth]{ 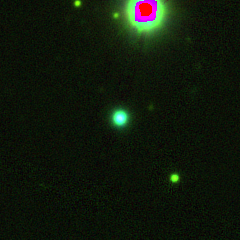 } 
            \put(5,5){\color{white}\bfseries\small giy }
        \end{overpic}%
        \includegraphics[width=0.45\textwidth]{ 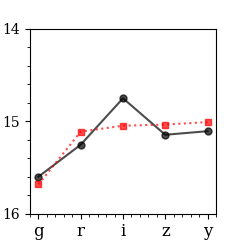 }
    \end{minipage}%
    \begin{minipage}[c]{0.23\textwidth}
        \centering
        \begin{overpic}[width=0.45\textwidth]{ 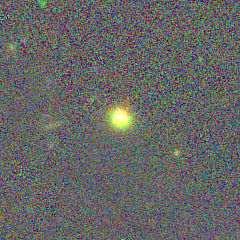 } 
            \put(5,5){\color{white}\bfseries\small giy }
        \end{overpic}%
        \includegraphics[width=0.45\textwidth]{ 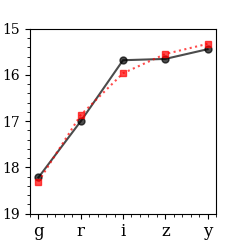 }
    \end{minipage}%
    \begin{minipage}[c]{0.23\textwidth}
        \centering
        \begin{overpic}[width=0.45\textwidth]{ 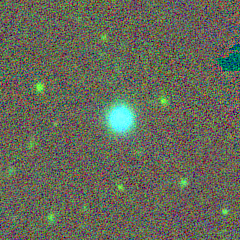 } 
            \put(5,5){\color{white}\bfseries\small giy }
        \end{overpic}%
        \includegraphics[width=0.45\textwidth]{ 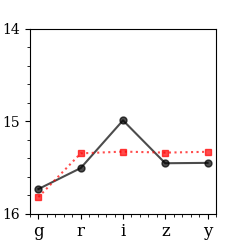 }
    \end{minipage}%
    \\[1ex] 
    \begin{minipage}[c]{0.23\textwidth}
        \centering
        \begin{overpic}[width=0.45\textwidth]{ 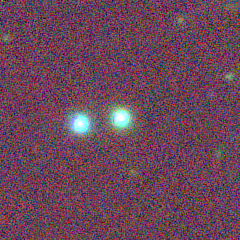 } 
            \put(5,5){\color{white}\bfseries\small giy }
        \end{overpic}%
        \includegraphics[width=0.45\textwidth]{ 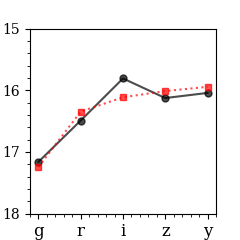 }
    \end{minipage}%
    \begin{minipage}[c]{0.23\textwidth}
        \centering
        \begin{overpic}[width=0.45\textwidth]{ 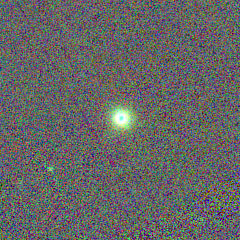 } 
            \put(5,5){\color{white}\bfseries\small giy }
        \end{overpic}%
        \includegraphics[width=0.45\textwidth]{ 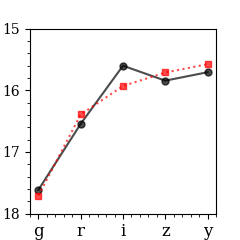 }
    \end{minipage}%
    \begin{minipage}[c]{0.23\textwidth}
        \centering
        \begin{overpic}[width=0.45\textwidth]{ 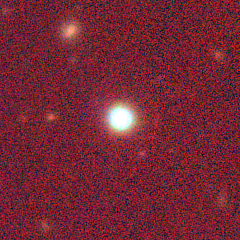 } 
            \put(5,5){\color{white}\bfseries\small giy }
        \end{overpic}%
        \includegraphics[width=0.45\textwidth]{ 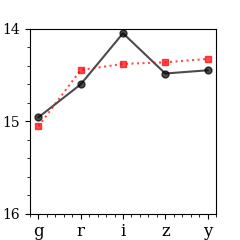 }
    \end{minipage}%
    \begin{minipage}[c]{0.23\textwidth}
        \centering
        \begin{overpic}[width=0.45\textwidth]{ 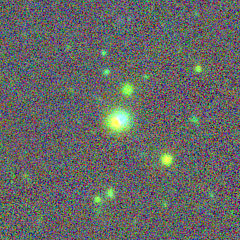 } 
            \put(5,5){\color{white}\bfseries\small giy }
        \end{overpic}%
        \includegraphics[width=0.45\textwidth]{ 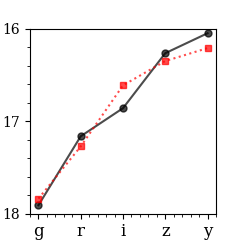 }
    \end{minipage}%
    \\[1ex] 
    \begin{minipage}[c]{0.23\textwidth}
        \centering
        \begin{overpic}[width=0.45\textwidth]{ 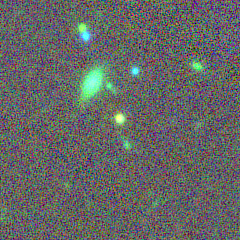 } 
            \put(5,5){\color{white}\bfseries\small giy }
        \end{overpic}%
        \includegraphics[width=0.45\textwidth]{ 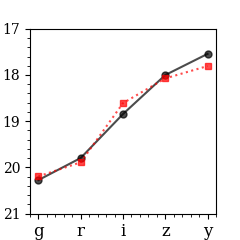 }
    \end{minipage}%
    \begin{minipage}[c]{0.23\textwidth}
        \centering
        \begin{overpic}[width=0.45\textwidth]{ 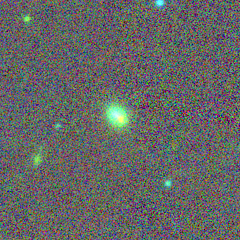 } 
            \put(5,5){\color{white}\bfseries\small giy }
        \end{overpic}%
        \includegraphics[width=0.45\textwidth]{ 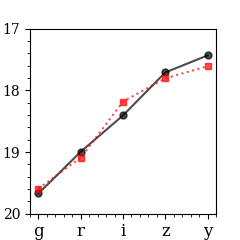 }
    \end{minipage}%
    \begin{minipage}[c]{0.23\textwidth}
        \centering
        \begin{overpic}[width=0.45\textwidth]{ 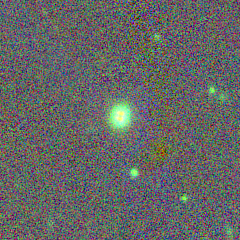 } 
            \put(5,5){\color{white}\bfseries\small giy }
        \end{overpic}%
        \includegraphics[width=0.45\textwidth]{ 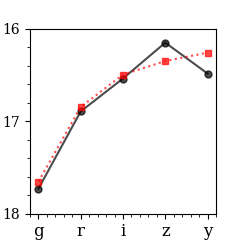 }
    \end{minipage}%
    \begin{minipage}[c]{0.23\textwidth}
        \centering
        \begin{overpic}[width=0.45\textwidth]{ 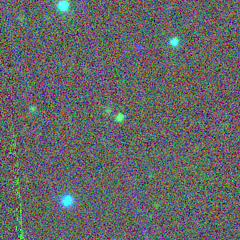 } 
            \put(5,5){\color{white}\bfseries\small giy }
        \end{overpic}%
        \includegraphics[width=0.45\textwidth]{ 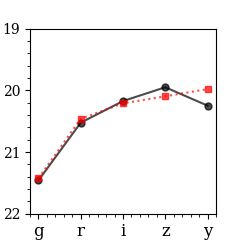 }
    \end{minipage}%
    \\[1ex] 
    \begin{minipage}[c]{0.23\textwidth}
        \centering
        \begin{overpic}[width=0.45\textwidth]{ 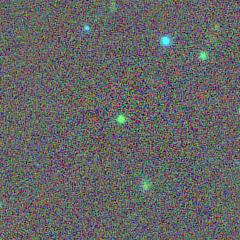 } 
            \put(5,5){\color{white}\bfseries\small giy }
        \end{overpic}%
        \includegraphics[width=0.45\textwidth]{ 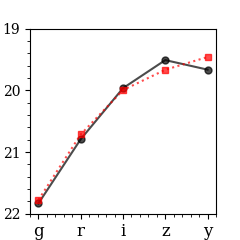 }
    \end{minipage}%
    \begin{minipage}[c]{0.23\textwidth}
        \centering
        \begin{overpic}[width=0.45\textwidth]{ 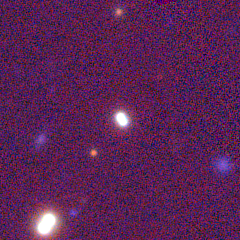 } 
            \put(5,5){\color{white}\bfseries\small giy }
        \end{overpic}%
        \includegraphics[width=0.45\textwidth]{ 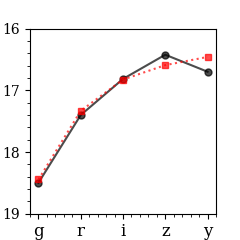 }
    \end{minipage}%
    \begin{minipage}[c]{0.23\textwidth}
        \centering
        \begin{overpic}[width=0.45\textwidth]{ 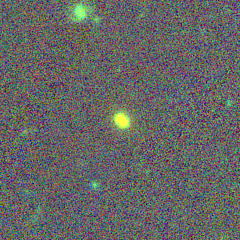 } 
            \put(5,5){\color{white}\bfseries\small giy }
        \end{overpic}%
        \includegraphics[width=0.45\textwidth]{ 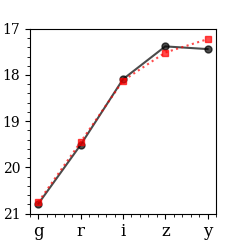 }
    \end{minipage}%
    \begin{minipage}[c]{0.23\textwidth}
        \centering
        \begin{overpic}[width=0.45\textwidth]{ 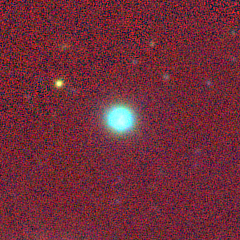 } 
            \put(5,5){\color{white}\bfseries\small riy }
        \end{overpic}%
        \includegraphics[width=0.45\textwidth]{ 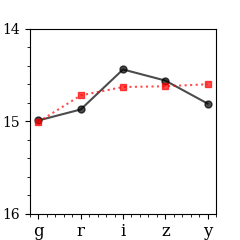 }
    \end{minipage}%
    \\[1ex] 
    \begin{minipage}[c]{0.23\textwidth}
        \centering
        \begin{overpic}[width=0.45\textwidth]{ 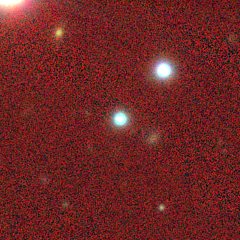 } 
            \put(5,5){\color{white}\bfseries\small riy }
        \end{overpic}%
        \includegraphics[width=0.45\textwidth]{ 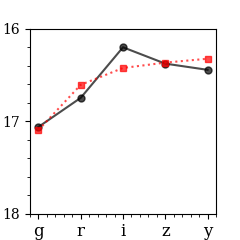 }
    \end{minipage}%
    \begin{minipage}[c]{0.23\textwidth}
        \centering
        \begin{overpic}[width=0.45\textwidth]{ 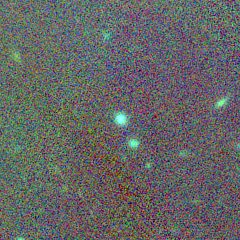 } 
            \put(5,5){\color{white}\bfseries\small riy }
        \end{overpic}%
        \includegraphics[width=0.45\textwidth]{ 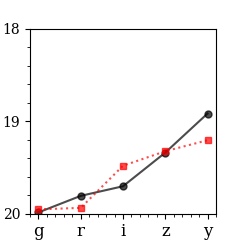 }
    \end{minipage}%
    \begin{minipage}[c]{0.23\textwidth}
        \centering
        \begin{overpic}[width=0.45\textwidth]{ 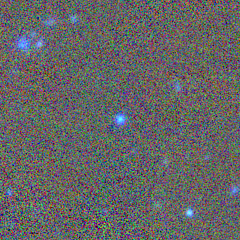 } 
            \put(5,5){\color{white}\bfseries\small izy }
        \end{overpic}%
        \includegraphics[width=0.45\textwidth]{ 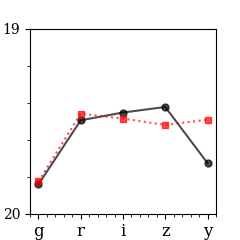 }
    \end{minipage}%
    \begin{minipage}[c]{0.23\textwidth}
        \centering
        \begin{overpic}[width=0.45\textwidth]{ anomaly_colored_by_riz_upper_43_r_187.0561_47.3451.png } 
            \put(5,5){\color{white}\bfseries\small riz }
        \end{overpic}%
        \includegraphics[width=0.45\textwidth]{ SED_upper_43.png }
    \end{minipage}%
    \\[1ex] 
    \begin{minipage}[c]{0.23\textwidth}
        \centering
        \begin{overpic}[width=0.45\textwidth]{ 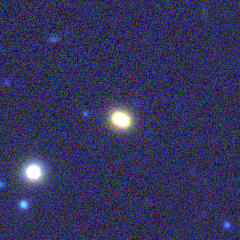 } 
            \put(5,5){\color{white}\bfseries\small riz }
        \end{overpic}%
        \includegraphics[width=0.45\textwidth]{ 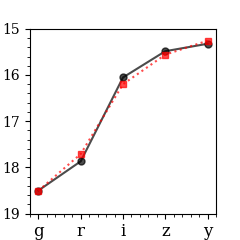 }
    \end{minipage}%
    \begin{minipage}[c]{0.23\textwidth}
        \centering
        \begin{overpic}[width=0.45\textwidth]{ 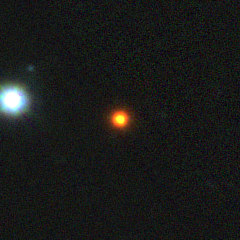 } 
            \put(5,5){\color{white}\bfseries\small riz }
        \end{overpic}%
        \includegraphics[width=0.45\textwidth]{ 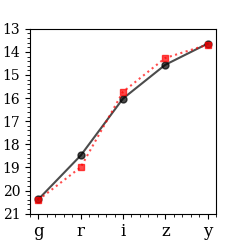 }
    \end{minipage}%
    \begin{minipage}[c]{0.23\textwidth}
        \centering
        \begin{overpic}[width=0.45\textwidth]{ 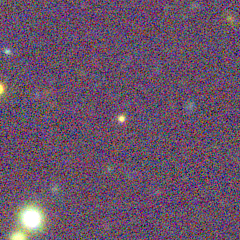 } 
            \put(5,5){\color{white}\bfseries\small giz }
        \end{overpic}%
        \includegraphics[width=0.45\textwidth]{ 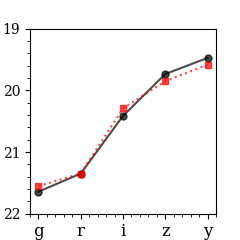 }
    \end{minipage}%
    \begin{minipage}[c]{0.23\textwidth}
        \centering
        \begin{overpic}[width=0.45\textwidth]{ anomaly_colored_by_giz_upper_11_g_160.817_25.2573.png } 
            \put(5,5){\color{white}\bfseries\small giz }
        \end{overpic}%
        \includegraphics[width=0.45\textwidth]{ 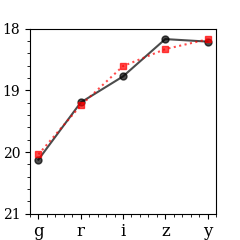 }
    \end{minipage}%
    \\[1ex] 
    \begin{minipage}[c]{0.23\textwidth}
        \centering
        \begin{overpic}[width=0.45\textwidth]{ 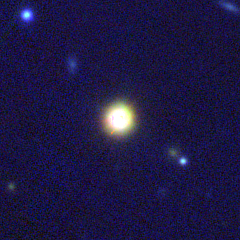 } 
            \put(5,5){\color{white}\bfseries\small giz }
        \end{overpic}%
        \includegraphics[width=0.45\textwidth]{ SED_upper_11.png }
    \end{minipage}%
    \begin{minipage}[c]{0.23\textwidth}
        \centering
        \begin{overpic}[width=0.45\textwidth]{ 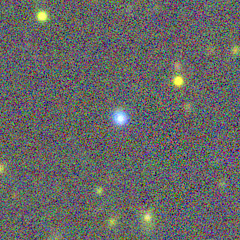 } 
            \put(5,5){\color{white}\bfseries\small giz }
        \end{overpic}%
        \includegraphics[width=0.45\textwidth]{ 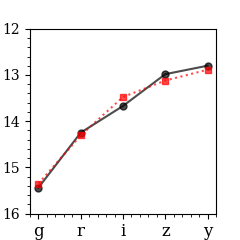 }
    \end{minipage}%
    \begin{minipage}[c]{0.23\textwidth}
        \centering
        \begin{overpic}[width=0.45\textwidth]{ 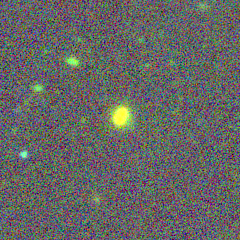 } 
            \put(5,5){\color{white}\bfseries\small giz }
        \end{overpic}%
        \includegraphics[width=0.45\textwidth]{ 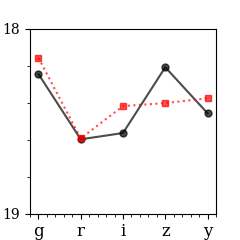 }
    \end{minipage}%
    \begin{minipage}[c]{0.23\textwidth}
        \centering
        \begin{overpic}[width=0.45\textwidth]{ 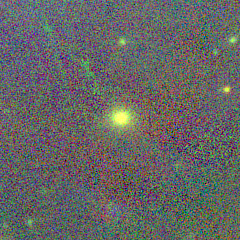 } 
            \put(5,5){\color{white}\bfseries\small giz }
        \end{overpic}%
        \includegraphics[width=0.45\textwidth]{ 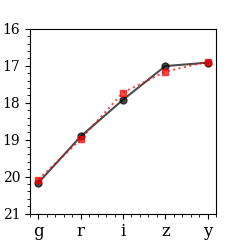 }
    \end{minipage}%
    \\[1ex] 
    \begin{minipage}[c]{0.23\textwidth}
        \centering
        \begin{overpic}[width=0.45\textwidth]{ 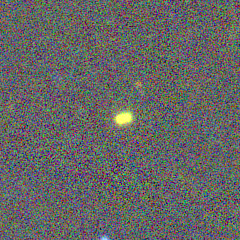 } 
            \put(5,5){\color{white}\bfseries\small giz }
        \end{overpic}%
        \includegraphics[width=0.45\textwidth]{ 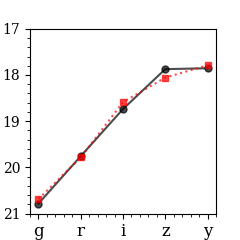 }
    \end{minipage}%
    \begin{minipage}[c]{0.23\textwidth}
        \centering
        \begin{overpic}[width=0.45\textwidth]{ 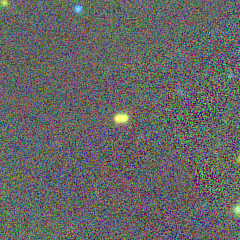 } 
            \put(5,5){\color{white}\bfseries\small giz }
        \end{overpic}%
        \includegraphics[width=0.45\textwidth]{ 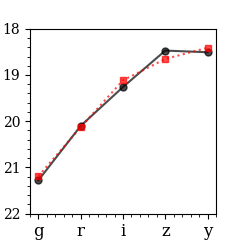 }
    \end{minipage}%
    \begin{minipage}[c]{0.23\textwidth}
        \centering
        \begin{overpic}[width=0.45\textwidth]{ 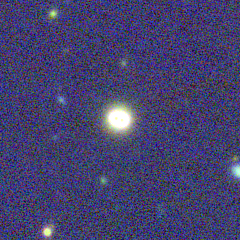 } 
            \put(5,5){\color{white}\bfseries\small giz }
        \end{overpic}%
        \includegraphics[width=0.45\textwidth]{ 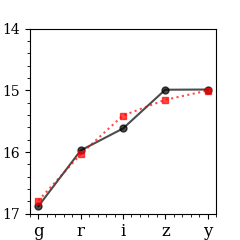 }
    \end{minipage}%
    \begin{minipage}[c]{0.23\textwidth}
        \centering
        \begin{overpic}[width=0.45\textwidth]{ 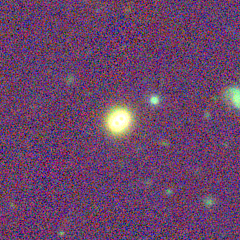 } 
            \put(5,5){\color{white}\bfseries\small giz }
        \end{overpic}%
        \includegraphics[width=0.45\textwidth]{ 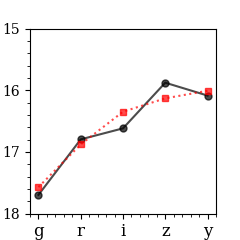 }
    \end{minipage}%
    \\[1ex] 
    \begin{minipage}[c]{0.23\textwidth}
        \centering
        \begin{overpic}[width=0.45\textwidth]{ 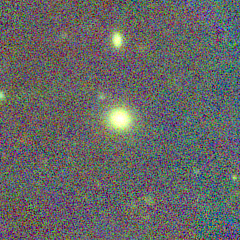 } 
            \put(5,5){\color{white}\bfseries\small giz }
        \end{overpic}%
        \includegraphics[width=0.45\textwidth]{ 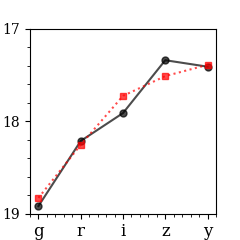 }
    \end{minipage}%
    \begin{minipage}[c]{0.23\textwidth}
        \centering
        \begin{overpic}[width=0.45\textwidth]{ 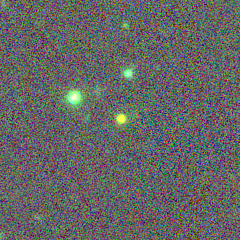 } 
            \put(5,5){\color{white}\bfseries\small giz }
        \end{overpic}%
        \includegraphics[width=0.45\textwidth]{ 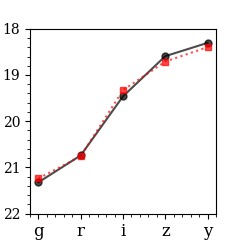 }
    \end{minipage}%
    \begin{minipage}[c]{0.23\textwidth}
        \centering
        \begin{overpic}[width=0.45\textwidth]{ 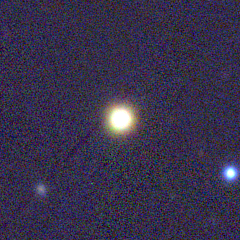 } 
            \put(5,5){\color{white}\bfseries\small giz }
        \end{overpic}%
        \includegraphics[width=0.45\textwidth]{ 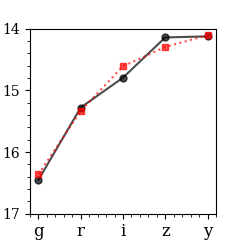 }
    \end{minipage}%
    \begin{minipage}[c]{0.23\textwidth}
        \centering
        \begin{overpic}[width=0.45\textwidth]{ 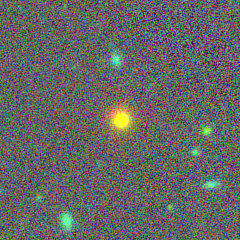 } 
            \put(5,5){\color{white}\bfseries\small giy }
        \end{overpic}%
        \includegraphics[width=0.45\textwidth]{ 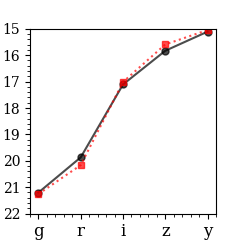 }
    \end{minipage}%
    \\[1ex] 
    \begin{minipage}[c]{0.23\textwidth}
        \centering
        \begin{overpic}[width=0.45\textwidth]{ 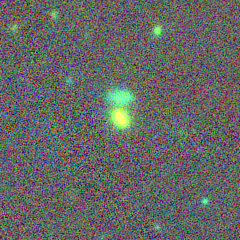 } 
            \put(5,5){\color{white}\bfseries\small giy }
        \end{overpic}%
        \includegraphics[width=0.45\textwidth]{ 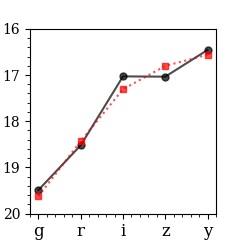 }
    \end{minipage}%
    \begin{minipage}[c]{0.23\textwidth}
        \centering
        \begin{overpic}[width=0.45\textwidth]{ 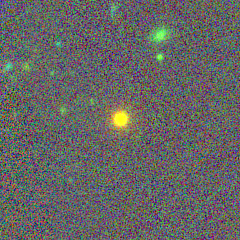 } 
            \put(5,5){\color{white}\bfseries\small giy }
        \end{overpic}%
        \includegraphics[width=0.45\textwidth]{ 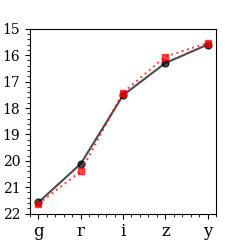 }
    \end{minipage}%
    \begin{minipage}[c]{0.23\textwidth}
        \centering
        \begin{overpic}[width=0.45\textwidth]{ 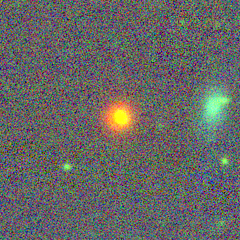 } 
            \put(5,5){\color{white}\bfseries\small giy }
        \end{overpic}%
        \includegraphics[width=0.45\textwidth]{ 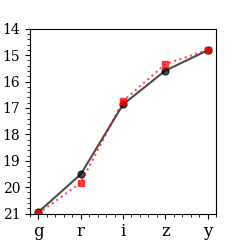 }
    \end{minipage}%
    \begin{minipage}[c]{0.23\textwidth}
        \centering
        \begin{overpic}[width=0.45\textwidth]{ 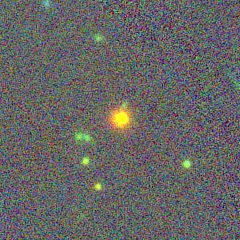 } 
            \put(5,5){\color{white}\bfseries\small giy }
        \end{overpic}%
        \includegraphics[width=0.45\textwidth]{ 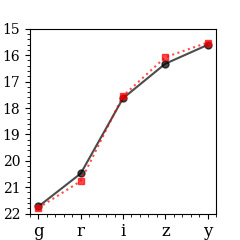 }
    \end{minipage}%
    \\[1ex] 
    \end{figure*}  

    \begin{figure*}[htbp]
        \centering
    \begin{minipage}[c]{0.23\textwidth}
        \centering
        \begin{overpic}[width=0.45\textwidth]{ 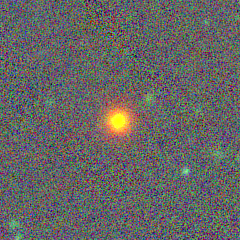 } 
            \put(5,5){\color{white}\bfseries\small giy }
        \end{overpic}%
        \includegraphics[width=0.45\textwidth]{ 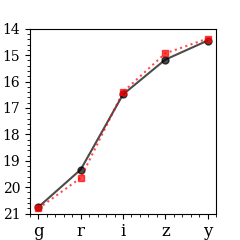 }
    \end{minipage}%
    \begin{minipage}[c]{0.23\textwidth}
        \centering
        \begin{overpic}[width=0.45\textwidth]{ 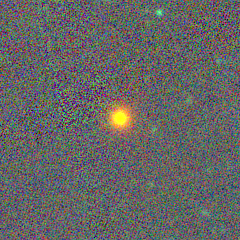 } 
            \put(5,5){\color{white}\bfseries\small giy }
        \end{overpic}%
        \includegraphics[width=0.45\textwidth]{ 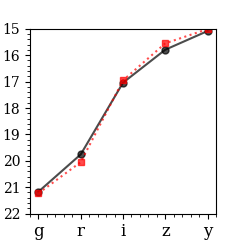 }
    \end{minipage}%
    \begin{minipage}[c]{0.23\textwidth}
        \centering
        \begin{overpic}[width=0.45\textwidth]{ 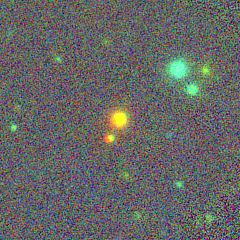 } 
            \put(5,5){\color{white}\bfseries\small giy }
        \end{overpic}%
        \includegraphics[width=0.45\textwidth]{ 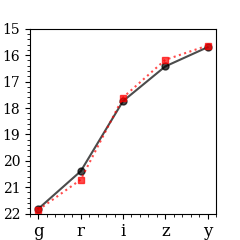 }
    \end{minipage}%
    \begin{minipage}[c]{0.23\textwidth}
        \centering
        \begin{overpic}[width=0.45\textwidth]{ 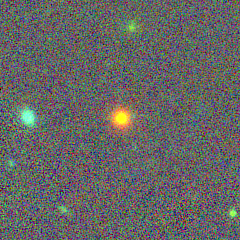 } 
            \put(5,5){\color{white}\bfseries\small giy }
        \end{overpic}%
        \includegraphics[width=0.45\textwidth]{ 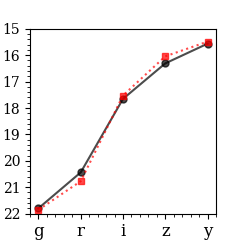 }
    \end{minipage}%
    \\[1ex] 
    \begin{minipage}[c]{0.23\textwidth}
        \centering
        \begin{overpic}[width=0.45\textwidth]{ 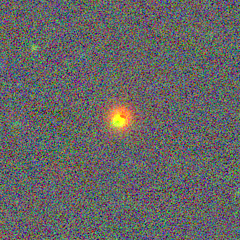 } 
            \put(5,5){\color{white}\bfseries\small giy }
        \end{overpic}%
        \includegraphics[width=0.45\textwidth]{ 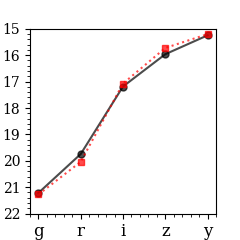 }
    \end{minipage}%
    \begin{minipage}[c]{0.23\textwidth}
        \centering
        \begin{overpic}[width=0.45\textwidth]{ 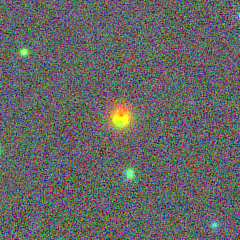 } 
            \put(5,5){\color{white}\bfseries\small giy }
        \end{overpic}%
        \includegraphics[width=0.45\textwidth]{ 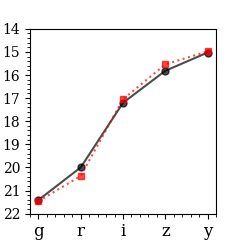 }
    \end{minipage}%
    \begin{minipage}[c]{0.23\textwidth}
        \centering
        \begin{overpic}[width=0.45\textwidth]{ 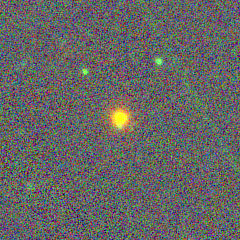 } 
            \put(5,5){\color{white}\bfseries\small giy }
        \end{overpic}%
        \includegraphics[width=0.45\textwidth]{ 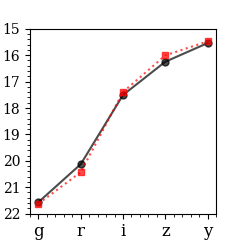 }
    \end{minipage}%
    \begin{minipage}[c]{0.23\textwidth}
        \centering
        \begin{overpic}[width=0.45\textwidth]{ 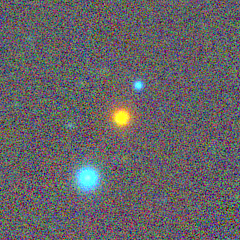 } 
            \put(5,5){\color{white}\bfseries\small giy }
        \end{overpic}%
        \includegraphics[width=0.45\textwidth]{ 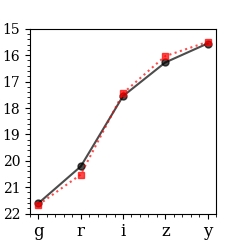 }
    \end{minipage}%
    \\[1ex] 
    \begin{minipage}[c]{0.23\textwidth}
        \centering
        \begin{overpic}[width=0.45\textwidth]{ 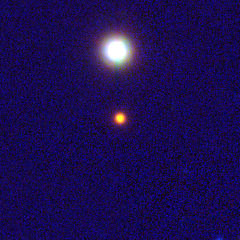 } 
            \put(5,5){\color{white}\bfseries\small giy }
        \end{overpic}%
        \includegraphics[width=0.45\textwidth]{ 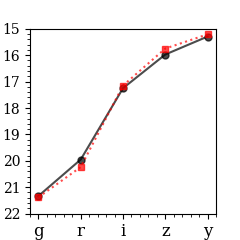 }
    \end{minipage}%
    \begin{minipage}[c]{0.23\textwidth}
        \centering
        \begin{overpic}[width=0.45\textwidth]{ 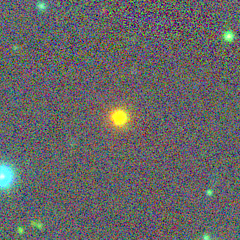 } 
            \put(5,5){\color{white}\bfseries\small giy }
        \end{overpic}%
        \includegraphics[width=0.45\textwidth]{ 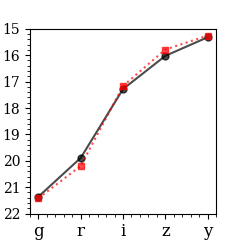 }
    \end{minipage}%
    \begin{minipage}[c]{0.23\textwidth}
        \centering
        \begin{overpic}[width=0.45\textwidth]{ 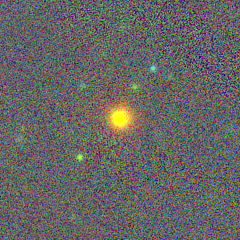 } 
            \put(5,5){\color{white}\bfseries\small giy }
        \end{overpic}%
        \includegraphics[width=0.45\textwidth]{ 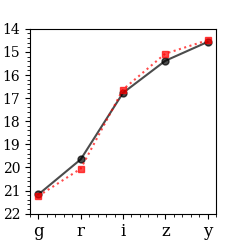 }
    \end{minipage}%
    \begin{minipage}[c]{0.23\textwidth}
        \centering
        \begin{overpic}[width=0.45\textwidth]{ 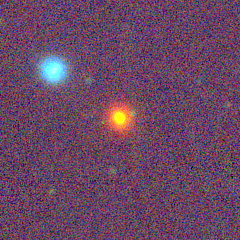 } 
            \put(5,5){\color{white}\bfseries\small giy }
        \end{overpic}%
        \includegraphics[width=0.45\textwidth]{ 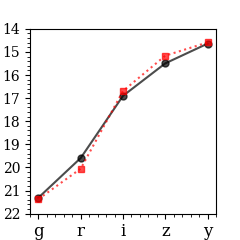 }
    \end{minipage}%
    \\[1ex] 
    \begin{minipage}[c]{0.23\textwidth}
        \centering
        \begin{overpic}[width=0.45\textwidth]{ 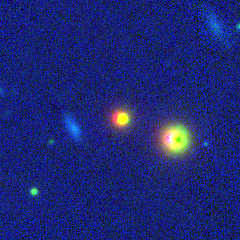 } 
            \put(5,5){\color{white}\bfseries\small giy }
        \end{overpic}%
        \includegraphics[width=0.45\textwidth]{ 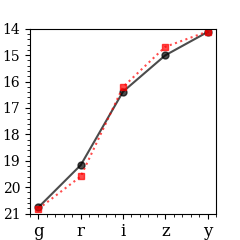 }
    \end{minipage}%
    \begin{minipage}[c]{0.23\textwidth}
        \centering
        \begin{overpic}[width=0.45\textwidth]{ 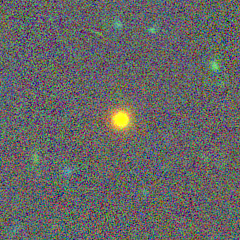 } 
            \put(5,5){\color{white}\bfseries\small giy }
        \end{overpic}%
        \includegraphics[width=0.45\textwidth]{ 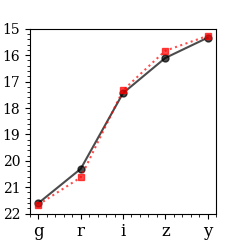 }
    \end{minipage}%
    \begin{minipage}[c]{0.23\textwidth}
        \centering
        \begin{overpic}[width=0.45\textwidth]{ 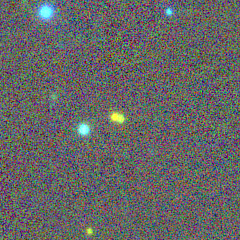 } 
            \put(5,5){\color{white}\bfseries\small giy }
        \end{overpic}%
        \includegraphics[width=0.45\textwidth]{ 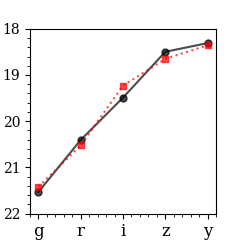 }
    \end{minipage}%
    \begin{minipage}[c]{0.23\textwidth}
        \centering
        \begin{overpic}[width=0.45\textwidth]{ 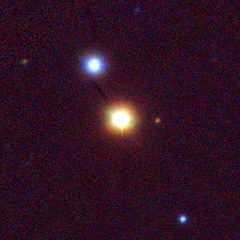 } 
            \put(5,5){\color{white}\bfseries\small giy }
        \end{overpic}%
        \includegraphics[width=0.45\textwidth]{ 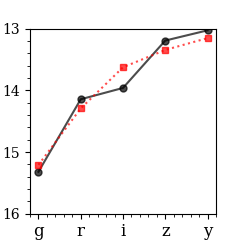 }
    \end{minipage}%
    \\[1ex] 
    \begin{minipage}[c]{0.23\textwidth}
        \centering
        \begin{overpic}[width=0.45\textwidth]{ 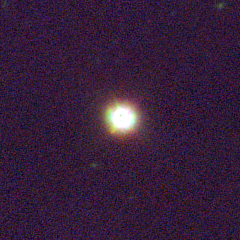 } 
            \put(5,5){\color{white}\bfseries\small giy }
        \end{overpic}%
        \includegraphics[width=0.45\textwidth]{ 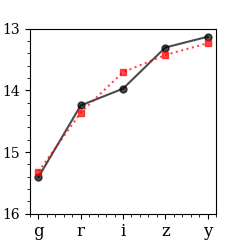 }
    \end{minipage}%
    \begin{minipage}[c]{0.23\textwidth}
        \centering
        \begin{overpic}[width=0.45\textwidth]{ 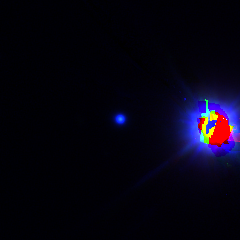 } 
            \put(5,5){\color{white}\bfseries\small giy }
        \end{overpic}%
        \includegraphics[width=0.45\textwidth]{ 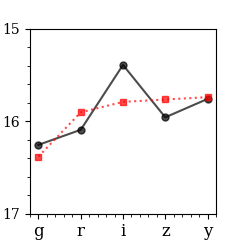 }
    \end{minipage}%
    \begin{minipage}[c]{0.23\textwidth}
        \centering
        \begin{overpic}[width=0.45\textwidth]{ 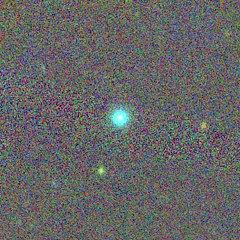 } 
            \put(5,5){\color{white}\bfseries\small giy }
        \end{overpic}%
        \includegraphics[width=0.45\textwidth]{ 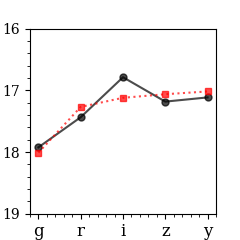 }
    \end{minipage}%
    \begin{minipage}[c]{0.23\textwidth}
        \centering
        \begin{overpic}[width=0.45\textwidth]{ 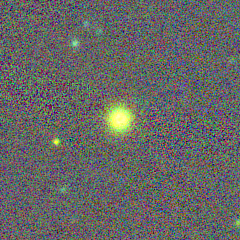 } 
            \put(5,5){\color{white}\bfseries\small giy }
        \end{overpic}%
        \includegraphics[width=0.45\textwidth]{ 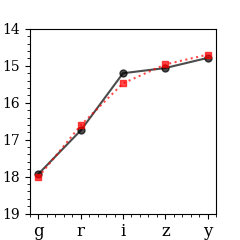 }
    \end{minipage}%
    \\[1ex] 
    \begin{minipage}[c]{0.23\textwidth}
        \centering
        \begin{overpic}[width=0.45\textwidth]{ 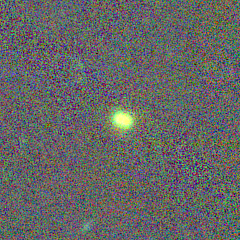 } 
            \put(5,5){\color{white}\bfseries\small giy }
        \end{overpic}%
        \includegraphics[width=0.45\textwidth]{ 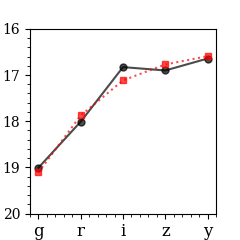 }
    \end{minipage}%
    \begin{minipage}[c]{0.23\textwidth}
        \centering
        \begin{overpic}[width=0.45\textwidth]{ 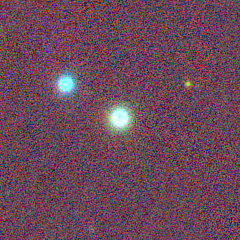 } 
            \put(5,5){\color{white}\bfseries\small giy }
        \end{overpic}%
        \includegraphics[width=0.45\textwidth]{ 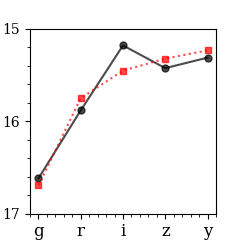 }
    \end{minipage}%
    \begin{minipage}[c]{0.23\textwidth}
        \centering
        \begin{overpic}[width=0.45\textwidth]{ anomaly_colored_by_giy_upper_73_r_188.2483_2.1041.png } 
            \put(5,5){\color{white}\bfseries\small giy }
        \end{overpic}%
        \includegraphics[width=0.45\textwidth]{ SED_upper_73.png }
    \end{minipage}%
    \begin{minipage}[c]{0.23\textwidth}
        \centering
        \begin{overpic}[width=0.45\textwidth]{ 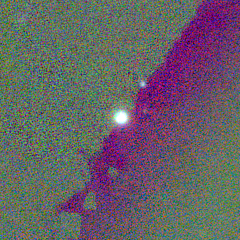 } 
            \put(5,5){\color{white}\bfseries\small giy }
        \end{overpic}%
        \includegraphics[width=0.45\textwidth]{ 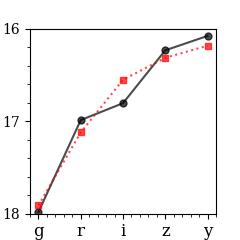 }
    \end{minipage}%
    \\[1ex] 
    \begin{minipage}[c]{0.23\textwidth}
        \centering
        \begin{overpic}[width=0.45\textwidth]{ 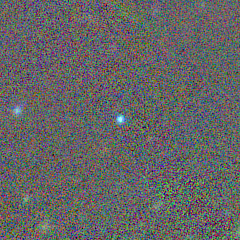 } 
            \put(5,5){\color{white}\bfseries\small giy }
        \end{overpic}%
        \includegraphics[width=0.45\textwidth]{ 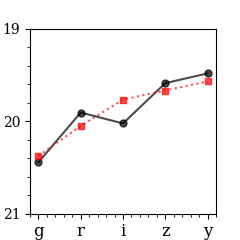 }
    \end{minipage}%
    \begin{minipage}[c]{0.23\textwidth}
        \centering
        \begin{overpic}[width=0.45\textwidth]{ 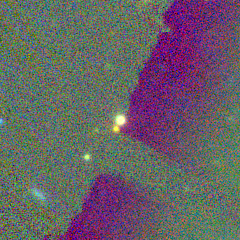 } 
            \put(5,5){\color{white}\bfseries\small giy }
        \end{overpic}%
        \includegraphics[width=0.45\textwidth]{ 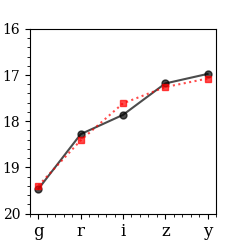 }
    \end{minipage}%
    \begin{minipage}[c]{0.23\textwidth}
        \centering
        \begin{overpic}[width=0.45\textwidth]{ 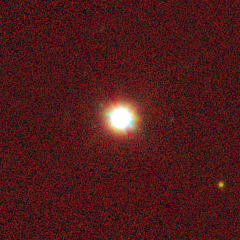 } 
            \put(5,5){\color{white}\bfseries\small giy }
        \end{overpic}%
        \includegraphics[width=0.45\textwidth]{ 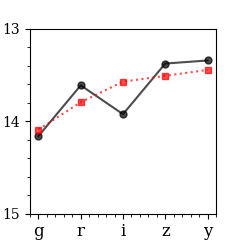 }
    \end{minipage}%
    \begin{minipage}[c]{0.23\textwidth}
        \centering
        \begin{overpic}[width=0.45\textwidth]{ 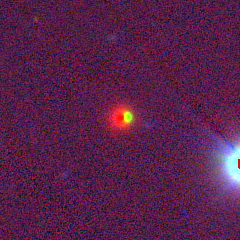 } 
            \put(5,5){\color{white}\bfseries\small giy }
        \end{overpic}%
        \includegraphics[width=0.45\textwidth]{ 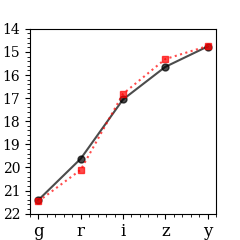 }
    \end{minipage}%
    \\[1ex] 
    \begin{minipage}[c]{0.23\textwidth}
        \centering
        \begin{overpic}[width=0.45\textwidth]{ 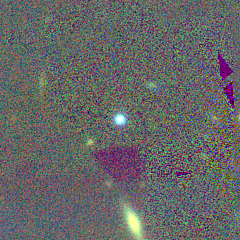 } 
            \put(5,5){\color{white}\bfseries\small giy }
        \end{overpic}%
        \includegraphics[width=0.45\textwidth]{ 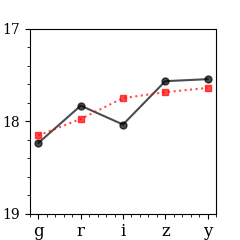 }
    \end{minipage}%
    \begin{minipage}[c]{0.23\textwidth}
        \centering
        \begin{overpic}[width=0.45\textwidth]{ anomaly_colored_by_giy_upper_99_z_218.2867_24.535.png } 
            \put(5,5){\color{white}\bfseries\small giy }
        \end{overpic}%
        \includegraphics[width=0.45\textwidth]{ SED_upper_99.png }
    \end{minipage}%
    \begin{minipage}[c]{0.23\textwidth}
        \centering
        \begin{overpic}[width=0.45\textwidth]{ 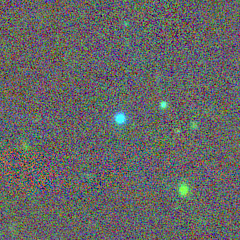 } 
            \put(5,5){\color{white}\bfseries\small giy }
        \end{overpic}%
        \includegraphics[width=0.45\textwidth]{ 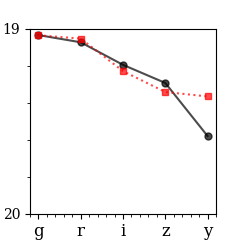 }
    \end{minipage}%
    \begin{minipage}[c]{0.23\textwidth}
        \centering
        \begin{overpic}[width=0.45\textwidth]{ 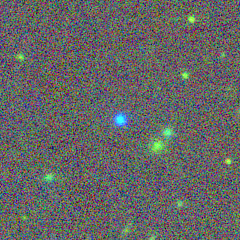 } 
            \put(5,5){\color{white}\bfseries\small giy }
        \end{overpic}%
        \includegraphics[width=0.45\textwidth]{ 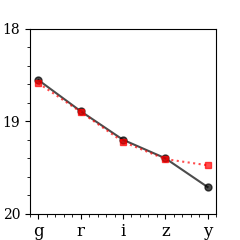 }
    \end{minipage}%
    \\[1ex] 
    \begin{minipage}[c]{0.23\textwidth}
        \centering
        \begin{overpic}[width=0.45\textwidth]{ 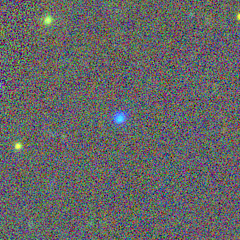 } 
            \put(5,5){\color{white}\bfseries\small giy }
        \end{overpic}%
        \includegraphics[width=0.45\textwidth]{ 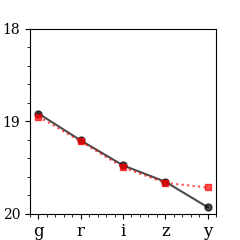 }
    \end{minipage}%
    \begin{minipage}[c]{0.23\textwidth}
        \centering
        \begin{overpic}[width=0.45\textwidth]{ 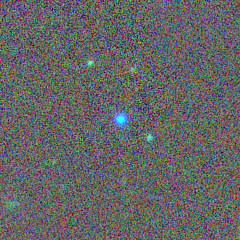 } 
            \put(5,5){\color{white}\bfseries\small giy }
        \end{overpic}%
        \includegraphics[width=0.45\textwidth]{ 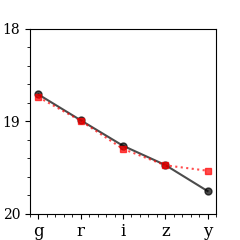 }
    \end{minipage}%
    \begin{minipage}[c]{0.23\textwidth}
        \centering
        \begin{overpic}[width=0.45\textwidth]{ 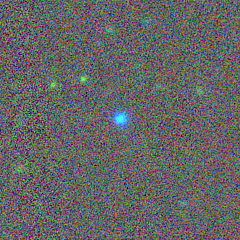 } 
            \put(5,5){\color{white}\bfseries\small giy }
        \end{overpic}%
        \includegraphics[width=0.45\textwidth]{ 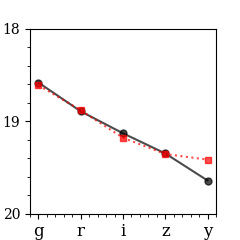 }
    \end{minipage}%
    \begin{minipage}[c]{0.23\textwidth}
        \centering
        \begin{overpic}[width=0.45\textwidth]{ 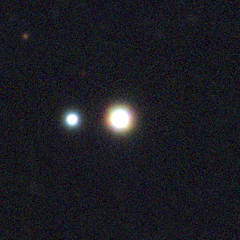 } 
            \put(5,5){\color{white}\bfseries\small gri }
        \end{overpic}%
        \includegraphics[width=0.45\textwidth]{ 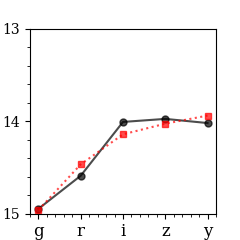 }
    \end{minipage}%
    \\[1ex] 
    \begin{minipage}[c]{0.23\textwidth}
        \centering
        \begin{overpic}[width=0.45\textwidth]{ 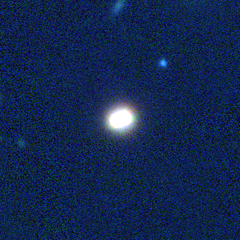 } 
            \put(5,5){\color{white}\bfseries\small gri }
        \end{overpic}%
        \includegraphics[width=0.45\textwidth]{ 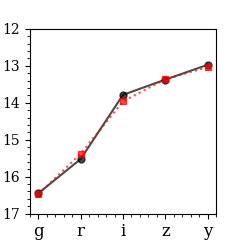 }
    \end{minipage}%
    \begin{minipage}[c]{0.23\textwidth}
        \centering
        \begin{overpic}[width=0.45\textwidth]{ anomaly_colored_by_gri_upper_68_r_167.8051_28.7102.png } 
            \put(5,5){\color{white}\bfseries\small gri }
        \end{overpic}%
        \includegraphics[width=0.45\textwidth]{ SED_upper_68.png }
    \end{minipage}%
    \begin{minipage}[c]{0.23\textwidth}
        \centering
        \begin{overpic}[width=0.45\textwidth]{ anomaly_colored_by_gri_upper_72_r_211.9128_39.0086.png } 
            \put(5,5){\color{white}\bfseries\small gri }
        \end{overpic}%
        \includegraphics[width=0.45\textwidth]{ SED_upper_72.png }
    \end{minipage}%
    \begin{minipage}[c]{0.23\textwidth}
        \centering
        \begin{overpic}[width=0.45\textwidth]{ 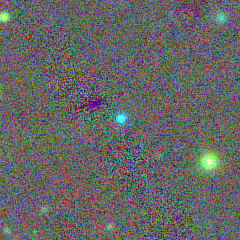 } 
            \put(5,5){\color{white}\bfseries\small giy }
        \end{overpic}%
        \includegraphics[width=0.45\textwidth]{ 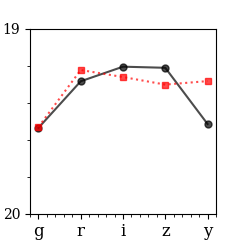 }
    \end{minipage}%
    \\[1ex] 
    \begin{minipage}[c]{0.23\textwidth}
        \centering
        \begin{overpic}[width=0.45\textwidth]{ anomaly_colored_by_giy_upper_143_y_198.178_18.1779.png } 
            \put(5,5){\color{white}\bfseries\small giy }
        \end{overpic}%
        \includegraphics[width=0.45\textwidth]{ SED_upper_143.png }
    \end{minipage}%
    \begin{minipage}[c]{0.23\textwidth}
        \centering
        \begin{overpic}[width=0.45\textwidth]{ 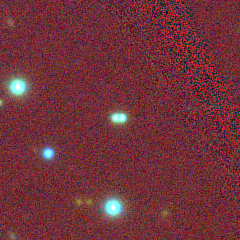 } 
            \put(5,5){\color{white}\bfseries\small giy }
        \end{overpic}%
        \includegraphics[width=0.45\textwidth]{ 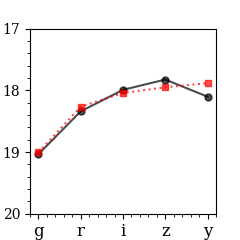 }
    \end{minipage}%
    \begin{minipage}[c]{0.23\textwidth}
        \centering
        \begin{overpic}[width=0.45\textwidth]{ 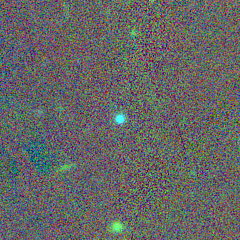 } 
            \put(5,5){\color{white}\bfseries\small giy }
        \end{overpic}%
        \includegraphics[width=0.45\textwidth]{ 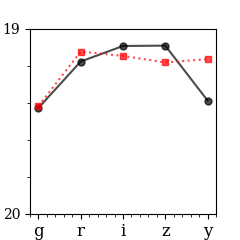 }
    \end{minipage}%
    \begin{minipage}[c]{0.23\textwidth}
        \centering
        \begin{overpic}[width=0.45\textwidth]{ 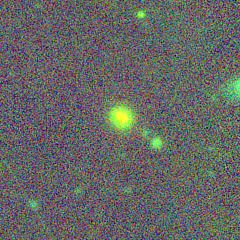 } 
            \put(5,5){\color{white}\bfseries\small giy }
        \end{overpic}%
        \includegraphics[width=0.45\textwidth]{ 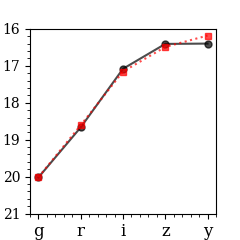 }
    \end{minipage}%
    \\[1ex] 
    \begin{minipage}[c]{0.23\textwidth}
        \centering
        \begin{overpic}[width=0.45\textwidth]{ 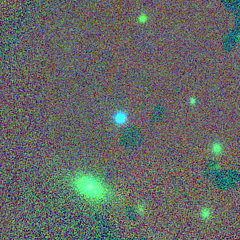 } 
            \put(5,5){\color{white}\bfseries\small giy }
        \end{overpic}%
        \includegraphics[width=0.45\textwidth]{ 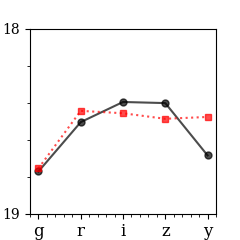 }
    \end{minipage}%
    \begin{minipage}[c]{0.23\textwidth}
        \centering
        \begin{overpic}[width=0.45\textwidth]{ 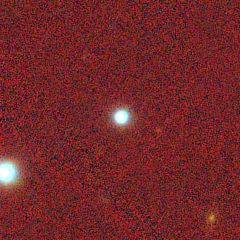 } 
            \put(5,5){\color{white}\bfseries\small giy }
        \end{overpic}%
        \includegraphics[width=0.45\textwidth]{ 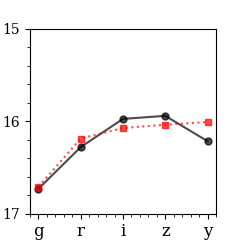 }
    \end{minipage}%
    \begin{minipage}[c]{0.23\textwidth}
        \centering
        \begin{overpic}[width=0.45\textwidth]{ 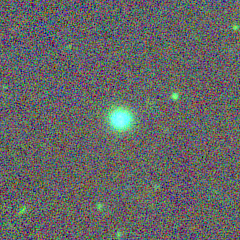 } 
            \put(5,5){\color{white}\bfseries\small giy }
        \end{overpic}%
        \includegraphics[width=0.45\textwidth]{ 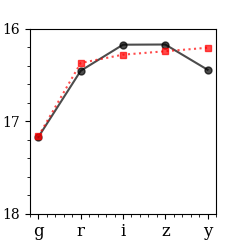 }
    \end{minipage}%
    \begin{minipage}[c]{0.23\textwidth}
        \centering
        \begin{overpic}[width=0.45\textwidth]{ 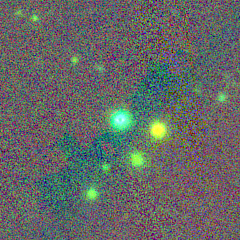 } 
            \put(5,5){\color{white}\bfseries\small giy }
        \end{overpic}%
        \includegraphics[width=0.45\textwidth]{ 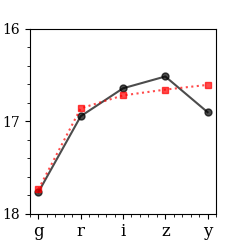 }
    \end{minipage}%
    \\[1ex] 
    \end{figure*}  

    \begin{figure*}[htbp]
        \centering
    \begin{minipage}[c]{0.23\textwidth}
        \centering
        \begin{overpic}[width=0.45\textwidth]{ 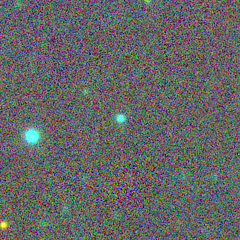 } 
            \put(5,5){\color{white}\bfseries\small giy }
        \end{overpic}%
        \includegraphics[width=0.45\textwidth]{ 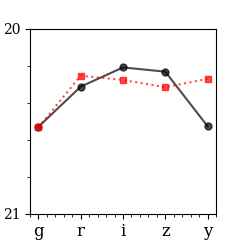 }
    \end{minipage}%
    \begin{minipage}[c]{0.23\textwidth}
        \centering
        \begin{overpic}[width=0.45\textwidth]{ 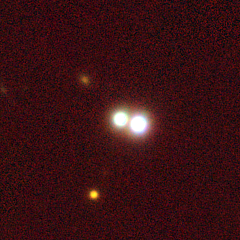 } 
            \put(5,5){\color{white}\bfseries\small giy }
        \end{overpic}%
        \includegraphics[width=0.45\textwidth]{ 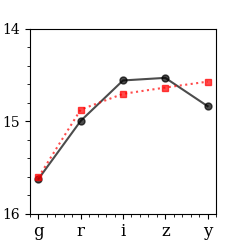 }
    \end{minipage}%
    \begin{minipage}[c]{0.23\textwidth}
        \centering
        \begin{overpic}[width=0.45\textwidth]{ 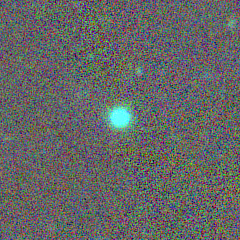 } 
            \put(5,5){\color{white}\bfseries\small giy }
        \end{overpic}%
        \includegraphics[width=0.45\textwidth]{ 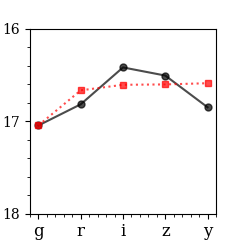 }
    \end{minipage}%
    \begin{minipage}[c]{0.23\textwidth}
        \centering
        \begin{overpic}[width=0.45\textwidth]{ 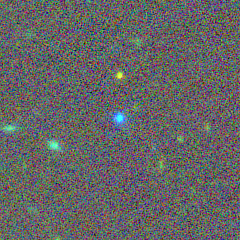 } 
            \put(5,5){\color{white}\bfseries\small giy }
        \end{overpic}%
        \includegraphics[width=0.45\textwidth]{ 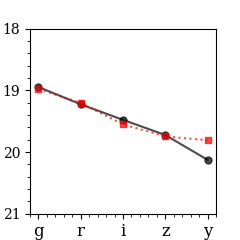 }
    \end{minipage}%
    \\[1ex] 
    \begin{minipage}[c]{0.23\textwidth}
        \centering
        \begin{overpic}[width=0.45\textwidth]{ 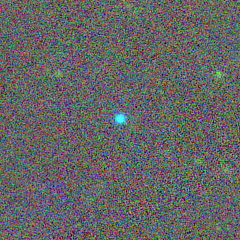 } 
            \put(5,5){\color{white}\bfseries\small giy }
        \end{overpic}%
        \includegraphics[width=0.45\textwidth]{ 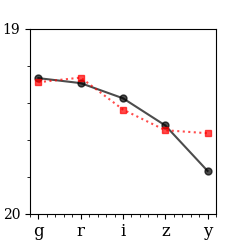 }
    \end{minipage}%
    \begin{minipage}[c]{0.23\textwidth}
        \centering
        \begin{overpic}[width=0.45\textwidth]{ 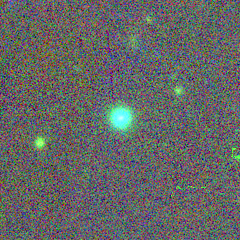 } 
            \put(5,5){\color{white}\bfseries\small giy }
        \end{overpic}%
        \includegraphics[width=0.45\textwidth]{ 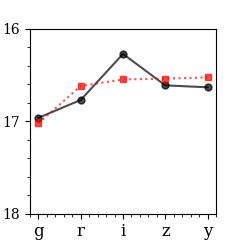 }
    \end{minipage}%
    \begin{minipage}[c]{0.23\textwidth}
        \centering
        \begin{overpic}[width=0.45\textwidth]{ 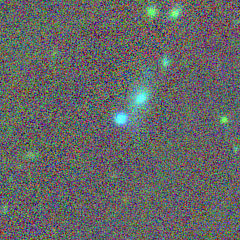 } 
            \put(5,5){\color{white}\bfseries\small giy }
        \end{overpic}%
        \includegraphics[width=0.45\textwidth]{ 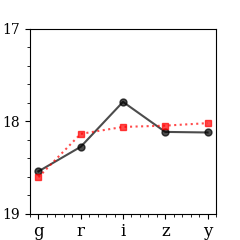 }
    \end{minipage}%
    \begin{minipage}[c]{0.23\textwidth}
        \centering
        \begin{overpic}[width=0.45\textwidth]{ 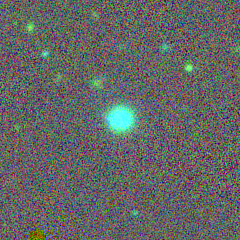 } 
            \put(5,5){\color{white}\bfseries\small giy }
        \end{overpic}%
        \includegraphics[width=0.45\textwidth]{ 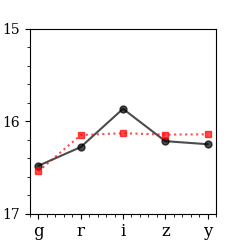 }
    \end{minipage}%
    \\[1ex] 
    \begin{minipage}[c]{0.23\textwidth}
        \centering
        \begin{overpic}[width=0.45\textwidth]{ 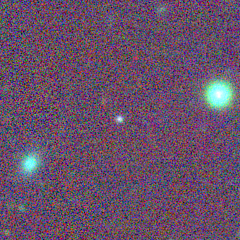 } 
            \put(5,5){\color{white}\bfseries\small giy }
        \end{overpic}%
        \includegraphics[width=0.45\textwidth]{ 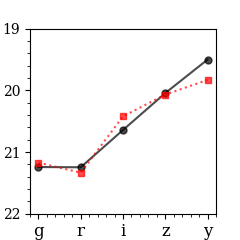 }
    \end{minipage}%
    \begin{minipage}[c]{0.23\textwidth}
        \centering
        \begin{overpic}[width=0.45\textwidth]{ 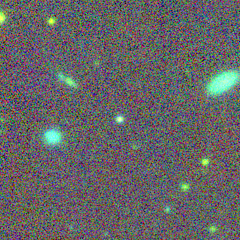 } 
            \put(5,5){\color{white}\bfseries\small giy }
        \end{overpic}%
        \includegraphics[width=0.45\textwidth]{ 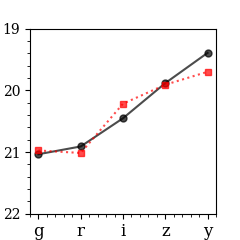 }
    \end{minipage}%
    \begin{minipage}[c]{0.23\textwidth}
        \centering
        \begin{overpic}[width=0.45\textwidth]{ 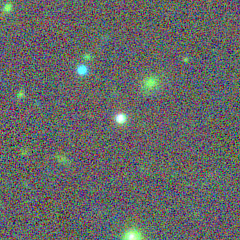 } 
            \put(5,5){\color{white}\bfseries\small giy }
        \end{overpic}%
        \includegraphics[width=0.45\textwidth]{ 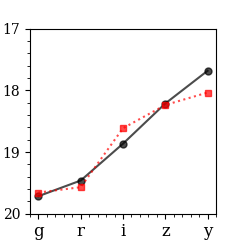 }
    \end{minipage}%
    \begin{minipage}[c]{0.23\textwidth}
        \centering
        \begin{overpic}[width=0.45\textwidth]{ 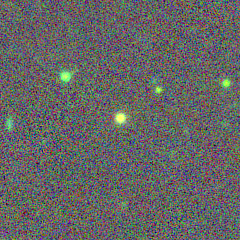 } 
            \put(5,5){\color{white}\bfseries\small giy }
        \end{overpic}%
        \includegraphics[width=0.45\textwidth]{ 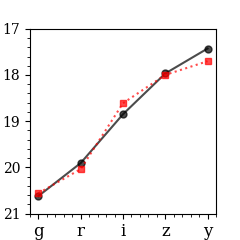 }
    \end{minipage}%
    \\[1ex] 
    \begin{minipage}[c]{0.23\textwidth}
        \centering
        \begin{overpic}[width=0.45\textwidth]{ 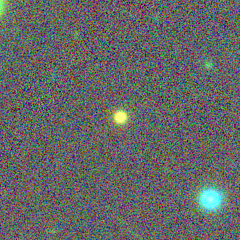 } 
            \put(5,5){\color{white}\bfseries\small giy }
        \end{overpic}%
        \includegraphics[width=0.45\textwidth]{ 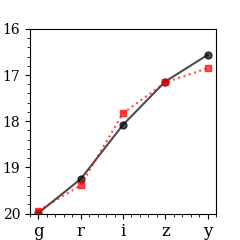 }
    \end{minipage}%
    \begin{minipage}[c]{0.23\textwidth}
        \centering
        \begin{overpic}[width=0.45\textwidth]{ 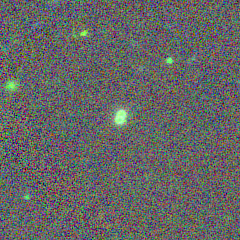 } 
            \put(5,5){\color{white}\bfseries\small giy }
        \end{overpic}%
        \includegraphics[width=0.45\textwidth]{ 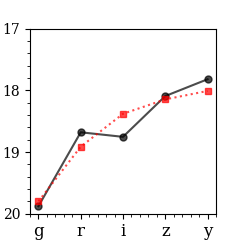 }
    \end{minipage}%
    \begin{minipage}[c]{0.23\textwidth}
        \centering
        \begin{overpic}[width=0.45\textwidth]{ 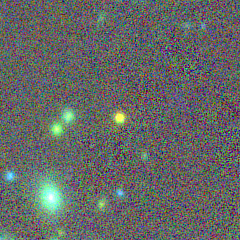 } 
            \put(5,5){\color{white}\bfseries\small giy }
        \end{overpic}%
        \includegraphics[width=0.45\textwidth]{ 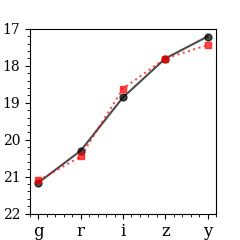 }
    \end{minipage}%
    \begin{minipage}[c]{0.23\textwidth}
        \centering
        \begin{overpic}[width=0.45\textwidth]{ anomaly_colored_by_giy_upper_89_i_179.5947_48.8696.png } 
            \put(5,5){\color{white}\bfseries\small giy }
        \end{overpic}%
        \includegraphics[width=0.45\textwidth]{ SED_upper_89.png }
    \end{minipage}%
    \\[1ex] 
    \begin{minipage}[c]{0.23\textwidth}
        \centering
        \begin{overpic}[width=0.45\textwidth]{ 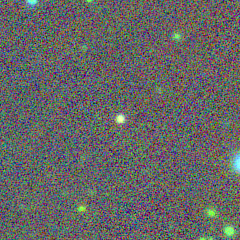 } 
            \put(5,5){\color{white}\bfseries\small giy }
        \end{overpic}%
        \includegraphics[width=0.45\textwidth]{ 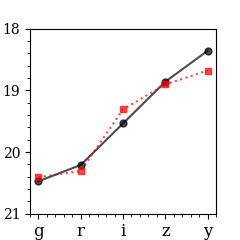 }
    \end{minipage}%
    \begin{minipage}[c]{0.23\textwidth}
        \centering
        \begin{overpic}[width=0.45\textwidth]{ 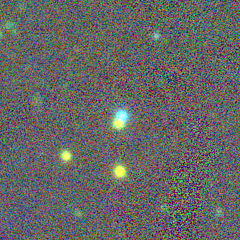 } 
            \put(5,5){\color{white}\bfseries\small giy }
        \end{overpic}%
        \includegraphics[width=0.45\textwidth]{ 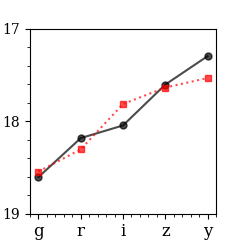 }
    \end{minipage}%
    \begin{minipage}[c]{0.23\textwidth}
        \centering
        \begin{overpic}[width=0.45\textwidth]{ 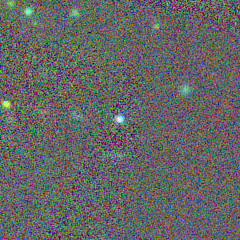 } 
            \put(5,5){\color{white}\bfseries\small giy }
        \end{overpic}%
        \includegraphics[width=0.45\textwidth]{ 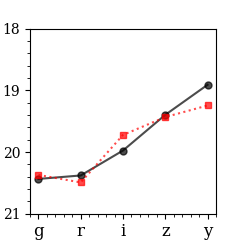 }
    \end{minipage}%
    \begin{minipage}[c]{0.23\textwidth}
        \centering
        \begin{overpic}[width=0.45\textwidth]{ 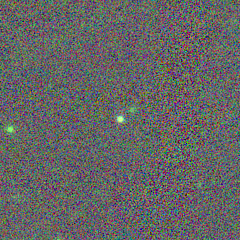 } 
            \put(5,5){\color{white}\bfseries\small giy }
        \end{overpic}%
        \includegraphics[width=0.45\textwidth]{ 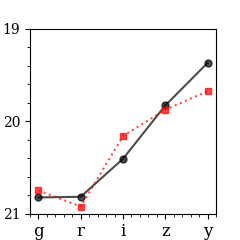 }
    \end{minipage}%
    \\[1ex] 
    \begin{minipage}[c]{0.23\textwidth}
        \centering
        \begin{overpic}[width=0.45\textwidth]{ 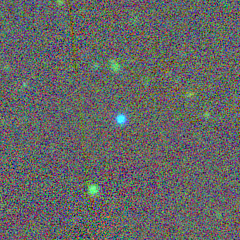 } 
            \put(5,5){\color{white}\bfseries\small giy }
        \end{overpic}%
        \includegraphics[width=0.45\textwidth]{ 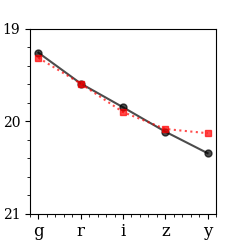 }
    \end{minipage}%
    \begin{minipage}[c]{0.23\textwidth}
        \centering
        \begin{overpic}[width=0.45\textwidth]{ 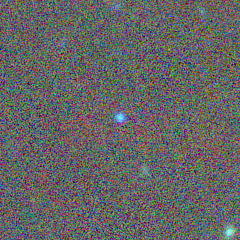 } 
            \put(5,5){\color{white}\bfseries\small rzy }
        \end{overpic}%
        \includegraphics[width=0.45\textwidth]{ 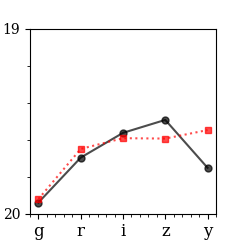 }
    \end{minipage}%
    \begin{minipage}[c]{0.23\textwidth}
        \centering
        \begin{overpic}[width=0.45\textwidth]{ 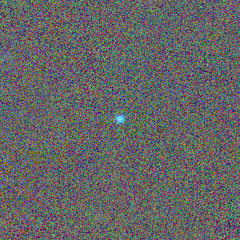 } 
            \put(5,5){\color{white}\bfseries\small rzy }
        \end{overpic}%
        \includegraphics[width=0.45\textwidth]{ 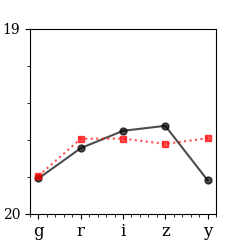 }
    \end{minipage}%
    \begin{minipage}[c]{0.23\textwidth}
        \centering
        \begin{overpic}[width=0.45\textwidth]{ 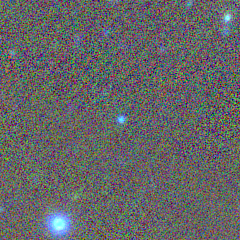 } 
            \put(5,5){\color{white}\bfseries\small rzy }
        \end{overpic}%
        \includegraphics[width=0.45\textwidth]{ 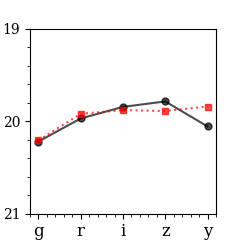 }
    \end{minipage}%
    \\[1ex] 
    \begin{minipage}[c]{0.23\textwidth}
        \centering
        \begin{overpic}[width=0.45\textwidth]{ 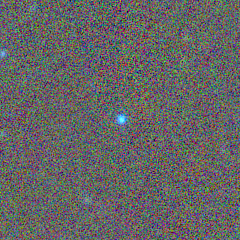 } 
            \put(5,5){\color{white}\bfseries\small rzy }
        \end{overpic}%
        \includegraphics[width=0.45\textwidth]{ 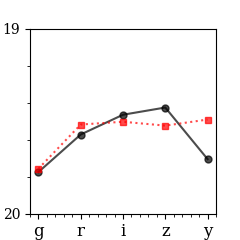 }
    \end{minipage}%

    \caption{PS1 images alongside observed photometry highlighting SED shapes of detected anomalies in which the LSTM-AE overestimates the observed PS magnitude ($m-m^\prime \geq 0.15$). Observed SED shapes in black, reconstructed SED shapes in red. Image Cutouts are 0.25" per pixel with a size of 240 pixels.}
\end{figure*}

\begin{figure*}[htbp]
    \centering
    \begin{minipage}[c]{0.23\textwidth}
        \centering
        \begin{overpic}[width=0.45\textwidth]{ 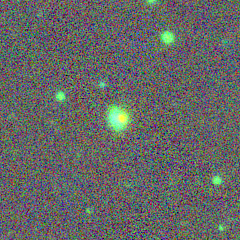 } 
            \put(5,5){\color{white}\bfseries\small giy }
        \end{overpic}%
        \includegraphics[width=0.45\textwidth]{ 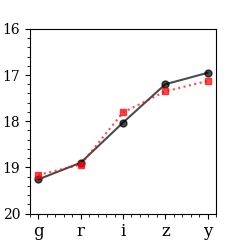 }
    \end{minipage}%
    \begin{minipage}[c]{0.23\textwidth}
        \centering
        \begin{overpic}[width=0.45\textwidth]{ 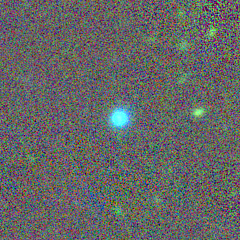 } 
            \put(5,5){\color{white}\bfseries\small giy }
        \end{overpic}%
        \includegraphics[width=0.45\textwidth]{ 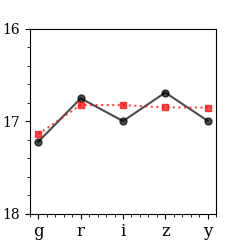 }
    \end{minipage}%
    \begin{minipage}[c]{0.23\textwidth}
        \centering
        \begin{overpic}[width=0.45\textwidth]{ 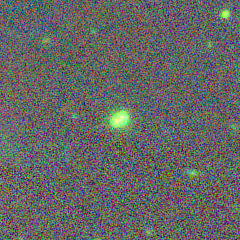 } 
            \put(5,5){\color{white}\bfseries\small giy }
        \end{overpic}%
        \includegraphics[width=0.45\textwidth]{ 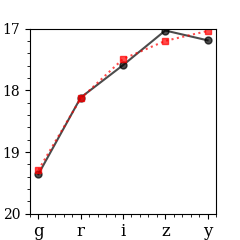 }
    \end{minipage}%
    \begin{minipage}[c]{0.23\textwidth}
        \centering
        \begin{overpic}[width=0.45\textwidth]{ 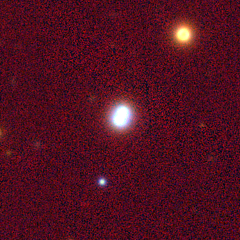 } 
            \put(5,5){\color{white}\bfseries\small giy }
        \end{overpic}%
        \includegraphics[width=0.45\textwidth]{ 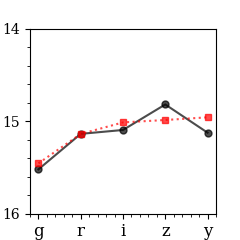 }
    \end{minipage}%
    \\[1ex] 
    \begin{minipage}[c]{0.23\textwidth}
        \centering
        \begin{overpic}[width=0.45\textwidth]{ 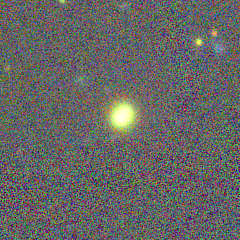 } 
            \put(5,5){\color{white}\bfseries\small giy }
        \end{overpic}%
        \includegraphics[width=0.45\textwidth]{ 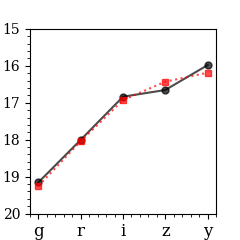 }
    \end{minipage}%
    \begin{minipage}[c]{0.23\textwidth}
        \centering
        \begin{overpic}[width=0.45\textwidth]{ 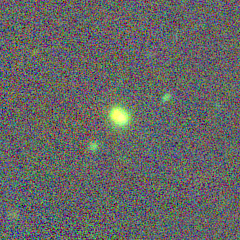 } 
            \put(5,5){\color{white}\bfseries\small giy }
        \end{overpic}%
        \includegraphics[width=0.45\textwidth]{ 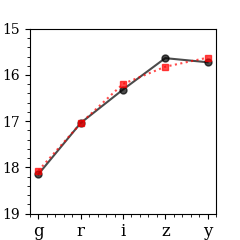 }
    \end{minipage}%
    \begin{minipage}[c]{0.23\textwidth}
        \centering
        \begin{overpic}[width=0.45\textwidth]{ 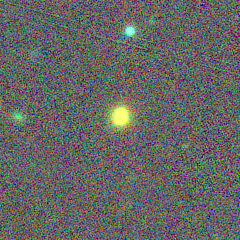 } 
            \put(5,5){\color{white}\bfseries\small giy }
        \end{overpic}%
        \includegraphics[width=0.45\textwidth]{ 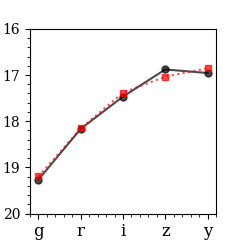 }
    \end{minipage}%
    \begin{minipage}[c]{0.23\textwidth}
        \centering
        \begin{overpic}[width=0.45\textwidth]{ 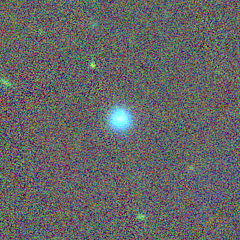 } 
            \put(5,5){\color{white}\bfseries\small giy }
        \end{overpic}%
        \includegraphics[width=0.45\textwidth]{ 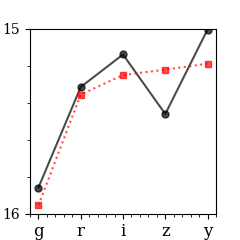 }
    \end{minipage}%
    \\[1ex] 
    \begin{minipage}[c]{0.23\textwidth}
        \centering
        \begin{overpic}[width=0.45\textwidth]{ 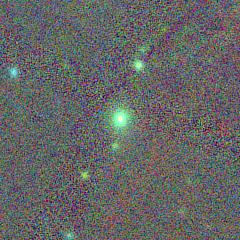 } 
            \put(5,5){\color{white}\bfseries\small giy }
        \end{overpic}%
        \includegraphics[width=0.45\textwidth]{ 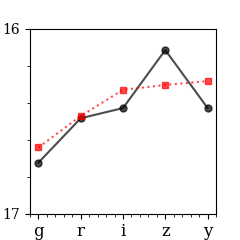 }
    \end{minipage}%
    \begin{minipage}[c]{0.23\textwidth}
        \centering
        \begin{overpic}[width=0.45\textwidth]{ 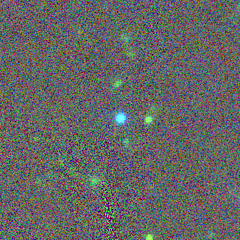 } 
            \put(5,5){\color{white}\bfseries\small giy }
        \end{overpic}%
        \includegraphics[width=0.45\textwidth]{ 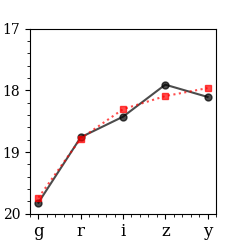 }
    \end{minipage}%
    \begin{minipage}[c]{0.23\textwidth}
        \centering
        \begin{overpic}[width=0.45\textwidth]{ 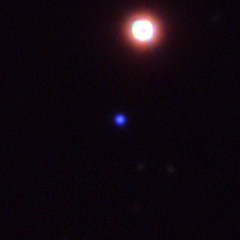 } 
            \put(5,5){\color{white}\bfseries\small giy }
        \end{overpic}%
        \includegraphics[width=0.45\textwidth]{ 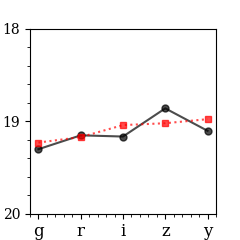 }
    \end{minipage}%
    \begin{minipage}[c]{0.23\textwidth}
        \centering
        \begin{overpic}[width=0.45\textwidth]{ 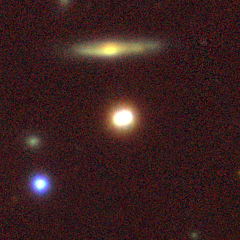 } 
            \put(5,5){\color{white}\bfseries\small giy }
        \end{overpic}%
        \includegraphics[width=0.45\textwidth]{ 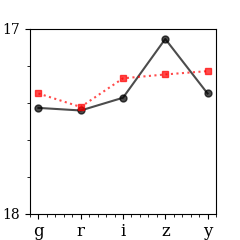 }
    \end{minipage}%
    \\[1ex] 
    \begin{minipage}[c]{0.23\textwidth}
        \centering
        \begin{overpic}[width=0.45\textwidth]{ 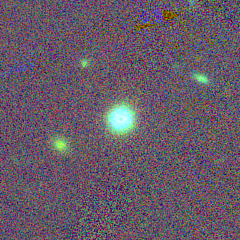 } 
            \put(5,5){\color{white}\bfseries\small giy }
        \end{overpic}%
        \includegraphics[width=0.45\textwidth]{ 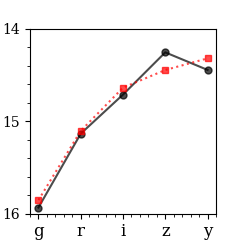 }
    \end{minipage}%
    \begin{minipage}[c]{0.23\textwidth}
        \centering
        \begin{overpic}[width=0.45\textwidth]{ 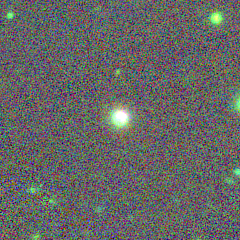 } 
            \put(5,5){\color{white}\bfseries\small giy }
        \end{overpic}%
        \includegraphics[width=0.45\textwidth]{ 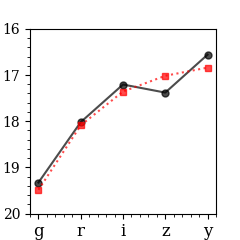 }
    \end{minipage}%
    \begin{minipage}[c]{0.23\textwidth}
        \centering
        \begin{overpic}[width=0.45\textwidth]{ 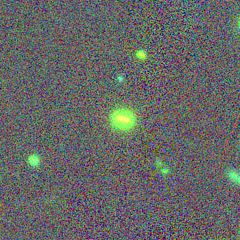 } 
            \put(5,5){\color{white}\bfseries\small giy }
        \end{overpic}%
        \includegraphics[width=0.45\textwidth]{ 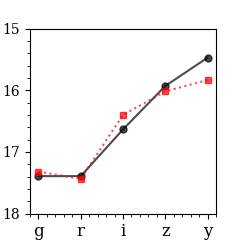 }
    \end{minipage}%
    \begin{minipage}[c]{0.23\textwidth}
        \centering
        \begin{overpic}[width=0.45\textwidth]{ 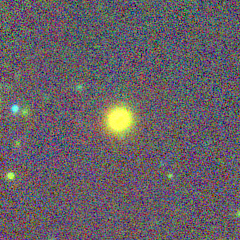 } 
            \put(5,5){\color{white}\bfseries\small giy }
        \end{overpic}%
        \includegraphics[width=0.45\textwidth]{ 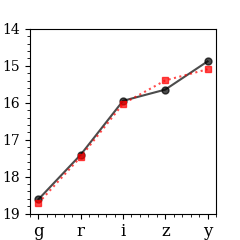 }
    \end{minipage}%
    \\[1ex] 
    \begin{minipage}[c]{0.23\textwidth}
        \centering
        \begin{overpic}[width=0.45\textwidth]{ 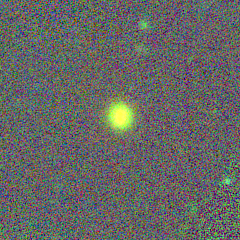 } 
            \put(5,5){\color{white}\bfseries\small giy }
        \end{overpic}%
        \includegraphics[width=0.45\textwidth]{ 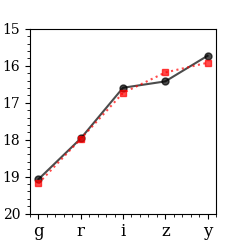 }
    \end{minipage}%
    \begin{minipage}[c]{0.23\textwidth}
        \centering
        \begin{overpic}[width=0.45\textwidth]{ 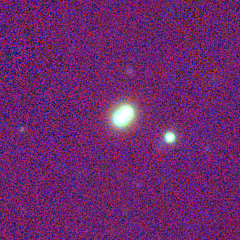 } 
            \put(5,5){\color{white}\bfseries\small giy }
        \end{overpic}%
        \includegraphics[width=0.45\textwidth]{ 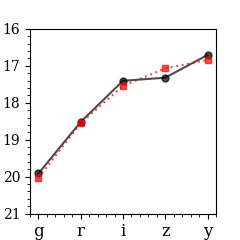 }
    \end{minipage}%
    \begin{minipage}[c]{0.23\textwidth}
        \centering
        \begin{overpic}[width=0.45\textwidth]{ 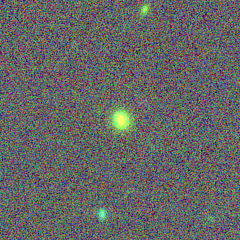 } 
            \put(5,5){\color{white}\bfseries\small giy }
        \end{overpic}%
        \includegraphics[width=0.45\textwidth]{ 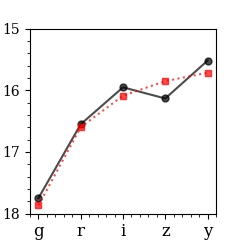 }
    \end{minipage}%
    \begin{minipage}[c]{0.23\textwidth}
        \centering
        \begin{overpic}[width=0.45\textwidth]{ 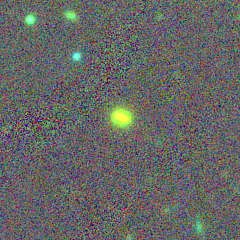 } 
            \put(5,5){\color{white}\bfseries\small giy }
        \end{overpic}%
        \includegraphics[width=0.45\textwidth]{ 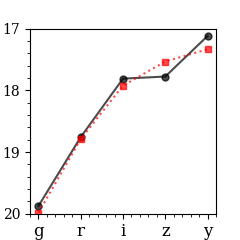 }
    \end{minipage}%
    \\[1ex] 
    \begin{minipage}[c]{0.23\textwidth}
        \centering
        \begin{overpic}[width=0.45\textwidth]{ 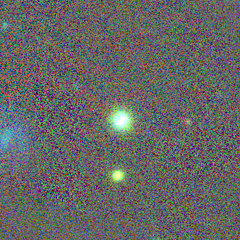 } 
            \put(5,5){\color{white}\bfseries\small giy }
        \end{overpic}%
        \includegraphics[width=0.45\textwidth]{ 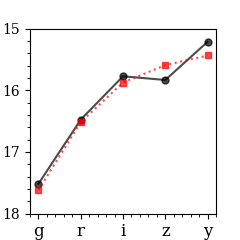 }
    \end{minipage}%
    \begin{minipage}[c]{0.23\textwidth}
        \centering
        \begin{overpic}[width=0.45\textwidth]{ 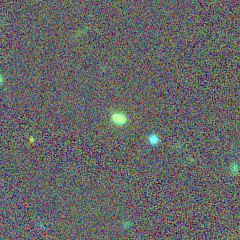 } 
            \put(5,5){\color{white}\bfseries\small giy }
        \end{overpic}%
        \includegraphics[width=0.45\textwidth]{ 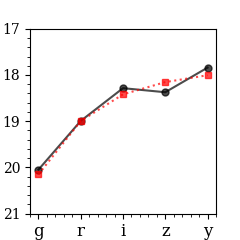 }
    \end{minipage}%
    \begin{minipage}[c]{0.23\textwidth}
        \centering
        \begin{overpic}[width=0.45\textwidth]{ 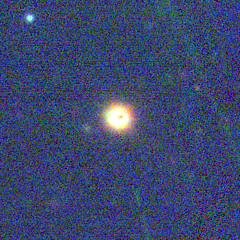 } 
            \put(5,5){\color{white}\bfseries\small gri }
        \end{overpic}%
        \includegraphics[width=0.45\textwidth]{ 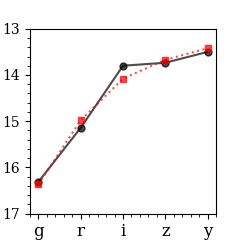 }
    \end{minipage}%
    \begin{minipage}[c]{0.23\textwidth}
        \centering
        \begin{overpic}[width=0.45\textwidth]{ 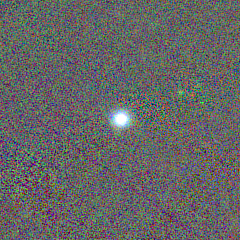 } 
            \put(5,5){\color{white}\bfseries\small gri }
        \end{overpic}%
        \includegraphics[width=0.45\textwidth]{ 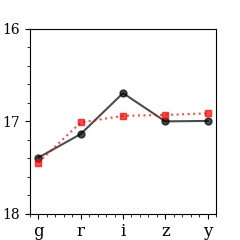 }
    \end{minipage}%
    \\[1ex] 
    \begin{minipage}[c]{0.23\textwidth}
        \centering
        \begin{overpic}[width=0.45\textwidth]{ 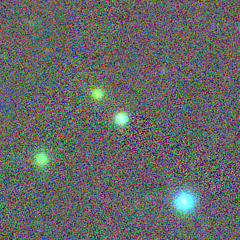 } 
            \put(5,5){\color{white}\bfseries\small giy }
        \end{overpic}%
        \includegraphics[width=0.45\textwidth]{ 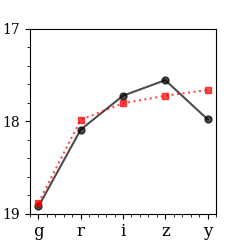 }
    \end{minipage}%
    \begin{minipage}[c]{0.23\textwidth}
        \centering
        \begin{overpic}[width=0.45\textwidth]{ 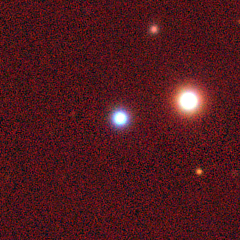 } 
            \put(5,5){\color{white}\bfseries\small giy }
        \end{overpic}%
        \includegraphics[width=0.45\textwidth]{ 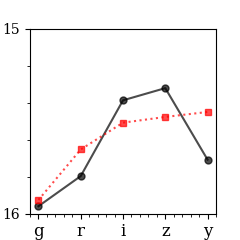 }
    \end{minipage}%
    \begin{minipage}[c]{0.23\textwidth}
        \centering
        \begin{overpic}[width=0.45\textwidth]{ 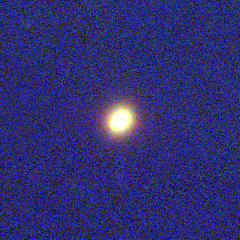 } 
            \put(5,5){\color{white}\bfseries\small giy }
        \end{overpic}%
        \includegraphics[width=0.45\textwidth]{ 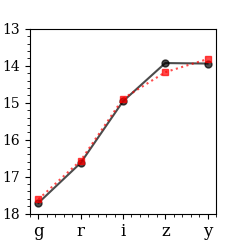 }
    \end{minipage}%
    \begin{minipage}[c]{0.23\textwidth}
        \centering
        \begin{overpic}[width=0.45\textwidth]{ 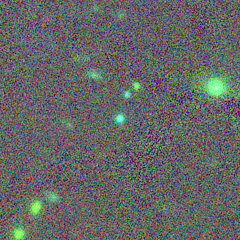 } 
            \put(5,5){\color{white}\bfseries\small giy }
        \end{overpic}%
        \includegraphics[width=0.45\textwidth]{ 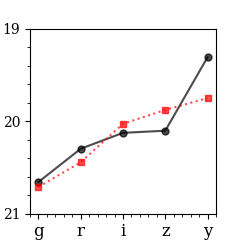 }
    \end{minipage}%
    \\[1ex] 
    \begin{minipage}[c]{0.23\textwidth}
        \centering
        \begin{overpic}[width=0.45\textwidth]{ 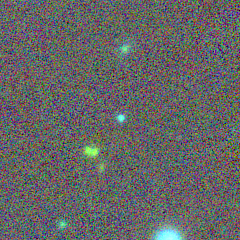 } 
            \put(5,5){\color{white}\bfseries\small giy }
        \end{overpic}%
        \includegraphics[width=0.45\textwidth]{ 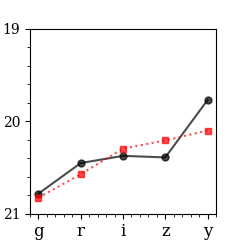 }
    \end{minipage}%
    \begin{minipage}[c]{0.23\textwidth}
        \centering
        \begin{overpic}[width=0.45\textwidth]{ 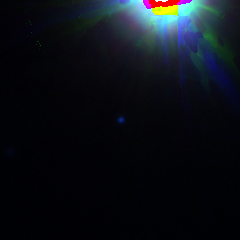 } 
            \put(5,5){\color{white}\bfseries\small giy }
        \end{overpic}%
        \includegraphics[width=0.45\textwidth]{ 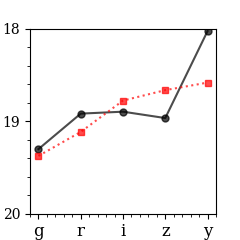 }
    \end{minipage}%
    \begin{minipage}[c]{0.23\textwidth}
        \centering
        \begin{overpic}[width=0.45\textwidth]{ 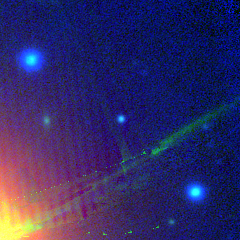 } 
            \put(5,5){\color{white}\bfseries\small giy }
        \end{overpic}%
        \includegraphics[width=0.45\textwidth]{ 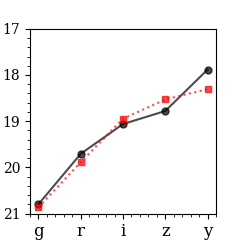 }
    \end{minipage}%
    \begin{minipage}[c]{0.23\textwidth}
        \centering
        \begin{overpic}[width=0.45\textwidth]{ 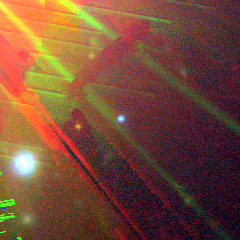 } 
            \put(5,5){\color{white}\bfseries\small giy }
        \end{overpic}%
        \includegraphics[width=0.45\textwidth]{ 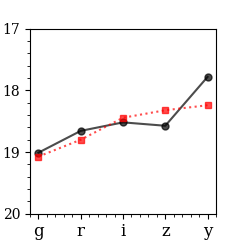 }
    \end{minipage}%
    \\[1ex] 
    \begin{minipage}[c]{0.23\textwidth}
        \centering
        \begin{overpic}[width=0.45\textwidth]{ 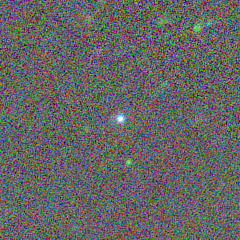 } 
            \put(5,5){\color{white}\bfseries\small giy }
        \end{overpic}%
        \includegraphics[width=0.45\textwidth]{ 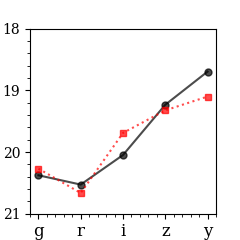 }
    \end{minipage}%
    \begin{minipage}[c]{0.23\textwidth}
        \centering
        \begin{overpic}[width=0.45\textwidth]{ 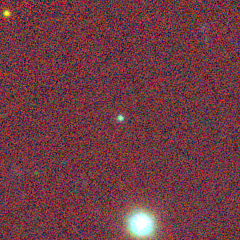 } 
            \put(5,5){\color{white}\bfseries\small giy }
        \end{overpic}%
        \includegraphics[width=0.45\textwidth]{ 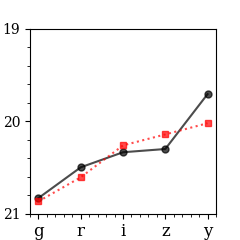 }
    \end{minipage}%
    \begin{minipage}[c]{0.23\textwidth}
        \centering
        \begin{overpic}[width=0.45\textwidth]{ 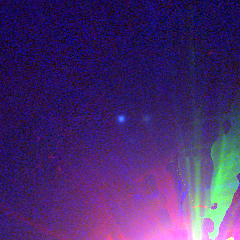 } 
            \put(5,5){\color{white}\bfseries\small giy }
        \end{overpic}%
        \includegraphics[width=0.45\textwidth]{ 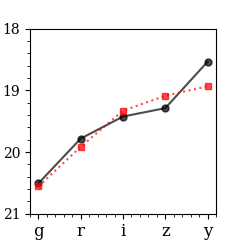 }
    \end{minipage}%
    \begin{minipage}[c]{0.23\textwidth}
        \centering
        \begin{overpic}[width=0.45\textwidth]{ 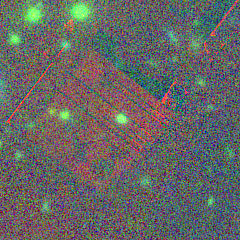 } 
            \put(5,5){\color{white}\bfseries\small giy }
        \end{overpic}%
        \includegraphics[width=0.45\textwidth]{ 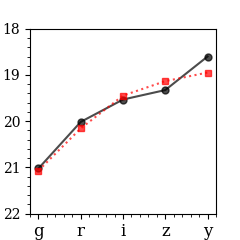 }
    \end{minipage}%
    \\[1ex] 
    \begin{minipage}[c]{0.23\textwidth}
        \centering
        \begin{overpic}[width=0.45\textwidth]{ 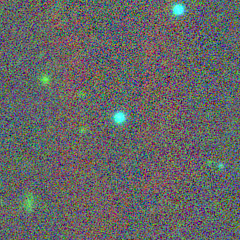 } 
            \put(5,5){\color{white}\bfseries\small giy }
        \end{overpic}%
        \includegraphics[width=0.45\textwidth]{ 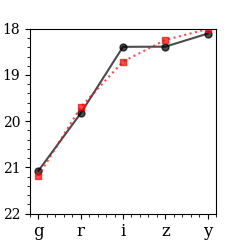 }
    \end{minipage}%
    \begin{minipage}[c]{0.23\textwidth}
        \centering
        \begin{overpic}[width=0.45\textwidth]{ 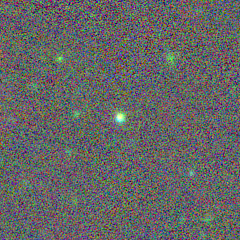 } 
            \put(5,5){\color{white}\bfseries\small giy }
        \end{overpic}%
        \includegraphics[width=0.45\textwidth]{ 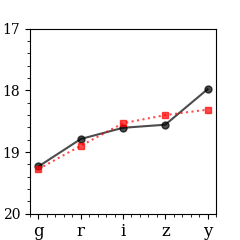 }
    \end{minipage}%
    \begin{minipage}[c]{0.23\textwidth}
        \centering
        \begin{overpic}[width=0.45\textwidth]{ 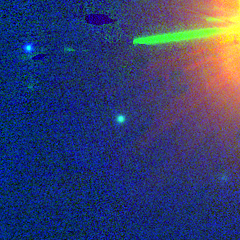 } 
            \put(5,5){\color{white}\bfseries\small giy }
        \end{overpic}%
        \includegraphics[width=0.45\textwidth]{ 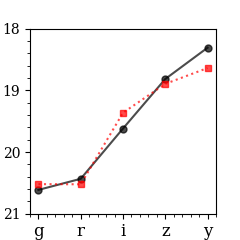 }
    \end{minipage}%
    \begin{minipage}[c]{0.23\textwidth}
        \centering
        \begin{overpic}[width=0.45\textwidth]{ 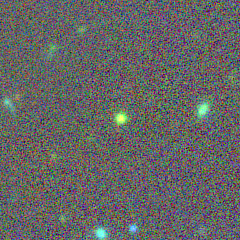 } 
            \put(5,5){\color{white}\bfseries\small giy }
        \end{overpic}%
        \includegraphics[width=0.45\textwidth]{ 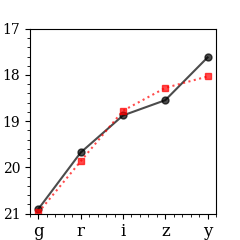 }
    \end{minipage}%
    \\[1ex] 
    \begin{minipage}[c]{0.23\textwidth}
        \centering
        \begin{overpic}[width=0.45\textwidth]{ 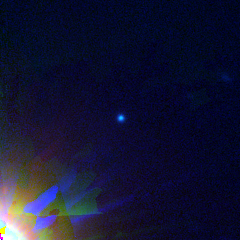 } 
            \put(5,5){\color{white}\bfseries\small giy }
        \end{overpic}%
        \includegraphics[width=0.45\textwidth]{ 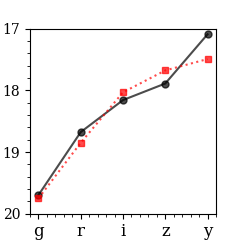 }
    \end{minipage}%
    \begin{minipage}[c]{0.23\textwidth}
        \centering
        \begin{overpic}[width=0.45\textwidth]{ 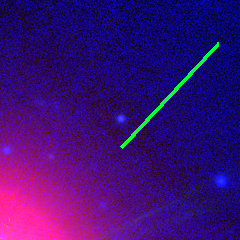 } 
            \put(5,5){\color{white}\bfseries\small giy }
        \end{overpic}%
        \includegraphics[width=0.45\textwidth]{ 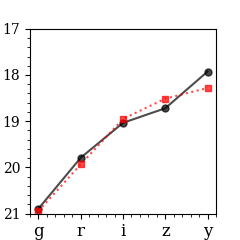 }
    \end{minipage}%
    \begin{minipage}[c]{0.23\textwidth}
        \centering
        \begin{overpic}[width=0.45\textwidth]{ 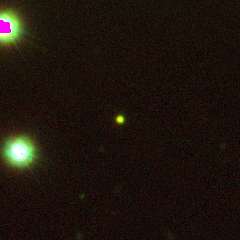 } 
            \put(5,5){\color{white}\bfseries\small giy }
        \end{overpic}%
        \includegraphics[width=0.45\textwidth]{ 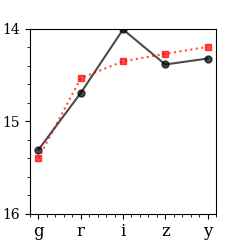 }
    \end{minipage}%
    \begin{minipage}[c]{0.23\textwidth}
        \centering
        \begin{overpic}[width=0.45\textwidth]{ 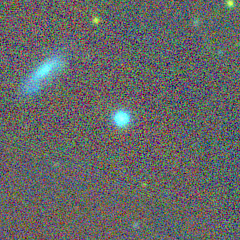 } 
            \put(5,5){\color{white}\bfseries\small giy }
        \end{overpic}%
        \includegraphics[width=0.45\textwidth]{ 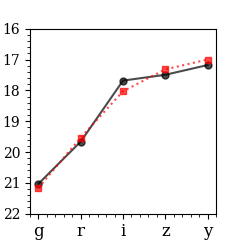 }
    \end{minipage}%
    \\[1ex] 
    \begin{minipage}[c]{0.23\textwidth}
        \centering
        \begin{overpic}[width=0.45\textwidth]{ 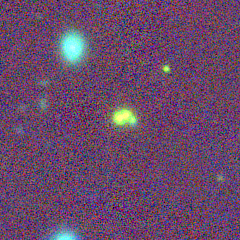 } 
            \put(5,5){\color{white}\bfseries\small giy }
        \end{overpic}%
        \includegraphics[width=0.45\textwidth]{ 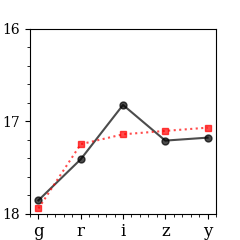 }
    \end{minipage}%
    \begin{minipage}[c]{0.23\textwidth}
        \centering
        \begin{overpic}[width=0.45\textwidth]{ 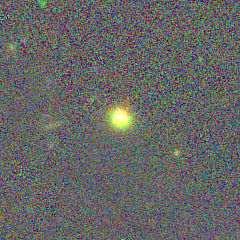 } 
            \put(5,5){\color{white}\bfseries\small giy }
        \end{overpic}%
        \includegraphics[width=0.45\textwidth]{ 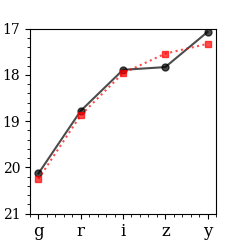 }
    \end{minipage}%
    \begin{minipage}[c]{0.23\textwidth}
        \centering
        \begin{overpic}[width=0.45\textwidth]{ 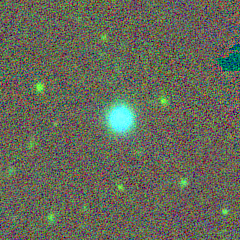 } 
            \put(5,5){\color{white}\bfseries\small giy }
        \end{overpic}%
        \includegraphics[width=0.45\textwidth]{ 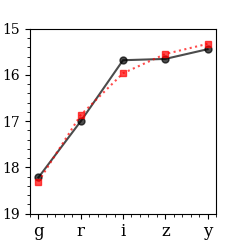 }
    \end{minipage}%
    \begin{minipage}[c]{0.23\textwidth}
        \centering
        \begin{overpic}[width=0.45\textwidth]{ 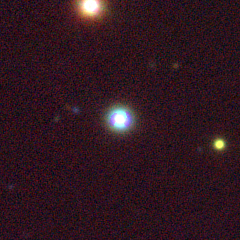 } 
            \put(5,5){\color{white}\bfseries\small giy }
        \end{overpic}%
        \includegraphics[width=0.45\textwidth]{ 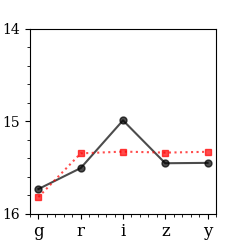 }
    \end{minipage}%
    \\[1ex] 
    \end{figure*}  

    \begin{figure*}[htbp]
        \centering
    \begin{minipage}[c]{0.23\textwidth}
        \centering
        \begin{overpic}[width=0.45\textwidth]{ 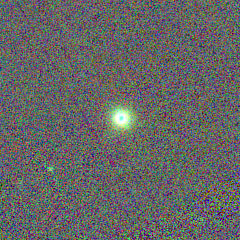 } 
            \put(5,5){\color{white}\bfseries\small giy }
        \end{overpic}%
        \includegraphics[width=0.45\textwidth]{ 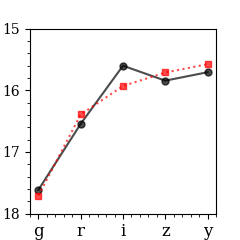 }
    \end{minipage}%
    \begin{minipage}[c]{0.23\textwidth}
        \centering
        \begin{overpic}[width=0.45\textwidth]{ 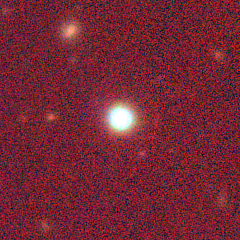 } 
            \put(5,5){\color{white}\bfseries\small giy }
        \end{overpic}%
        \includegraphics[width=0.45\textwidth]{ 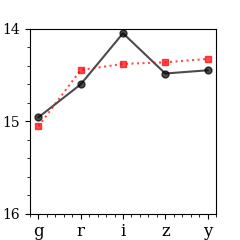 }
    \end{minipage}%
    \begin{minipage}[c]{0.23\textwidth}
        \centering
        \begin{overpic}[width=0.45\textwidth]{ 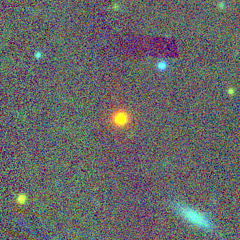 } 
            \put(5,5){\color{white}\bfseries\small giy }
        \end{overpic}%
        \includegraphics[width=0.45\textwidth]{ 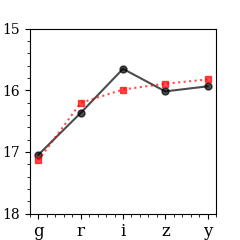 }
    \end{minipage}%
    \begin{minipage}[c]{0.23\textwidth}
        \centering
        \begin{overpic}[width=0.45\textwidth]{ 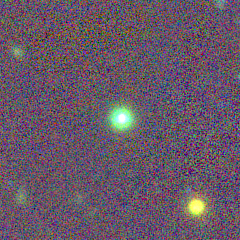 } 
            \put(5,5){\color{white}\bfseries\small giy }
        \end{overpic}%
        \includegraphics[width=0.45\textwidth]{ 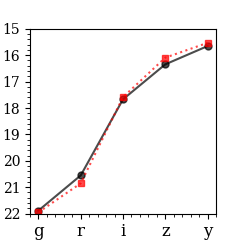 }
    \end{minipage}%
    \\[1ex] 
    \begin{minipage}[c]{0.23\textwidth}
        \centering
        \begin{overpic}[width=0.45\textwidth]{ 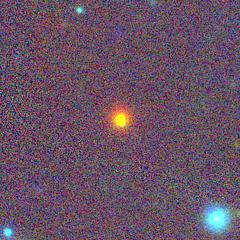 } 
            \put(5,5){\color{white}\bfseries\small giy }
        \end{overpic}%
        \includegraphics[width=0.45\textwidth]{ 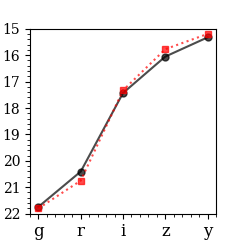 }
    \end{minipage}%
    \begin{minipage}[c]{0.23\textwidth}
        \centering
        \begin{overpic}[width=0.45\textwidth]{ 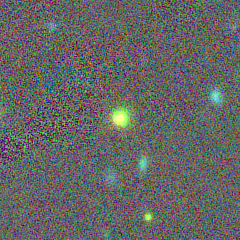 } 
            \put(5,5){\color{white}\bfseries\small giy }
        \end{overpic}%
        \includegraphics[width=0.45\textwidth]{ 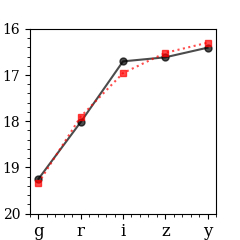 }
    \end{minipage}%
    \begin{minipage}[c]{0.23\textwidth}
        \centering
        \begin{overpic}[width=0.45\textwidth]{ 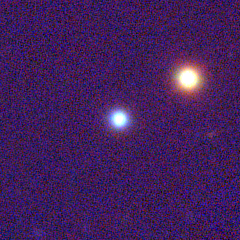 } 
            \put(5,5){\color{white}\bfseries\small giy }
        \end{overpic}%
        \includegraphics[width=0.45\textwidth]{ 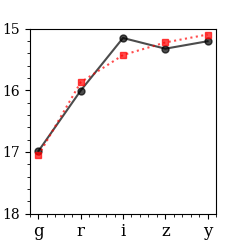 }
    \end{minipage}%
    \begin{minipage}[c]{0.23\textwidth}
        \centering
        \begin{overpic}[width=0.45\textwidth]{ 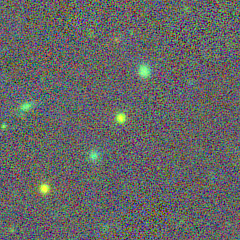 } 
            \put(5,5){\color{white}\bfseries\small giy }
        \end{overpic}%
        \includegraphics[width=0.45\textwidth]{ 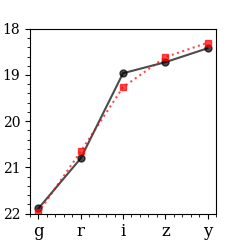 }
    \end{minipage}%
    \\[1ex] 
    \begin{minipage}[c]{0.23\textwidth}
        \centering
        \begin{overpic}[width=0.45\textwidth]{ 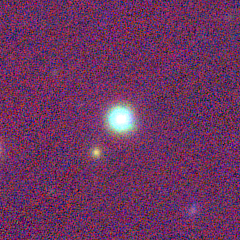 } 
            \put(5,5){\color{white}\bfseries\small giy }
        \end{overpic}%
        \includegraphics[width=0.45\textwidth]{ 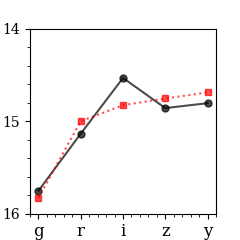 }
    \end{minipage}%
    \begin{minipage}[c]{0.23\textwidth}
        \centering
        \begin{overpic}[width=0.45\textwidth]{ 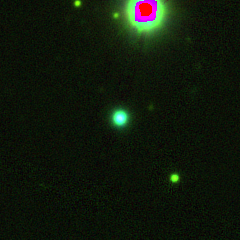 } 
            \put(5,5){\color{white}\bfseries\small giy }
        \end{overpic}%
        \includegraphics[width=0.45\textwidth]{ 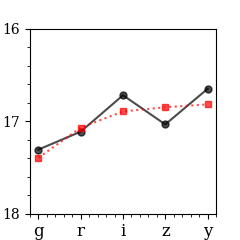 }
    \end{minipage}%
    \begin{minipage}[c]{0.23\textwidth}
        \centering
        \begin{overpic}[width=0.45\textwidth]{ 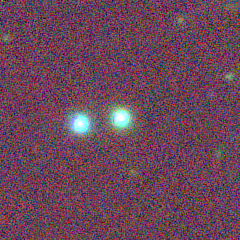 } 
            \put(5,5){\color{white}\bfseries\small giy }
        \end{overpic}%
        \includegraphics[width=0.45\textwidth]{ 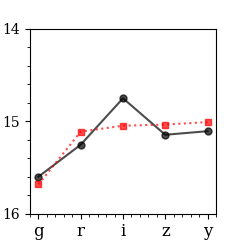 }
    \end{minipage}%
    \begin{minipage}[c]{0.23\textwidth}
        \centering
        \begin{overpic}[width=0.45\textwidth]{ 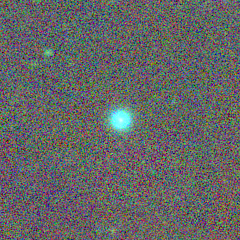 } 
            \put(5,5){\color{white}\bfseries\small giy }
        \end{overpic}%
        \includegraphics[width=0.45\textwidth]{ 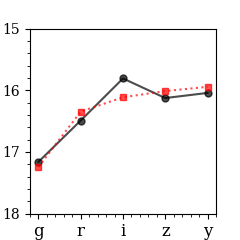 }
    \end{minipage}%
    \\[1ex] 
    \begin{minipage}[c]{0.23\textwidth}
        \centering
        \begin{overpic}[width=0.45\textwidth]{ anomaly_colored_by_izy_lower_113_z_190.71_55.8425.png } 
            \put(5,5){\color{white}\bfseries\small izy }
        \end{overpic}%
        \includegraphics[width=0.45\textwidth]{ 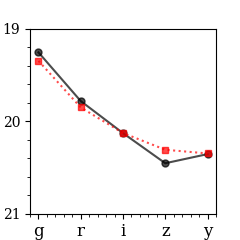 }
    \end{minipage}%
    \begin{minipage}[c]{0.23\textwidth}
        \centering
        \begin{overpic}[width=0.45\textwidth]{ 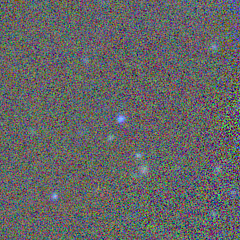 } 
            \put(5,5){\color{white}\bfseries\small izy }
        \end{overpic}%
        \includegraphics[width=0.45\textwidth]{ SED_lower_113.png }
    \end{minipage}%
    \begin{minipage}[c]{0.23\textwidth}
        \centering
        \begin{overpic}[width=0.45\textwidth]{ 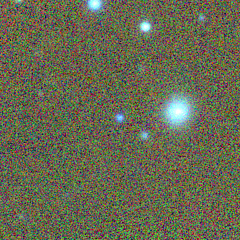 } 
            \put(5,5){\color{white}\bfseries\small izy }
        \end{overpic}%
        \includegraphics[width=0.45\textwidth]{ 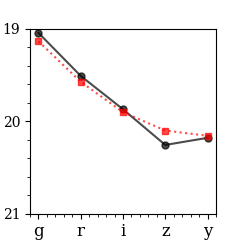 }
    \end{minipage}%
    \begin{minipage}[c]{0.23\textwidth}
        \centering
        \begin{overpic}[width=0.45\textwidth]{ 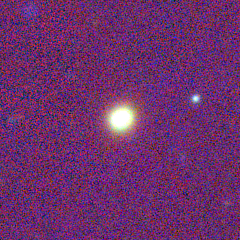 } 
            \put(5,5){\color{white}\bfseries\small giy }
        \end{overpic}%
        \includegraphics[width=0.45\textwidth]{ 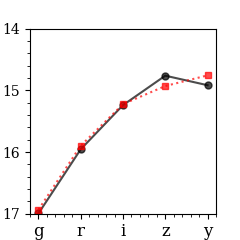 }
    \end{minipage}%
    \\[1ex] 
    \begin{minipage}[c]{0.23\textwidth}
        \centering
        \begin{overpic}[width=0.45\textwidth]{ 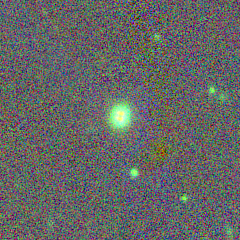 } 
            \put(5,5){\color{white}\bfseries\small giy }
        \end{overpic}%
        \includegraphics[width=0.45\textwidth]{ 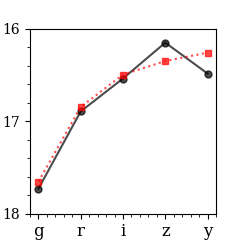 }
    \end{minipage}%
    \begin{minipage}[c]{0.23\textwidth}
        \centering
        \begin{overpic}[width=0.45\textwidth]{ 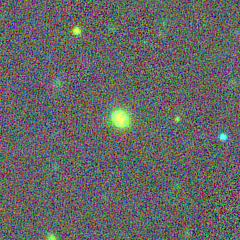 } 
            \put(5,5){\color{white}\bfseries\small giy }
        \end{overpic}%
        \includegraphics[width=0.45\textwidth]{ 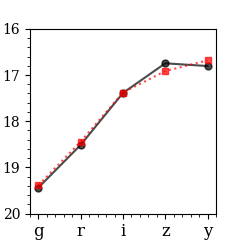 }
    \end{minipage}%
    \begin{minipage}[c]{0.23\textwidth}
        \centering
        \begin{overpic}[width=0.45\textwidth]{ 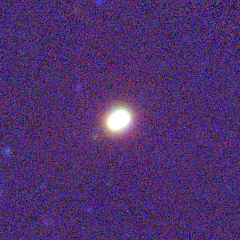 } 
            \put(5,5){\color{white}\bfseries\small giy }
        \end{overpic}%
        \includegraphics[width=0.45\textwidth]{ 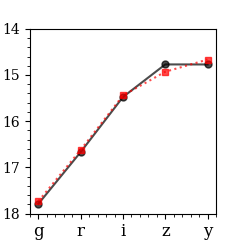 }
    \end{minipage}%
    \begin{minipage}[c]{0.23\textwidth}
        \centering
        \begin{overpic}[width=0.45\textwidth]{ 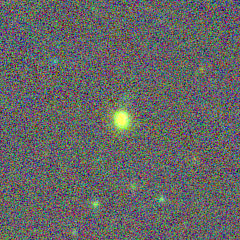 } 
            \put(5,5){\color{white}\bfseries\small giy }
        \end{overpic}%
        \includegraphics[width=0.45\textwidth]{ 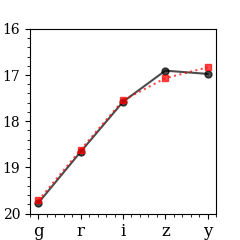 }
    \end{minipage}%
    \\[1ex] 
    \begin{minipage}[c]{0.23\textwidth}
        \centering
        \begin{overpic}[width=0.45\textwidth]{ 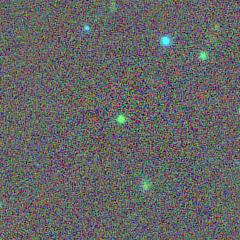 } 
            \put(5,5){\color{white}\bfseries\small giy }
        \end{overpic}%
        \includegraphics[width=0.45\textwidth]{ 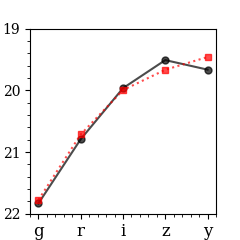 }
    \end{minipage}%
    \begin{minipage}[c]{0.23\textwidth}
        \centering
        \begin{overpic}[width=0.45\textwidth]{ 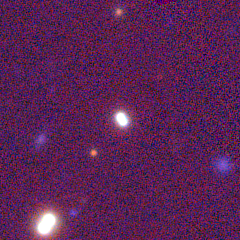 } 
            \put(5,5){\color{white}\bfseries\small giy }
        \end{overpic}%
        \includegraphics[width=0.45\textwidth]{ 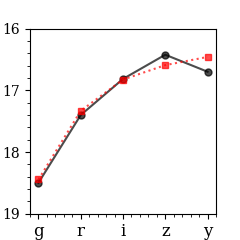 }
    \end{minipage}%
    \begin{minipage}[c]{0.23\textwidth}
        \centering
        \begin{overpic}[width=0.45\textwidth]{ 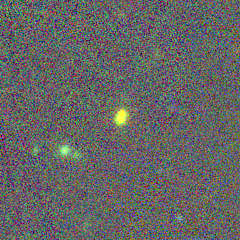 } 
            \put(5,5){\color{white}\bfseries\small giy }
        \end{overpic}%
        \includegraphics[width=0.45\textwidth]{ 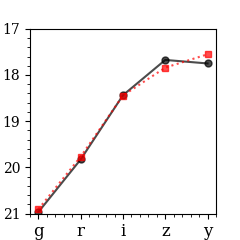 }
    \end{minipage}%
    \begin{minipage}[c]{0.23\textwidth}
        \centering
        \begin{overpic}[width=0.45\textwidth]{ 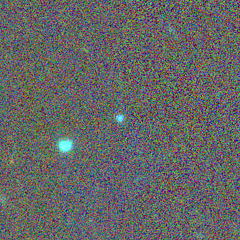 } 
            \put(5,5){\color{white}\bfseries\small giy }
        \end{overpic}%
        \includegraphics[width=0.45\textwidth]{ 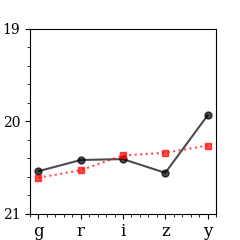 }
    \end{minipage}%
    \\[1ex] 
    \begin{minipage}[c]{0.23\textwidth}
        \centering
        \begin{overpic}[width=0.45\textwidth]{ 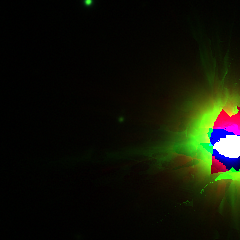 } 
            \put(5,5){\color{white}\bfseries\small giy }
        \end{overpic}%
        \includegraphics[width=0.45\textwidth]{ 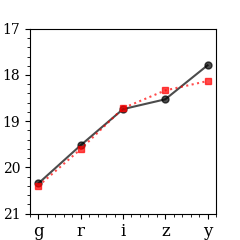 }
    \end{minipage}%
    \begin{minipage}[c]{0.23\textwidth}
        \centering
        \begin{overpic}[width=0.45\textwidth]{ 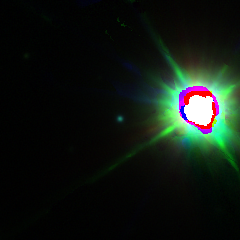 } 
            \put(5,5){\color{white}\bfseries\small giy }
        \end{overpic}%
        \includegraphics[width=0.45\textwidth]{ 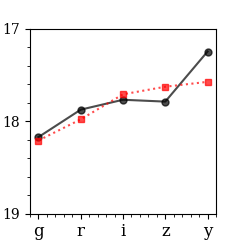 }
    \end{minipage}%
    \begin{minipage}[c]{0.23\textwidth}
        \centering
        \begin{overpic}[width=0.45\textwidth]{ 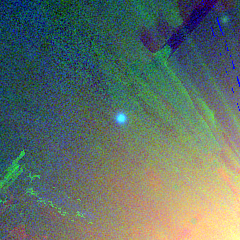 } 
            \put(5,5){\color{white}\bfseries\small giy }
        \end{overpic}%
        \includegraphics[width=0.45\textwidth]{ 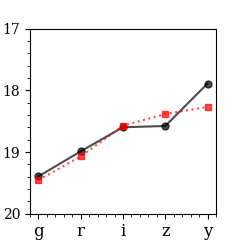 }
    \end{minipage}%
    \begin{minipage}[c]{0.23\textwidth}
        \centering
        \begin{overpic}[width=0.45\textwidth]{ 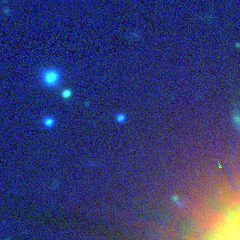 } 
            \put(5,5){\color{white}\bfseries\small giy }
        \end{overpic}%
        \includegraphics[width=0.45\textwidth]{ 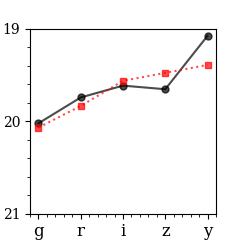 }
    \end{minipage}%
    \\[1ex] 
    \begin{minipage}[c]{0.23\textwidth}
        \centering
        \begin{overpic}[width=0.45\textwidth]{ 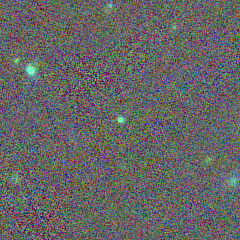 } 
            \put(5,5){\color{white}\bfseries\small giy }
        \end{overpic}%
        \includegraphics[width=0.45\textwidth]{ 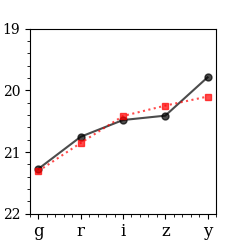 }
    \end{minipage}%
    \begin{minipage}[c]{0.23\textwidth}
        \centering
        \begin{overpic}[width=0.45\textwidth]{ 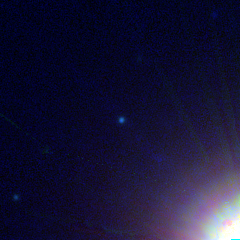 } 
            \put(5,5){\color{white}\bfseries\small giy }
        \end{overpic}%
        \includegraphics[width=0.45\textwidth]{ 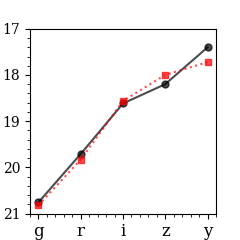 }
    \end{minipage}%
    \begin{minipage}[c]{0.23\textwidth}
        \centering
        \begin{overpic}[width=0.45\textwidth]{ 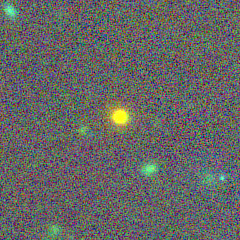 } 
            \put(5,5){\color{white}\bfseries\small giy }
        \end{overpic}%
        \includegraphics[width=0.45\textwidth]{ 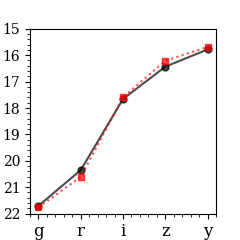 }
    \end{minipage}%
    \begin{minipage}[c]{0.23\textwidth}
        \centering
        \begin{overpic}[width=0.45\textwidth]{ 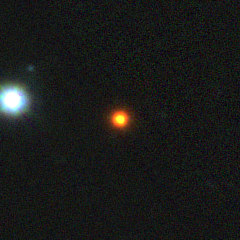 } 
            \put(5,5){\color{white}\bfseries\small riz }
        \end{overpic}%
        \includegraphics[width=0.45\textwidth]{ 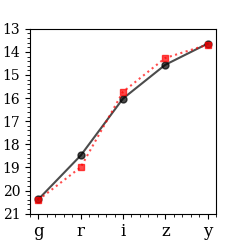 }
    \end{minipage}%
    \\[1ex] 
    \begin{minipage}[c]{0.23\textwidth}
        \centering
        \begin{overpic}[width=0.45\textwidth]{ 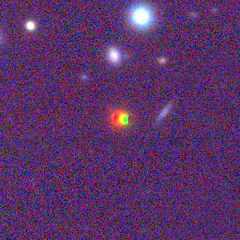 } 
            \put(5,5){\color{white}\bfseries\small riz }
        \end{overpic}%
        \includegraphics[width=0.45\textwidth]{ 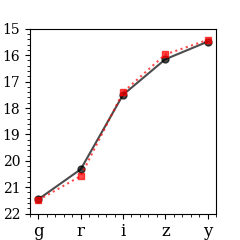 }
    \end{minipage}%
    \begin{minipage}[c]{0.23\textwidth}
        \centering
        \begin{overpic}[width=0.45\textwidth]{ 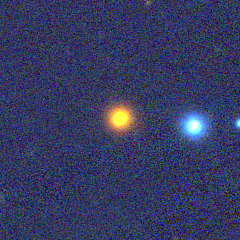 } 
            \put(5,5){\color{white}\bfseries\small riz }
        \end{overpic}%
        \includegraphics[width=0.45\textwidth]{ 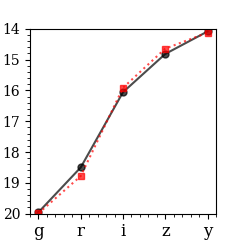 }
    \end{minipage}%
    \begin{minipage}[c]{0.23\textwidth}
        \centering
        \begin{overpic}[width=0.45\textwidth]{ 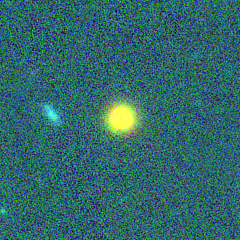 } 
            \put(5,5){\color{white}\bfseries\small riz }
        \end{overpic}%
        \includegraphics[width=0.45\textwidth]{ 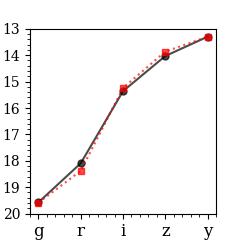 }
    \end{minipage}%
    \begin{minipage}[c]{0.23\textwidth}
        \centering
        \begin{overpic}[width=0.45\textwidth]{ 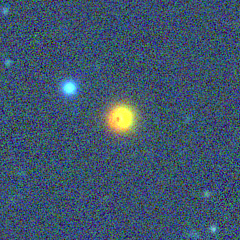 } 
            \put(5,5){\color{white}\bfseries\small riz }
        \end{overpic}%
        \includegraphics[width=0.45\textwidth]{ 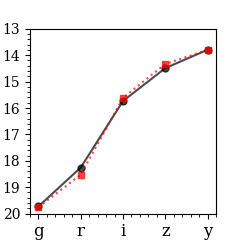 }
    \end{minipage}%
    \\[1ex] 
    \begin{minipage}[c]{0.23\textwidth}
        \centering
        \begin{overpic}[width=0.45\textwidth]{ 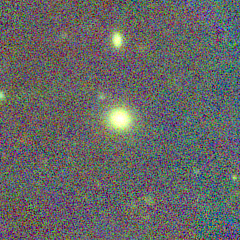 } 
            \put(5,5){\color{white}\bfseries\small giz }
        \end{overpic}%
        \includegraphics[width=0.45\textwidth]{ 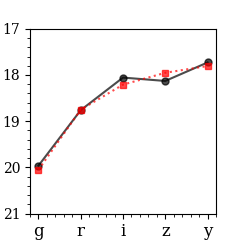 }
    \end{minipage}%
    \begin{minipage}[c]{0.23\textwidth}
        \centering
        \begin{overpic}[width=0.45\textwidth]{ 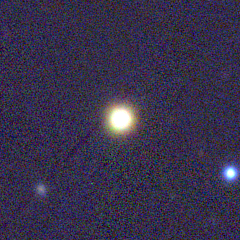 } 
            \put(5,5){\color{white}\bfseries\small giz }
        \end{overpic}%
        \includegraphics[width=0.45\textwidth]{ 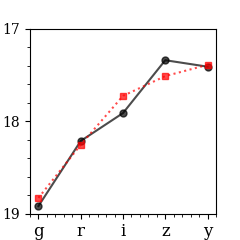 }
    \end{minipage}%
    \begin{minipage}[c]{0.23\textwidth}
        \centering
        \begin{overpic}[width=0.45\textwidth]{ 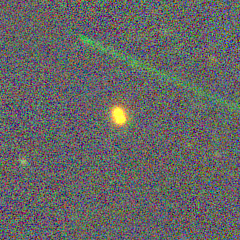 } 
            \put(5,5){\color{white}\bfseries\small giz }
        \end{overpic}%
        \includegraphics[width=0.45\textwidth]{ 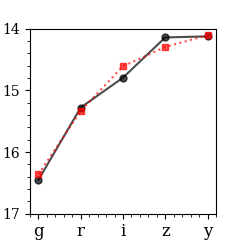 }
    \end{minipage}%
    \begin{minipage}[c]{0.23\textwidth}
        \centering
        \begin{overpic}[width=0.45\textwidth]{ anomaly_colored_by_giz_lower_116_z_185.309_9.0607.png } 
            \put(5,5){\color{white}\bfseries\small giz }
        \end{overpic}%
        \includegraphics[width=0.45\textwidth]{ 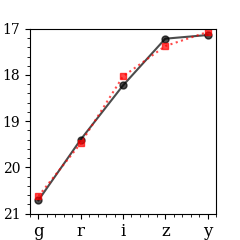 }
    \end{minipage}%
    \\[1ex] 
    \begin{minipage}[c]{0.23\textwidth}
        \centering
        \begin{overpic}[width=0.45\textwidth]{ 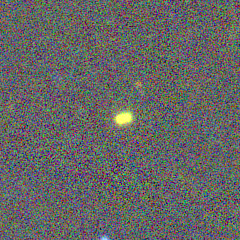 } 
            \put(5,5){\color{white}\bfseries\small giz }
        \end{overpic}%
        \includegraphics[width=0.45\textwidth]{ SED_lower_116.png }
    \end{minipage}%
    \begin{minipage}[c]{0.23\textwidth}
        \centering
        \begin{overpic}[width=0.45\textwidth]{ 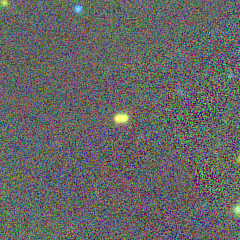 } 
            \put(5,5){\color{white}\bfseries\small giz }
        \end{overpic}%
        \includegraphics[width=0.45\textwidth]{ 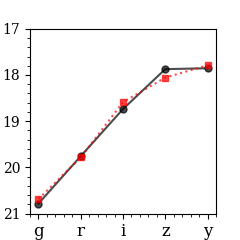 }
    \end{minipage}%
    \begin{minipage}[c]{0.23\textwidth}
        \centering
        \begin{overpic}[width=0.45\textwidth]{ 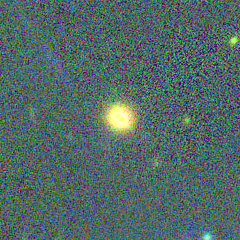 } 
            \put(5,5){\color{white}\bfseries\small giz }
        \end{overpic}%
        \includegraphics[width=0.45\textwidth]{ 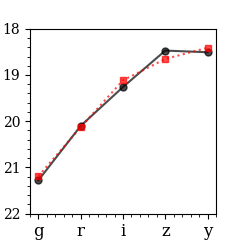 }
    \end{minipage}%
    \begin{minipage}[c]{0.23\textwidth}
        \centering
        \begin{overpic}[width=0.45\textwidth]{ 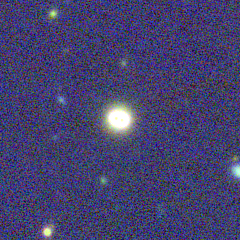 } 
            \put(5,5){\color{white}\bfseries\small giz }
        \end{overpic}%
        \includegraphics[width=0.45\textwidth]{ 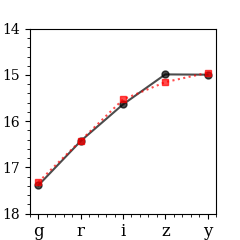 }
    \end{minipage}%
    \\[1ex] 
    \begin{minipage}[c]{0.23\textwidth}
        \centering
        \begin{overpic}[width=0.45\textwidth]{ 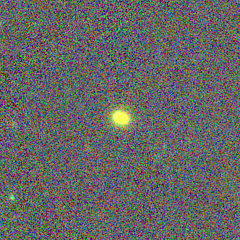 } 
            \put(5,5){\color{white}\bfseries\small giz }
        \end{overpic}%
        \includegraphics[width=0.45\textwidth]{ 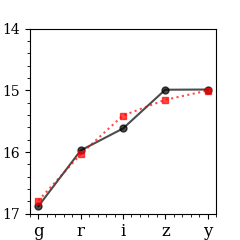 }
    \end{minipage}%
    \begin{minipage}[c]{0.23\textwidth}
        \centering
        \begin{overpic}[width=0.45\textwidth]{ 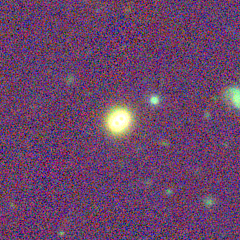 } 
            \put(5,5){\color{white}\bfseries\small giz }
        \end{overpic}%
        \includegraphics[width=0.45\textwidth]{ 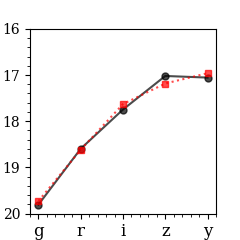 }
    \end{minipage}%
    \begin{minipage}[c]{0.23\textwidth}
        \centering
        \begin{overpic}[width=0.45\textwidth]{ 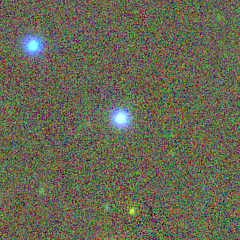 } 
            \put(5,5){\color{white}\bfseries\small giz }
        \end{overpic}%
        \includegraphics[width=0.45\textwidth]{ 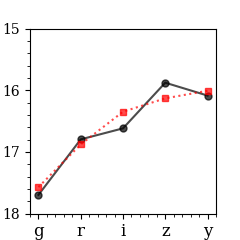 }
    \end{minipage}%
    \begin{minipage}[c]{0.23\textwidth}
        \centering
        \begin{overpic}[width=0.45\textwidth]{ 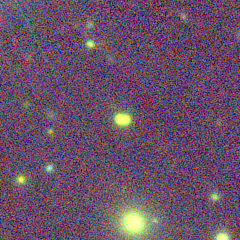 } 
            \put(5,5){\color{white}\bfseries\small giz }
        \end{overpic}%
        \includegraphics[width=0.45\textwidth]{ 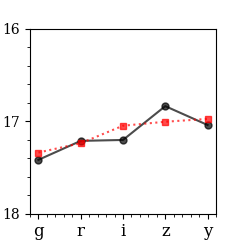 }
    \end{minipage}%
    \\[1ex] 
    \end{figure*}  

    \begin{figure*}[htbp]
        \centering
    \begin{minipage}[c]{0.23\textwidth}
        \centering
        \begin{overpic}[width=0.45\textwidth]{ 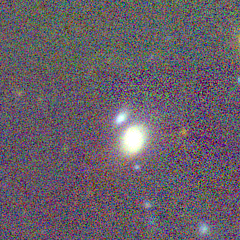 } 
            \put(5,5){\color{white}\bfseries\small giz }
        \end{overpic}%
        \includegraphics[width=0.45\textwidth]{ 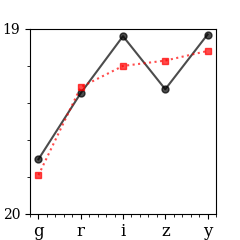 }
    \end{minipage}%
    \begin{minipage}[c]{0.23\textwidth}
        \centering
        \begin{overpic}[width=0.45\textwidth]{ 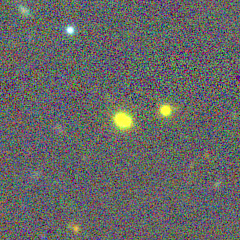 } 
            \put(5,5){\color{white}\bfseries\small giz }
        \end{overpic}%
        \includegraphics[width=0.45\textwidth]{ 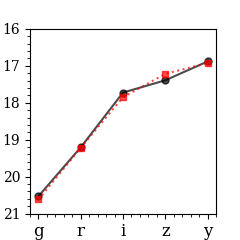 }
    \end{minipage}%
    \begin{minipage}[c]{0.23\textwidth}
        \centering
        \begin{overpic}[width=0.45\textwidth]{ 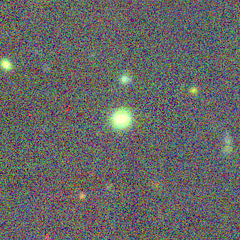 } 
            \put(5,5){\color{white}\bfseries\small giz }
        \end{overpic}%
        \includegraphics[width=0.45\textwidth]{ 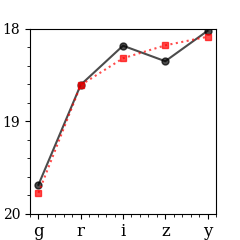 }
    \end{minipage}%
    \begin{minipage}[c]{0.23\textwidth}
        \centering
        \begin{overpic}[width=0.45\textwidth]{ 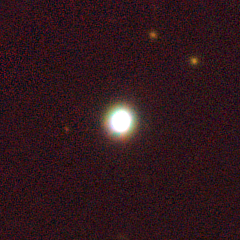 } 
            \put(5,5){\color{white}\bfseries\small giz }
        \end{overpic}%
        \includegraphics[width=0.45\textwidth]{ 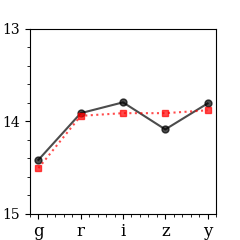 }
    \end{minipage}%
    \\[1ex] 
    \begin{minipage}[c]{0.23\textwidth}
        \centering
        \begin{overpic}[width=0.45\textwidth]{ 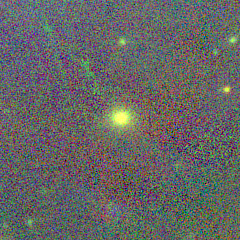 } 
            \put(5,5){\color{white}\bfseries\small giz }
        \end{overpic}%
        \includegraphics[width=0.45\textwidth]{ 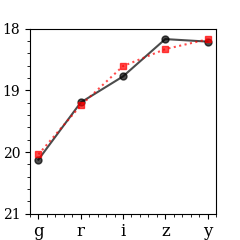 }
    \end{minipage}%
    \begin{minipage}[c]{0.23\textwidth}
        \centering
        \begin{overpic}[width=0.45\textwidth]{ 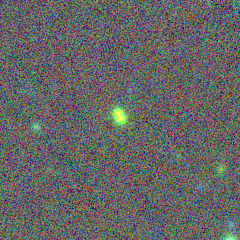 } 
            \put(5,5){\color{white}\bfseries\small giy }
        \end{overpic}%
        \includegraphics[width=0.45\textwidth]{ 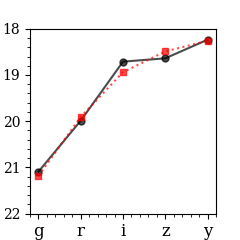 }
    \end{minipage}%
    \begin{minipage}[c]{0.23\textwidth}
        \centering
        \begin{overpic}[width=0.45\textwidth]{ 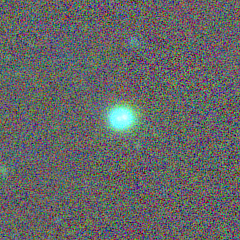 } 
            \put(5,5){\color{white}\bfseries\small giy }
        \end{overpic}%
        \includegraphics[width=0.45\textwidth]{ 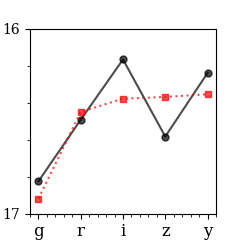 }
    \end{minipage}%
    \begin{minipage}[c]{0.23\textwidth}
        \centering
        \begin{overpic}[width=0.45\textwidth]{ 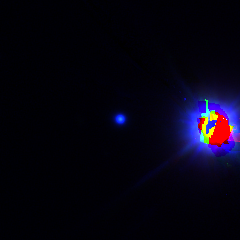 } 
            \put(5,5){\color{white}\bfseries\small giy }
        \end{overpic}%
        \includegraphics[width=0.45\textwidth]{ 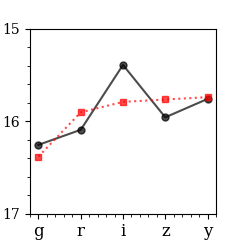 }
    \end{minipage}%
    \\[1ex] 
    \begin{minipage}[c]{0.23\textwidth}
        \centering
        \begin{overpic}[width=0.45\textwidth]{ 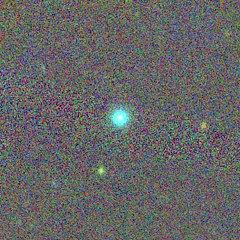 } 
            \put(5,5){\color{white}\bfseries\small giy }
        \end{overpic}%
        \includegraphics[width=0.45\textwidth]{ 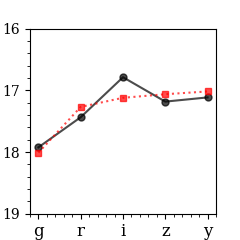 }
    \end{minipage}%
    \begin{minipage}[c]{0.23\textwidth}
        \centering
        \begin{overpic}[width=0.45\textwidth]{ 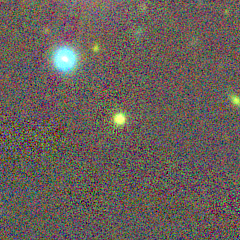 } 
            \put(5,5){\color{white}\bfseries\small giy }
        \end{overpic}%
        \includegraphics[width=0.45\textwidth]{ 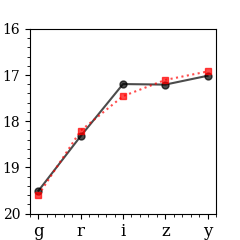 }
    \end{minipage}%
    \begin{minipage}[c]{0.23\textwidth}
        \centering
        \begin{overpic}[width=0.45\textwidth]{ 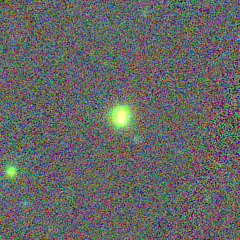 } 
            \put(5,5){\color{white}\bfseries\small giy }
        \end{overpic}%
        \includegraphics[width=0.45\textwidth]{ 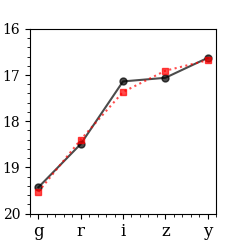 }
    \end{minipage}%
    \begin{minipage}[c]{0.23\textwidth}
        \centering
        \begin{overpic}[width=0.45\textwidth]{ 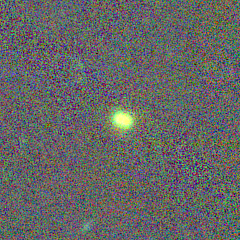 } 
            \put(5,5){\color{white}\bfseries\small giy }
        \end{overpic}%
        \includegraphics[width=0.45\textwidth]{ 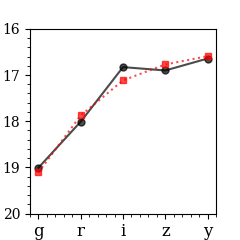 }
    \end{minipage}%
    \\[1ex] 
    \begin{minipage}[c]{0.23\textwidth}
        \centering
        \begin{overpic}[width=0.45\textwidth]{ 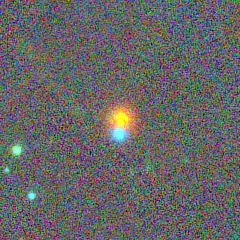 } 
            \put(5,5){\color{white}\bfseries\small giy }
        \end{overpic}%
        \includegraphics[width=0.45\textwidth]{ 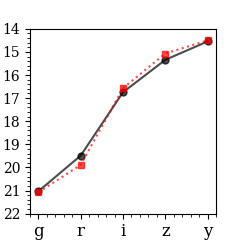 }
    \end{minipage}%
    \begin{minipage}[c]{0.23\textwidth}
        \centering
        \begin{overpic}[width=0.45\textwidth]{ 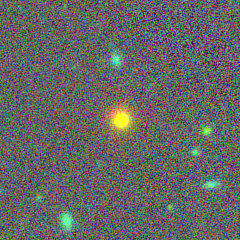 } 
            \put(5,5){\color{white}\bfseries\small giy }
        \end{overpic}%
        \includegraphics[width=0.45\textwidth]{ 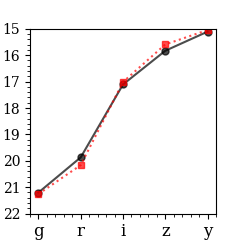 }
    \end{minipage}%
    \begin{minipage}[c]{0.23\textwidth}
        \centering
        \begin{overpic}[width=0.45\textwidth]{ 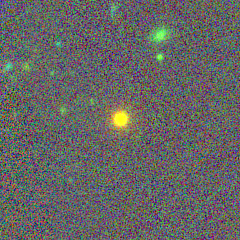 } 
            \put(5,5){\color{white}\bfseries\small giy }
        \end{overpic}%
        \includegraphics[width=0.45\textwidth]{ 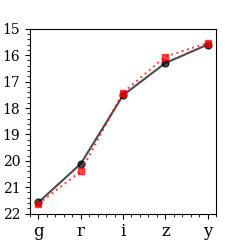 }
    \end{minipage}%
    \begin{minipage}[c]{0.23\textwidth}
        \centering
        \begin{overpic}[width=0.45\textwidth]{ 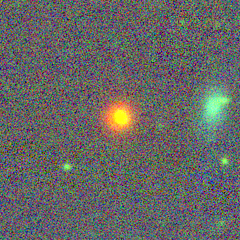 } 
            \put(5,5){\color{white}\bfseries\small giy }
        \end{overpic}%
        \includegraphics[width=0.45\textwidth]{ 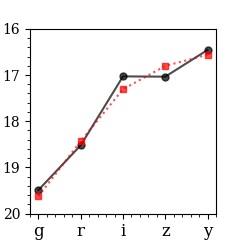 }
    \end{minipage}%
    \\[1ex] 
    \begin{minipage}[c]{0.23\textwidth}
        \centering
        \begin{overpic}[width=0.45\textwidth]{ 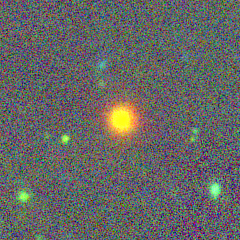 } 
            \put(5,5){\color{white}\bfseries\small giy }
        \end{overpic}%
        \includegraphics[width=0.45\textwidth]{ 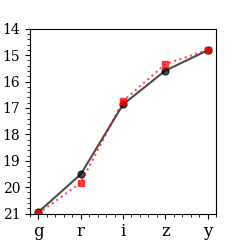 }
    \end{minipage}%
    \begin{minipage}[c]{0.23\textwidth}
        \centering
        \begin{overpic}[width=0.45\textwidth]{ 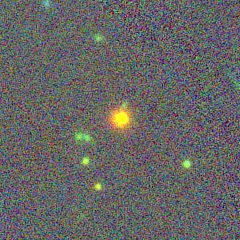 } 
            \put(5,5){\color{white}\bfseries\small giy }
        \end{overpic}%
        \includegraphics[width=0.45\textwidth]{ 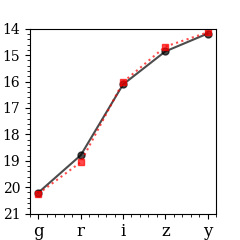 }
    \end{minipage}%
    \begin{minipage}[c]{0.23\textwidth}
        \centering
        \begin{overpic}[width=0.45\textwidth]{ 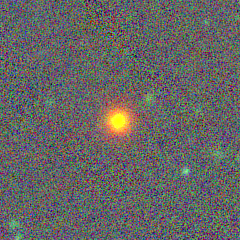 } 
            \put(5,5){\color{white}\bfseries\small giy }
        \end{overpic}%
        \includegraphics[width=0.45\textwidth]{ 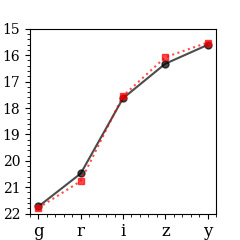 }
    \end{minipage}%
    \begin{minipage}[c]{0.23\textwidth}
        \centering
        \begin{overpic}[width=0.45\textwidth]{ 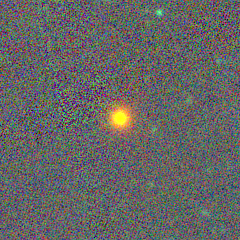 } 
            \put(5,5){\color{white}\bfseries\small giy }
        \end{overpic}%
        \includegraphics[width=0.45\textwidth]{ 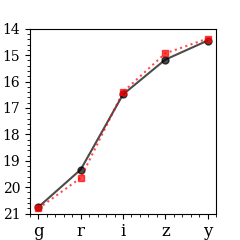 }
    \end{minipage}%
    \\[1ex] 
    \begin{minipage}[c]{0.23\textwidth}
        \centering
        \begin{overpic}[width=0.45\textwidth]{ 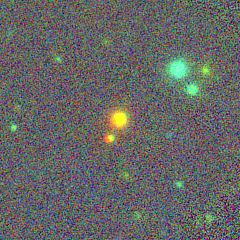 } 
            \put(5,5){\color{white}\bfseries\small giy }
        \end{overpic}%
        \includegraphics[width=0.45\textwidth]{ 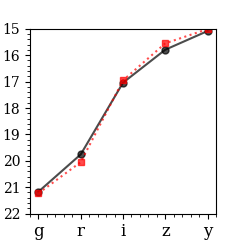 }
    \end{minipage}%
    \begin{minipage}[c]{0.23\textwidth}
        \centering
        \begin{overpic}[width=0.45\textwidth]{ 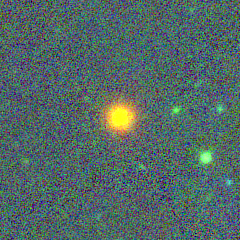 } 
            \put(5,5){\color{white}\bfseries\small giy }
        \end{overpic}%
        \includegraphics[width=0.45\textwidth]{ 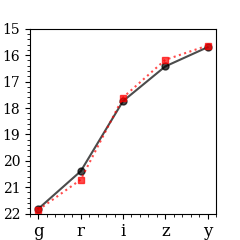 }
    \end{minipage}%
    \begin{minipage}[c]{0.23\textwidth}
        \centering
        \begin{overpic}[width=0.45\textwidth]{ 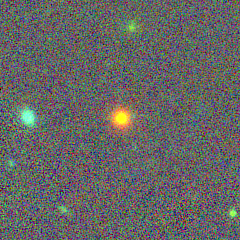 } 
            \put(5,5){\color{white}\bfseries\small giy }
        \end{overpic}%
        \includegraphics[width=0.45\textwidth]{ 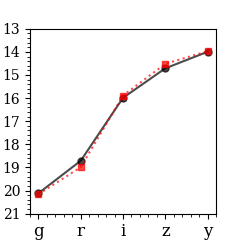 }
    \end{minipage}%
    \begin{minipage}[c]{0.23\textwidth}
        \centering
        \begin{overpic}[width=0.45\textwidth]{ 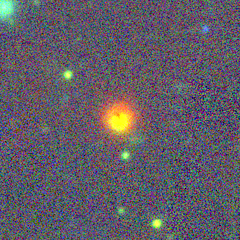 } 
            \put(5,5){\color{white}\bfseries\small giy }
        \end{overpic}%
        \includegraphics[width=0.45\textwidth]{ 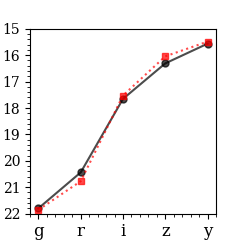 }
    \end{minipage}%
    \\[1ex] 
    \begin{minipage}[c]{0.23\textwidth}
        \centering
        \begin{overpic}[width=0.45\textwidth]{ 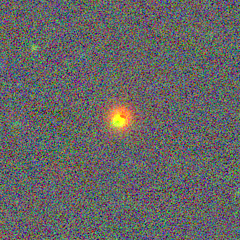 } 
            \put(5,5){\color{white}\bfseries\small giy }
        \end{overpic}%
        \includegraphics[width=0.45\textwidth]{ 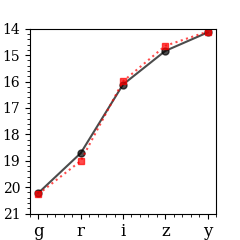 }
    \end{minipage}%
    \begin{minipage}[c]{0.23\textwidth}
        \centering
        \begin{overpic}[width=0.45\textwidth]{ 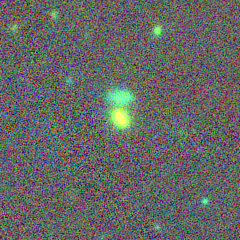 } 
            \put(5,5){\color{white}\bfseries\small giy }
        \end{overpic}%
        \includegraphics[width=0.45\textwidth]{ 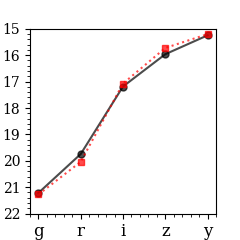 }
    \end{minipage}%
    \begin{minipage}[c]{0.23\textwidth}
        \centering
        \begin{overpic}[width=0.45\textwidth]{ 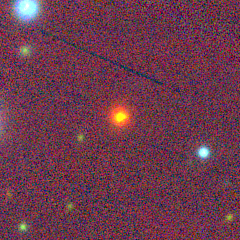 } 
            \put(5,5){\color{white}\bfseries\small giy }
        \end{overpic}%
        \includegraphics[width=0.45\textwidth]{ 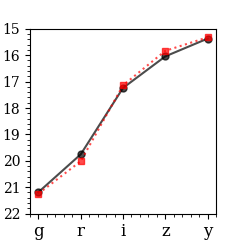 }
    \end{minipage}%
    \begin{minipage}[c]{0.23\textwidth}
        \centering
        \begin{overpic}[width=0.45\textwidth]{ 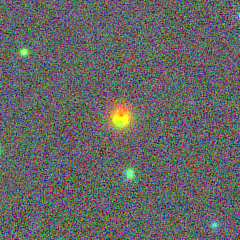 } 
            \put(5,5){\color{white}\bfseries\small giy }
        \end{overpic}%
        \includegraphics[width=0.45\textwidth]{ 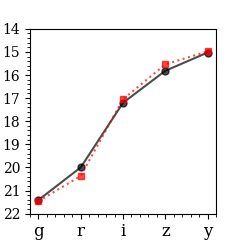 }
    \end{minipage}%
    \\[1ex] 
    \begin{minipage}[c]{0.23\textwidth}
        \centering
        \begin{overpic}[width=0.45\textwidth]{ 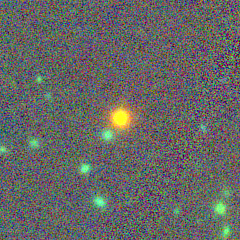 } 
            \put(5,5){\color{white}\bfseries\small giy }
        \end{overpic}%
        \includegraphics[width=0.45\textwidth]{ 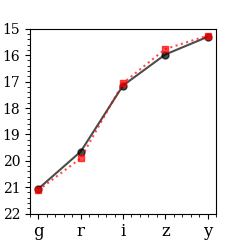 }
    \end{minipage}%
    \begin{minipage}[c]{0.23\textwidth}
        \centering
        \begin{overpic}[width=0.45\textwidth]{ 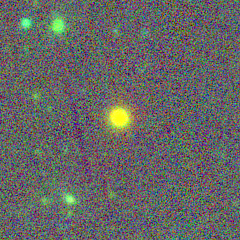 } 
            \put(5,5){\color{white}\bfseries\small giy }
        \end{overpic}%
        \includegraphics[width=0.45\textwidth]{ 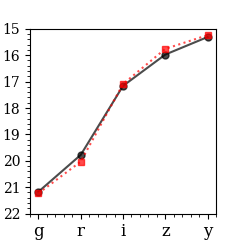 }
    \end{minipage}%
    \begin{minipage}[c]{0.23\textwidth}
        \centering
        \begin{overpic}[width=0.45\textwidth]{ 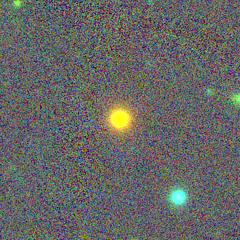 } 
            \put(5,5){\color{white}\bfseries\small giy }
        \end{overpic}%
        \includegraphics[width=0.45\textwidth]{ 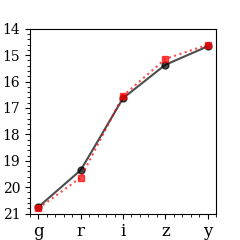 }
    \end{minipage}%
    \begin{minipage}[c]{0.23\textwidth}
        \centering
        \begin{overpic}[width=0.45\textwidth]{ 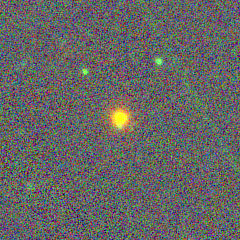 } 
            \put(5,5){\color{white}\bfseries\small giy }
        \end{overpic}%
        \includegraphics[width=0.45\textwidth]{ 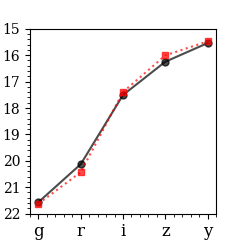 }
    \end{minipage}%
    \\[1ex] 
    \begin{minipage}[c]{0.23\textwidth}
        \centering
        \begin{overpic}[width=0.45\textwidth]{ 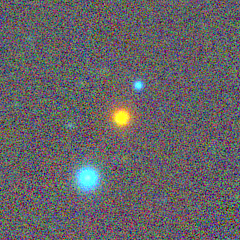 } 
            \put(5,5){\color{white}\bfseries\small giy }
        \end{overpic}%
        \includegraphics[width=0.45\textwidth]{ 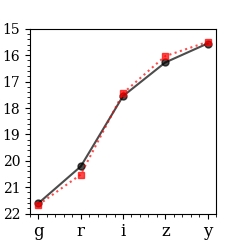 }
    \end{minipage}%
    \begin{minipage}[c]{0.23\textwidth}
        \centering
        \begin{overpic}[width=0.45\textwidth]{ 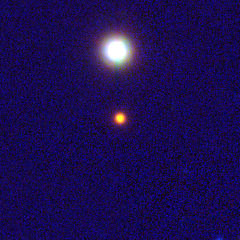 } 
            \put(5,5){\color{white}\bfseries\small giy }
        \end{overpic}%
        \includegraphics[width=0.45\textwidth]{ 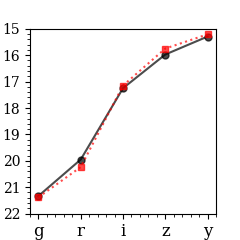 }
    \end{minipage}%
    \begin{minipage}[c]{0.23\textwidth}
        \centering
        \begin{overpic}[width=0.45\textwidth]{ 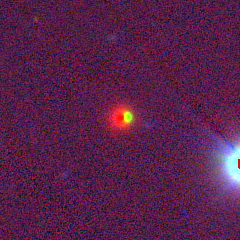 } 
            \put(5,5){\color{white}\bfseries\small giy }
        \end{overpic}%
        \includegraphics[width=0.45\textwidth]{ 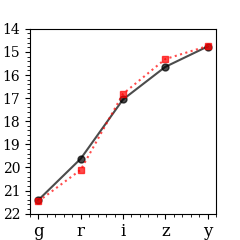 }
    \end{minipage}%
    \begin{minipage}[c]{0.23\textwidth}
        \centering
        \begin{overpic}[width=0.45\textwidth]{ 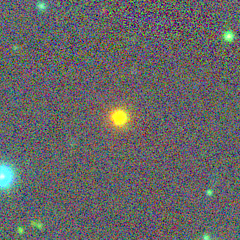 } 
            \put(5,5){\color{white}\bfseries\small giy }
        \end{overpic}%
        \includegraphics[width=0.45\textwidth]{ 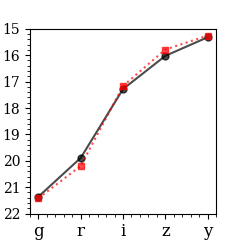 }
    \end{minipage}%
    \\[1ex] 
    \begin{minipage}[c]{0.23\textwidth}
        \centering
        \begin{overpic}[width=0.45\textwidth]{ 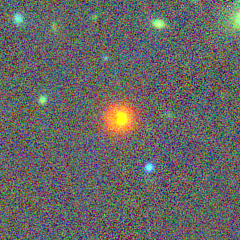 } 
            \put(5,5){\color{white}\bfseries\small giy }
        \end{overpic}%
        \includegraphics[width=0.45\textwidth]{ 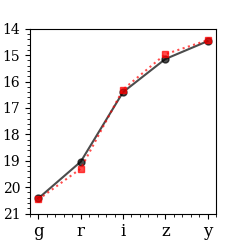 }
    \end{minipage}%
    \begin{minipage}[c]{0.23\textwidth}
        \centering
        \begin{overpic}[width=0.45\textwidth]{ 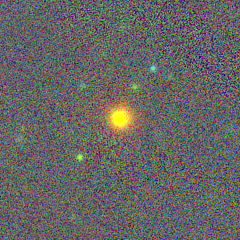 } 
            \put(5,5){\color{white}\bfseries\small giy }
        \end{overpic}%
        \includegraphics[width=0.45\textwidth]{ 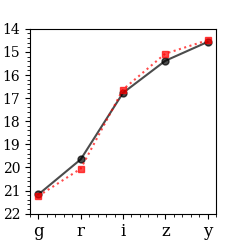 }
    \end{minipage}%
    \begin{minipage}[c]{0.23\textwidth}
        \centering
        \begin{overpic}[width=0.45\textwidth]{ 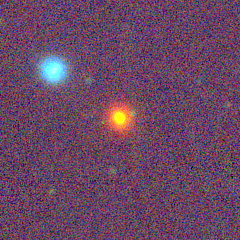 } 
            \put(5,5){\color{white}\bfseries\small giy }
        \end{overpic}%
        \includegraphics[width=0.45\textwidth]{ 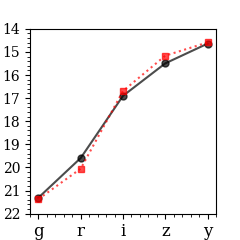 }
    \end{minipage}%
    \begin{minipage}[c]{0.23\textwidth}
        \centering
        \begin{overpic}[width=0.45\textwidth]{ 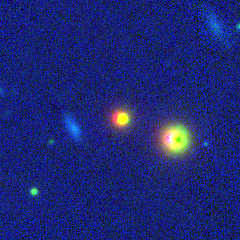 } 
            \put(5,5){\color{white}\bfseries\small giy }
        \end{overpic}%
        \includegraphics[width=0.45\textwidth]{ 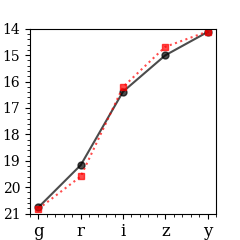 }
    \end{minipage}%
    \\[1ex] 
    \begin{minipage}[c]{0.23\textwidth}
        \centering
        \begin{overpic}[width=0.45\textwidth]{ 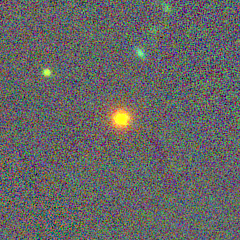 } 
            \put(5,5){\color{white}\bfseries\small giy }
        \end{overpic}%
        \includegraphics[width=0.45\textwidth]{ 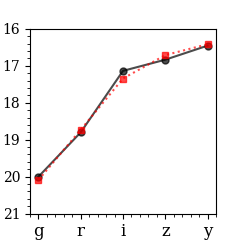 }
    \end{minipage}%
    \begin{minipage}[c]{0.23\textwidth}
        \centering
        \begin{overpic}[width=0.45\textwidth]{ 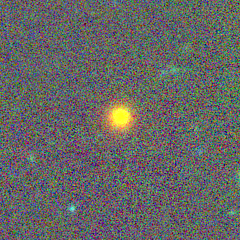 } 
            \put(5,5){\color{white}\bfseries\small giy }
        \end{overpic}%
        \includegraphics[width=0.45\textwidth]{ 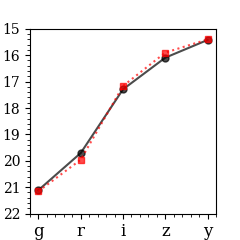 }
    \end{minipage}%
    \begin{minipage}[c]{0.23\textwidth}
        \centering
        \begin{overpic}[width=0.45\textwidth]{ 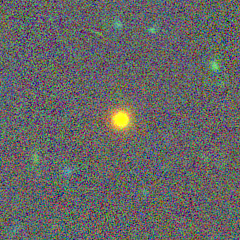 } 
            \put(5,5){\color{white}\bfseries\small giy }
        \end{overpic}%
        \includegraphics[width=0.45\textwidth]{ 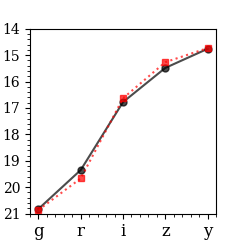 }
    \end{minipage}%
    \begin{minipage}[c]{0.23\textwidth}
        \centering
        \begin{overpic}[width=0.45\textwidth]{ 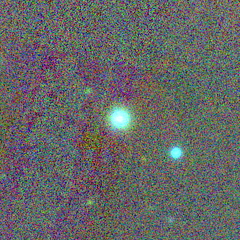 } 
            \put(5,5){\color{white}\bfseries\small giy }
        \end{overpic}%
        \includegraphics[width=0.45\textwidth]{ 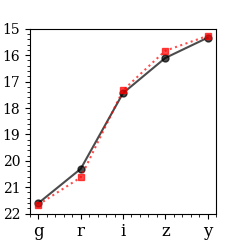 }
    \end{minipage}%
    \\[1ex] 
    \begin{minipage}[c]{0.23\textwidth}
        \centering
        \begin{overpic}[width=0.45\textwidth]{ 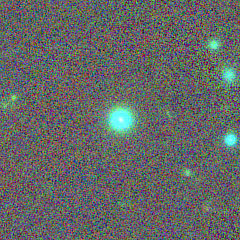 } 
            \put(5,5){\color{white}\bfseries\small giy }
        \end{overpic}%
        \includegraphics[width=0.45\textwidth]{ 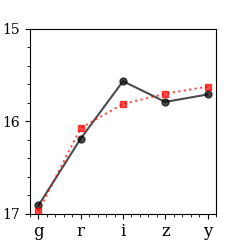 }
    \end{minipage}%
    \begin{minipage}[c]{0.23\textwidth}
        \centering
        \begin{overpic}[width=0.45\textwidth]{ 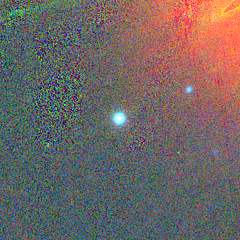 } 
            \put(5,5){\color{white}\bfseries\small giy }
        \end{overpic}%
        \includegraphics[width=0.45\textwidth]{ 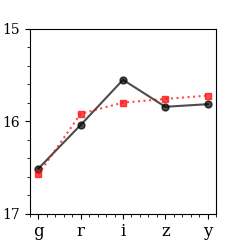 }
    \end{minipage}%
    \begin{minipage}[c]{0.23\textwidth}
        \centering
        \begin{overpic}[width=0.45\textwidth]{ 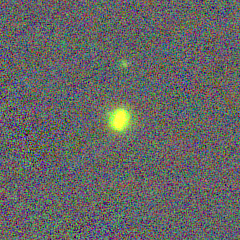 } 
            \put(5,5){\color{white}\bfseries\small giy }
        \end{overpic}%
        \includegraphics[width=0.45\textwidth]{ 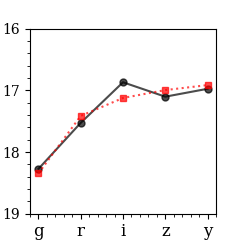 }
    \end{minipage}%
    \begin{minipage}[c]{0.23\textwidth}
        \centering
        \begin{overpic}[width=0.45\textwidth]{ 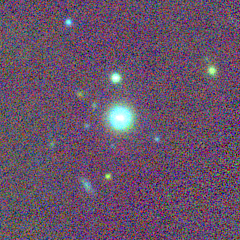 } 
            \put(5,5){\color{white}\bfseries\small giy }
        \end{overpic}%
        \includegraphics[width=0.45\textwidth]{ 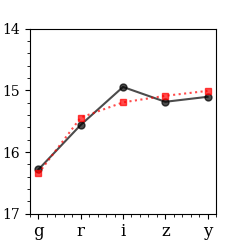 }
    \end{minipage}%
    \\[1ex] 
    \end{figure*}  

    \begin{figure*}[htbp]
        \centering
    \begin{minipage}[c]{0.23\textwidth}
        \centering
        \begin{overpic}[width=0.45\textwidth]{ 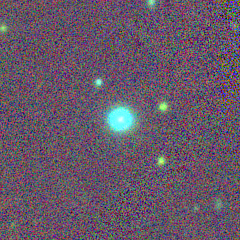 } 
            \put(5,5){\color{white}\bfseries\small giy }
        \end{overpic}%
        \includegraphics[width=0.45\textwidth]{ 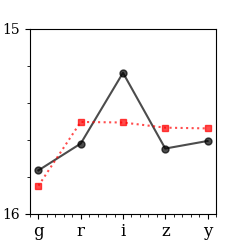 }
    \end{minipage}%
    \begin{minipage}[c]{0.23\textwidth}
        \centering
        \begin{overpic}[width=0.45\textwidth]{ 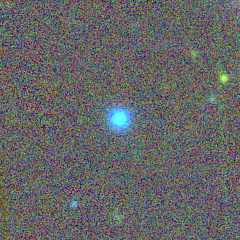 } 
            \put(5,5){\color{white}\bfseries\small giy }
        \end{overpic}%
        \includegraphics[width=0.45\textwidth]{ 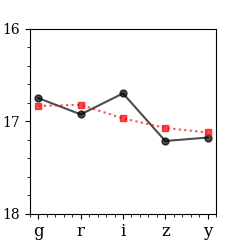 }
    \end{minipage}%
    \begin{minipage}[c]{0.23\textwidth}
        \centering
        \begin{overpic}[width=0.45\textwidth]{ 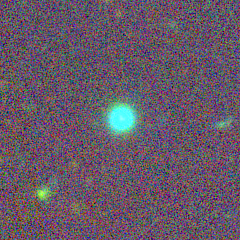 } 
            \put(5,5){\color{white}\bfseries\small giy }
        \end{overpic}%
        \includegraphics[width=0.45\textwidth]{ 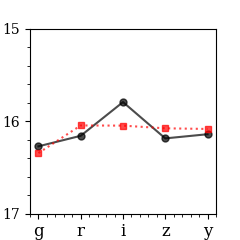 }
    \end{minipage}%
    \begin{minipage}[c]{0.23\textwidth}
        \centering
        \begin{overpic}[width=0.45\textwidth]{ 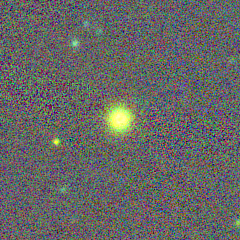 } 
            \put(5,5){\color{white}\bfseries\small giy }
        \end{overpic}%
        \includegraphics[width=0.45\textwidth]{ 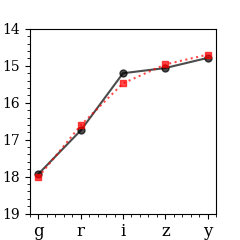 }
    \end{minipage}%
    \\[1ex] 
    \begin{minipage}[c]{0.23\textwidth}
        \centering
        \begin{overpic}[width=0.45\textwidth]{ 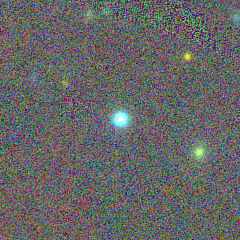 } 
            \put(5,5){\color{white}\bfseries\small giy }
        \end{overpic}%
        \includegraphics[width=0.45\textwidth]{ 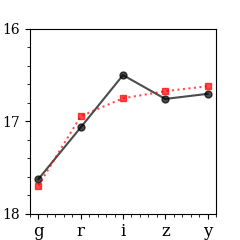 }
    \end{minipage}%
    \begin{minipage}[c]{0.23\textwidth}
        \centering
        \begin{overpic}[width=0.45\textwidth]{ 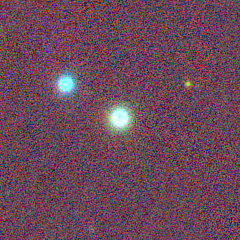 } 
            \put(5,5){\color{white}\bfseries\small giy }
        \end{overpic}%
        \includegraphics[width=0.45\textwidth]{ 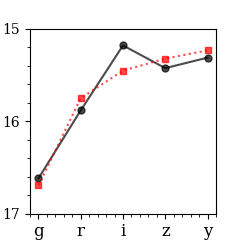 }
    \end{minipage}%
    \begin{minipage}[c]{0.23\textwidth}
        \centering
        \begin{overpic}[width=0.45\textwidth]{ 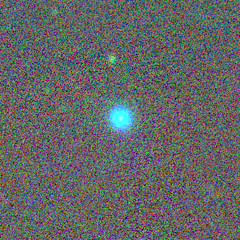 } 
            \put(5,5){\color{white}\bfseries\small giy }
        \end{overpic}%
        \includegraphics[width=0.45\textwidth]{ 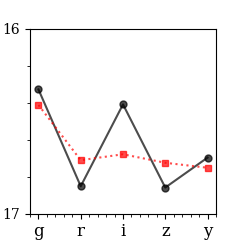 }
    \end{minipage}%
    \begin{minipage}[c]{0.23\textwidth}
        \centering
        \begin{overpic}[width=0.45\textwidth]{ 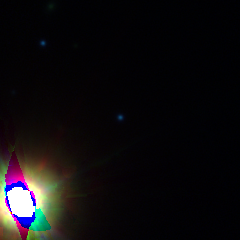 } 
            \put(5,5){\color{white}\bfseries\small giy }
        \end{overpic}%
        \includegraphics[width=0.45\textwidth]{ 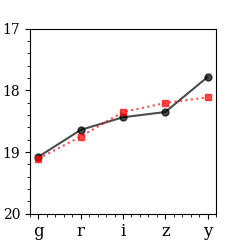 }
    \end{minipage}%
    \\[1ex] 
    \begin{minipage}[c]{0.23\textwidth}
        \centering
        \begin{overpic}[width=0.45\textwidth]{ 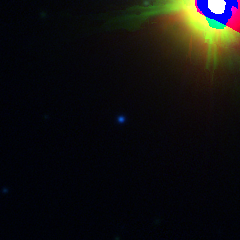 } 
            \put(5,5){\color{white}\bfseries\small giy }
        \end{overpic}%
        \includegraphics[width=0.45\textwidth]{ 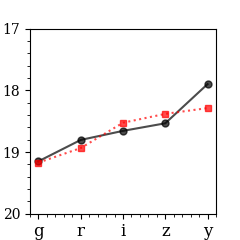 }
    \end{minipage}%
    \begin{minipage}[c]{0.23\textwidth}
        \centering
        \begin{overpic}[width=0.45\textwidth]{ 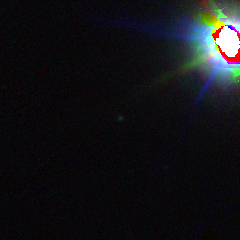 } 
            \put(5,5){\color{white}\bfseries\small giy }
        \end{overpic}%
        \includegraphics[width=0.45\textwidth]{ 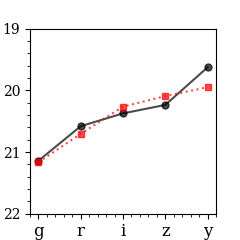 }
    \end{minipage}%
    \begin{minipage}[c]{0.23\textwidth}
        \centering
        \begin{overpic}[width=0.45\textwidth]{ 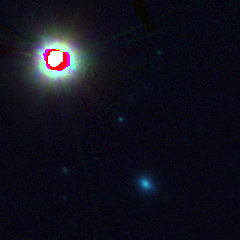 } 
            \put(5,5){\color{white}\bfseries\small giy }
        \end{overpic}%
        \includegraphics[width=0.45\textwidth]{ 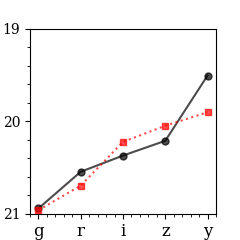 }
    \end{minipage}%
    \begin{minipage}[c]{0.23\textwidth}
        \centering
        \begin{overpic}[width=0.45\textwidth]{ 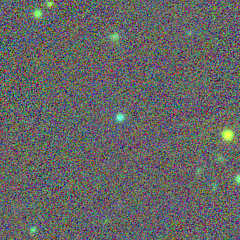 } 
            \put(5,5){\color{white}\bfseries\small giy }
        \end{overpic}%
        \includegraphics[width=0.45\textwidth]{ 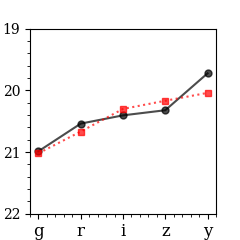 }
    \end{minipage}%
    \\[1ex] 
    \begin{minipage}[c]{0.23\textwidth}
        \centering
        \begin{overpic}[width=0.45\textwidth]{ 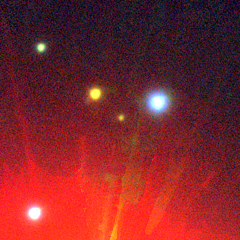 } 
            \put(5,5){\color{white}\bfseries\small giy }
        \end{overpic}%
        \includegraphics[width=0.45\textwidth]{ 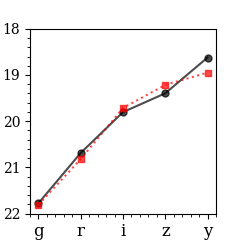 }
    \end{minipage}%
    \begin{minipage}[c]{0.23\textwidth}
        \centering
        \begin{overpic}[width=0.45\textwidth]{ anomaly_colored_by_giy_lower_154_y_208.2826_38.3051.png } 
            \put(5,5){\color{white}\bfseries\small giy }
        \end{overpic}%
        \includegraphics[width=0.45\textwidth]{ SED_lower_154.png }
    \end{minipage}%
    \begin{minipage}[c]{0.23\textwidth}
        \centering
        \begin{overpic}[width=0.45\textwidth]{ 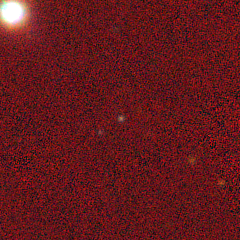 } 
            \put(5,5){\color{white}\bfseries\small giy }
        \end{overpic}%
        \includegraphics[width=0.45\textwidth]{ 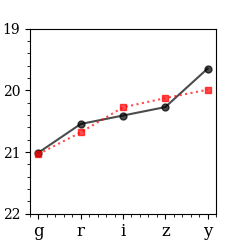 }
    \end{minipage}%
    \begin{minipage}[c]{0.23\textwidth}
        \centering
        \begin{overpic}[width=0.45\textwidth]{ 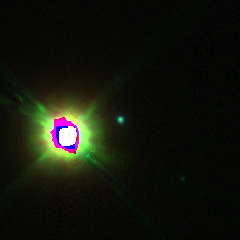 } 
            \put(5,5){\color{white}\bfseries\small giy }
        \end{overpic}%
        \includegraphics[width=0.45\textwidth]{ 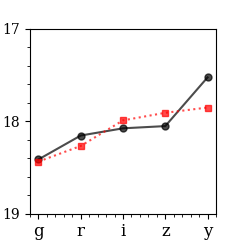 }
    \end{minipage}%
    \\[1ex] 
    \begin{minipage}[c]{0.23\textwidth}
        \centering
        \begin{overpic}[width=0.45\textwidth]{ 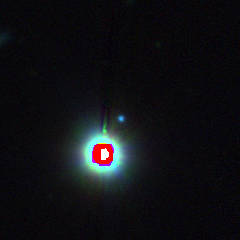 } 
            \put(5,5){\color{white}\bfseries\small giy }
        \end{overpic}%
        \includegraphics[width=0.45\textwidth]{ 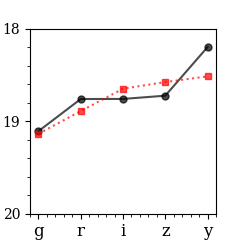 }
    \end{minipage}%
    \begin{minipage}[c]{0.23\textwidth}
        \centering
        \begin{overpic}[width=0.45\textwidth]{ 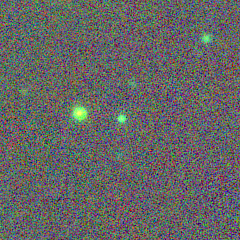 } 
            \put(5,5){\color{white}\bfseries\small giy }
        \end{overpic}%
        \includegraphics[width=0.45\textwidth]{ 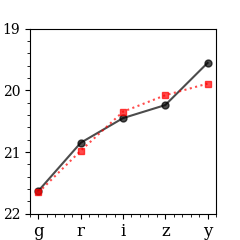 }
    \end{minipage}%
    \begin{minipage}[c]{0.23\textwidth}
        \centering
        \begin{overpic}[width=0.45\textwidth]{ 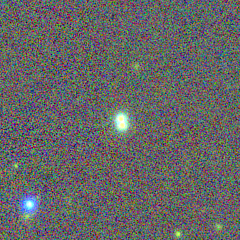 } 
            \put(5,5){\color{white}\bfseries\small gzy }
        \end{overpic}%
        \includegraphics[width=0.45\textwidth]{ 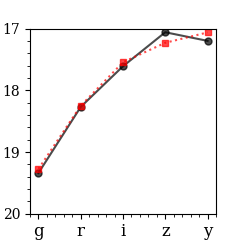 }
    \end{minipage}%
    \begin{minipage}[c]{0.23\textwidth}
        \centering
        \begin{overpic}[width=0.45\textwidth]{ 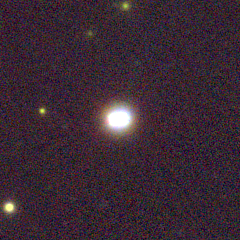 } 
            \put(5,5){\color{white}\bfseries\small gzy }
        \end{overpic}%
        \includegraphics[width=0.45\textwidth]{ 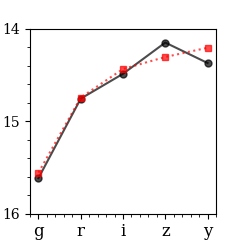 }
    \end{minipage}%
    \\[1ex] 
    \begin{minipage}[c]{0.23\textwidth}
        \centering
        \begin{overpic}[width=0.45\textwidth]{ 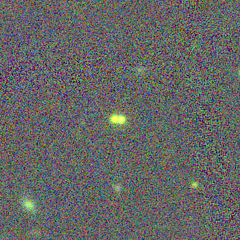 } 
            \put(5,5){\color{white}\bfseries\small gzy }
        \end{overpic}%
        \includegraphics[width=0.45\textwidth]{ 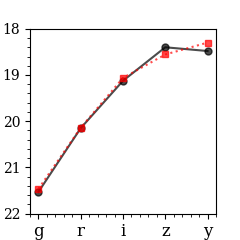 }
    \end{minipage}%
    \begin{minipage}[c]{0.23\textwidth}
        \centering
        \begin{overpic}[width=0.45\textwidth]{ 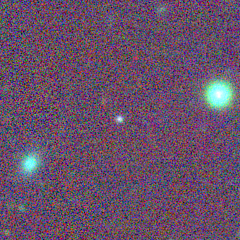 } 
            \put(5,5){\color{white}\bfseries\small giy }
        \end{overpic}%
        \includegraphics[width=0.45\textwidth]{ 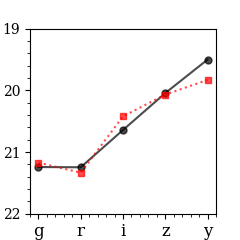 }
    \end{minipage}%
    \begin{minipage}[c]{0.23\textwidth}
        \centering
        \begin{overpic}[width=0.45\textwidth]{ 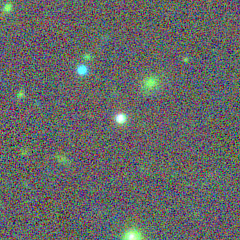 } 
            \put(5,5){\color{white}\bfseries\small giy }
        \end{overpic}%
        \includegraphics[width=0.45\textwidth]{ 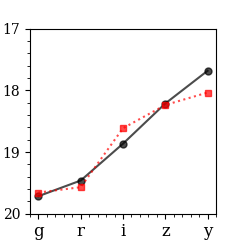 }
    \end{minipage}%
    \begin{minipage}[c]{0.23\textwidth}
        \centering
        \begin{overpic}[width=0.45\textwidth]{ 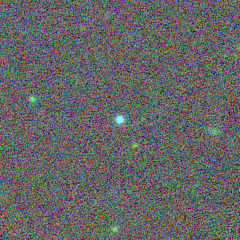 } 
            \put(5,5){\color{white}\bfseries\small giy }
        \end{overpic}%
        \includegraphics[width=0.45\textwidth]{ 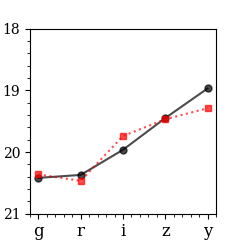 }
    \end{minipage}%
    \\[1ex] 
    \begin{minipage}[c]{0.23\textwidth}
        \centering
        \begin{overpic}[width=0.45\textwidth]{ 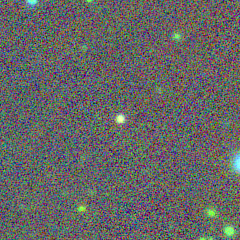 } 
            \put(5,5){\color{white}\bfseries\small giy }
        \end{overpic}%
        \includegraphics[width=0.45\textwidth]{ 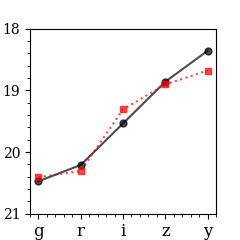 }
    \end{minipage}%
    \begin{minipage}[c]{0.23\textwidth}
        \centering
        \begin{overpic}[width=0.45\textwidth]{ 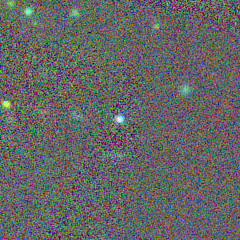 } 
            \put(5,5){\color{white}\bfseries\small giy }
        \end{overpic}%
        \includegraphics[width=0.45\textwidth]{ 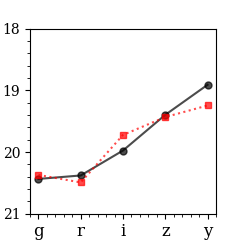 }
    \end{minipage}%
    \begin{minipage}[c]{0.23\textwidth}
        \centering
        \begin{overpic}[width=0.45\textwidth]{ 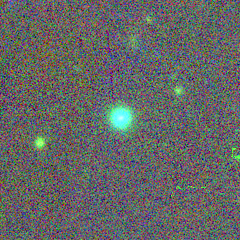 } 
            \put(5,5){\color{white}\bfseries\small giy }
        \end{overpic}%
        \includegraphics[width=0.45\textwidth]{ 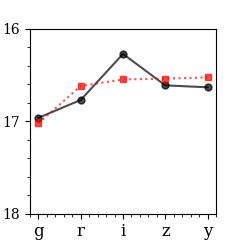 }
    \end{minipage}%
    \begin{minipage}[c]{0.23\textwidth}
        \centering
        \begin{overpic}[width=0.45\textwidth]{ 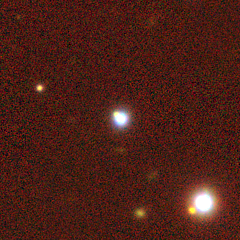 } 
            \put(5,5){\color{white}\bfseries\small giy }
        \end{overpic}%
        \includegraphics[width=0.45\textwidth]{ 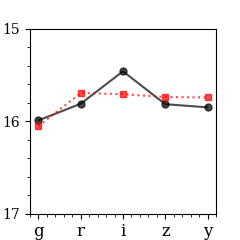 }
    \end{minipage}%
    \\[1ex] 
    \begin{minipage}[c]{0.23\textwidth}
        \centering
        \begin{overpic}[width=0.45\textwidth]{ 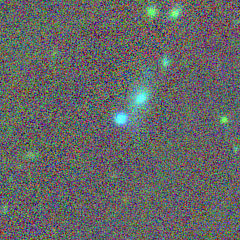 } 
            \put(5,5){\color{white}\bfseries\small giy }
        \end{overpic}%
        \includegraphics[width=0.45\textwidth]{ 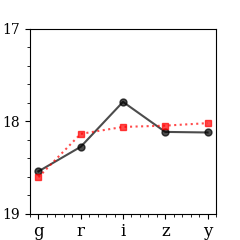 }
    \end{minipage}%
    \begin{minipage}[c]{0.23\textwidth}
        \centering
        \begin{overpic}[width=0.45\textwidth]{ 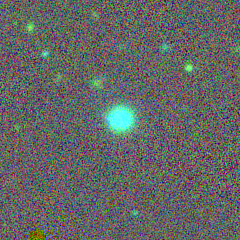 } 
            \put(5,5){\color{white}\bfseries\small giy }
        \end{overpic}%
        \includegraphics[width=0.45\textwidth]{ 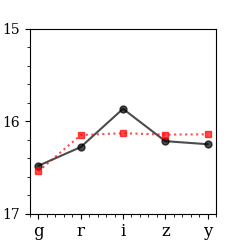 }
    \end{minipage}%
    \begin{minipage}[c]{0.23\textwidth}
        \centering
        \begin{overpic}[width=0.45\textwidth]{ 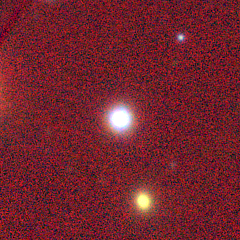 } 
            \put(5,5){\color{white}\bfseries\small giy }
        \end{overpic}%
        \includegraphics[width=0.45\textwidth]{ 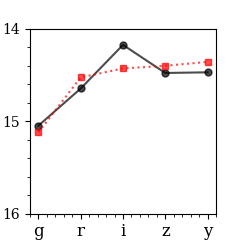 }
    \end{minipage}%

    \caption{PS1 images alongside observed photometry highlighting SED shapes of detected anomalies in which the LSTM-AE underestimates the observed PS magnitude ($m-m^\prime \leq -0.15$). Observed SED shapes in black, reconstructed SED shapes in red. Image Cutouts are 0.25" per pixel with a size of 240 pixels.}
\end{figure*}

\end{document}